\documentclass[acmtog,nonacm]{acmart}

\usepackage{subcaption}
\usepackage{amsmath}
\usepackage{microtype}
\graphicspath{ {images/} }
\usepackage{multirow}
\usepackage{wrapfig}

\usepackage[utf8]{inputenc}
\usepackage{xargs}
\usepackage[colorinlistoftodos,prependcaption,textsize=tiny]{todonotes}
\usepackage{eso-pic}

\usepackage{cleveref}
\crefname{figure}{Fig.}{Figs.}
\crefname{section}{Section}{Sections}
\crefname{subsection}{Sec.}{Secs.}
\crefname{subsubsection}{Sec.}{Secs.}
\crefname{table}{Tab.}{Tabs.}

\AtBeginDocument{%
  \providecommand\BibTeX{{%
    \normalfont B\kern-0.5em{\scshape i\kern-0.25em b}\kern-0.8em\TeX}}}

\setcopyright{acmcopyright}
\copyrightyear{2021}
\acmYear{2021}
\acmDOI{}

\acmJournal{TOG}
\acmVolume{40}
\acmNumber{4}
\acmArticle{0}
\acmMonth{8}

\acmSubmissionID{450}

\usepackage{hyperref}

\usepackage{graphicx}
\usepackage{epstopdf}
\begin{document}

\title{PAVEL: Decorative Patterns with Packed Volumetric Elements}

\author{Filippo Andrea Fanni}
\affiliation{%
  \department{Dept.\ of Computer Science}
  \institution{University of Verona}
  \city{Verona}
  \country{Italy}}
\email{filippoandrea.fanni@univr.it}
\author{Fabio Pellacini}
\affiliation{%
  \department{Dept.\ of Computer Science}
  \institution{Sapienza University of Rome}
  \city{Rome}
  \country{Italy}}
\email{pellacini@di.uniroma1.it}
\author{Riccardo Scateni}
\affiliation{%
  \department{Dept.\ of Mathematics and Computer Science}
  \institution{University of Cagliari}
  \city{Cagliari}
  \country{Italy}}
\email{riccardo@unica.it}
\author{Andrea Giachetti}
\affiliation{%
  \department{Dept.\ of Computer Science}
  \institution{University of Verona}
  \city{Verona}
  \country{Italy}}
\email{andrea.giachetti@univr.it}
%

%\renewcommand{\shortauthors}{, et al.}

\begin{abstract}
Many real-world hand-crafted objects are decorated with elements that are packed onto the object's surface and deformed to cover it as much as possible. Examples are artisanal ceramics and metal jewelry. Inspired by these objects, we present a method to enrich surfaces with packed volumetric decorations. Our algorithm works by first determining the locations in which to add the decorative elements and then removing the non-physical overlap between them while preserving the decoration volume. For the placement, we support several strategies depending on the desired overall motif. To remove the overlap, we use an approach based on implicit deformable models creating the qualitative effect of plastic warping while avoiding expensive and hard-to-control physical simulations. Our decorative elements can be used to enhance virtual surfaces, as well as 3D-printed pieces, by assembling the decorations onto real-surfaces to obtain tangible reproductions.
\end{abstract}




\begin{teaserfigure}
\centering
\includegraphics[width=\textwidth]{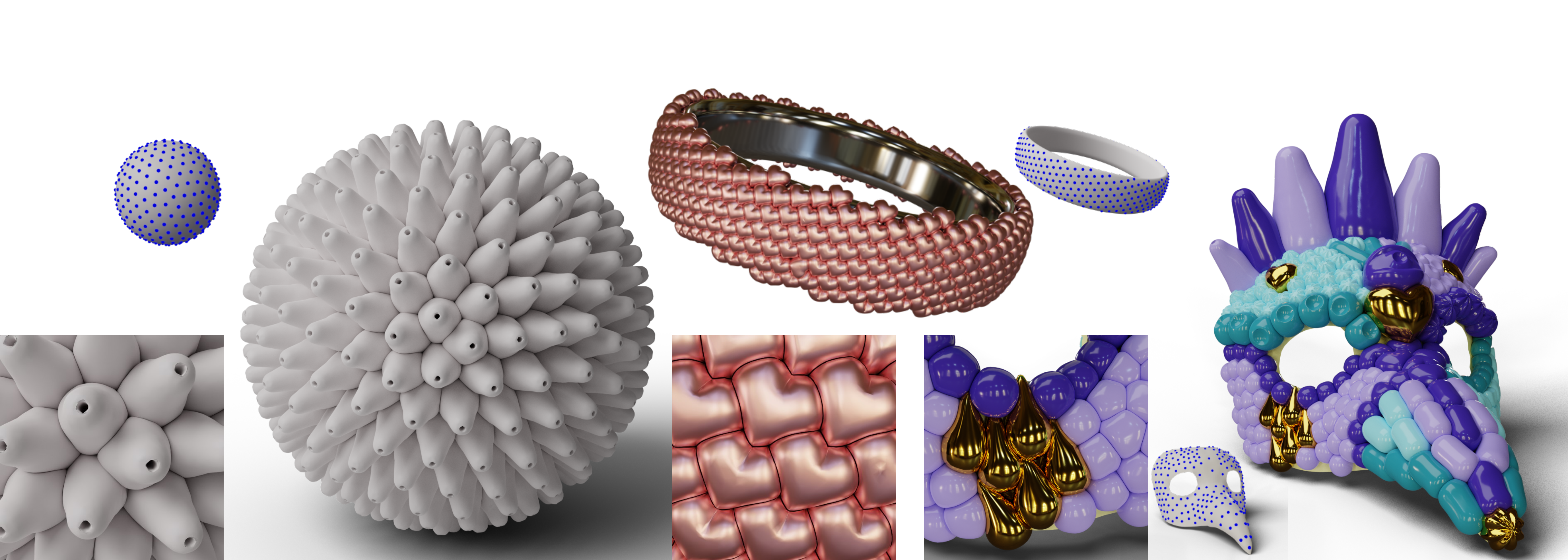}
\caption{Three examples of objects decorated with packed volumetric elements generated by \emph{PAVEL}. On the left, a reproduction of a real-world ceramic artwork (see \cref{fig:examples}, right), in the center, a bracelet, and on the right, a plague doctor masque, inspired by the Venetian carnival tradition. In \emph{PAVEL}, decorations are placed on a surface in an overlapped configuration. We then deforms the decorative elements in a manner that mimics the plastic deformations of real-world artworks. For each model, we show the base shape with, in blue, the positions for the decorative elements. The placement of decorations is automatic in the first two examples, and can also be done manually for better artistic control, as shown in the latter example.}
\label{fig:teaser}
\end{teaserfigure}

\maketitle

\section{Introduction}
\label{sec:intro}

The quest for representing reality is one of the main goals of computer graphics since its origins. Due to the advances in modeling, rendering, and AI, it is often possible to virtually reproduce the look of real-world objects, characters, and scenes, in a manner that is indistinguishable from reality.
Recent advances in digital manufacturing allow us to physically reproduce realistic objects that were initially modeled as digital shapes. This is becoming an important field of application for 3D geometry processing algorithms and techniques.

The difference between a model that appears realistic and one that does not is often the complex surface details typical of real-world objects. In graphics, surfaces are often enriched by varying their material properties, texture maps, or generating relief patterns.
As pointed out in a recent survey \cite{zhang2019computer}, most of the current methods used to generate reliefs patterns work in image-space, are applied mainly on planar objects and cannot reproduce details coming from 3D decorative elements placed onto the surface.

\begin{figure}[tb]
   \includegraphics[height=23mm]{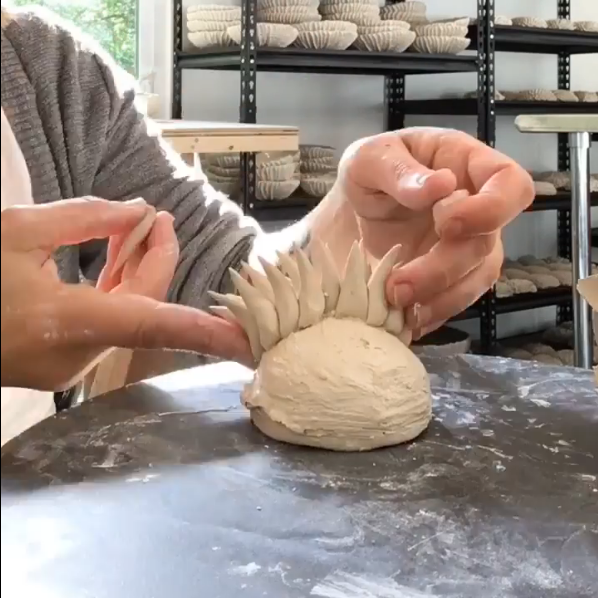}
   \includegraphics[height=23mm]{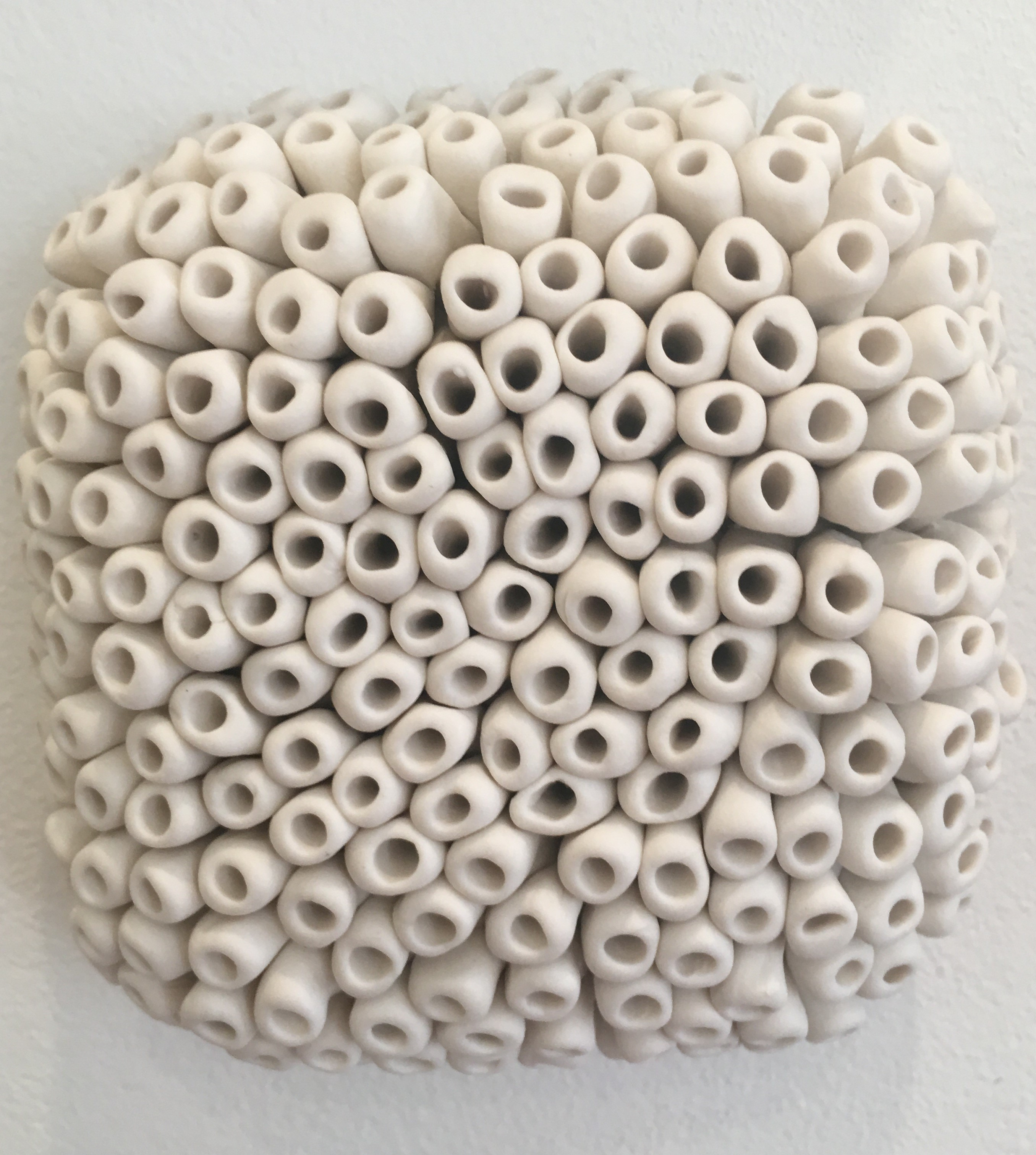}
   \includegraphics[height=23mm]{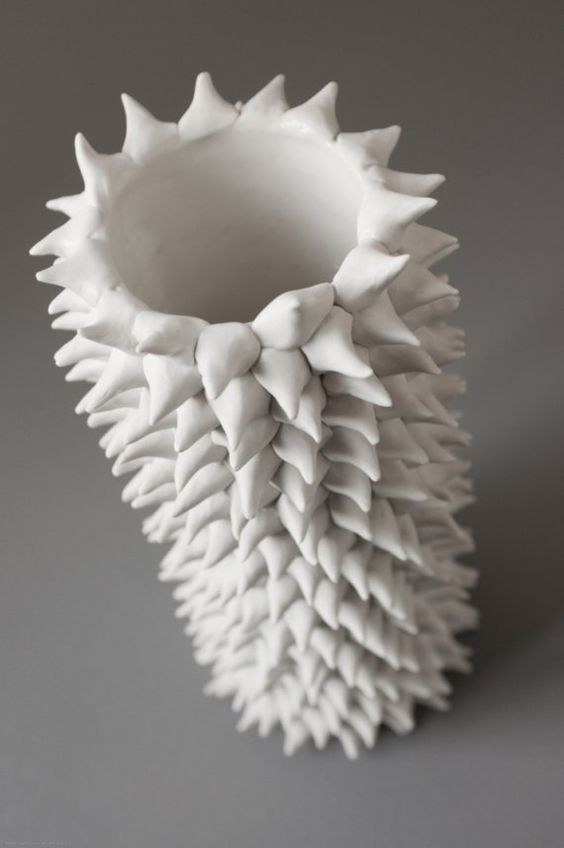}
   \includegraphics[height=23mm]{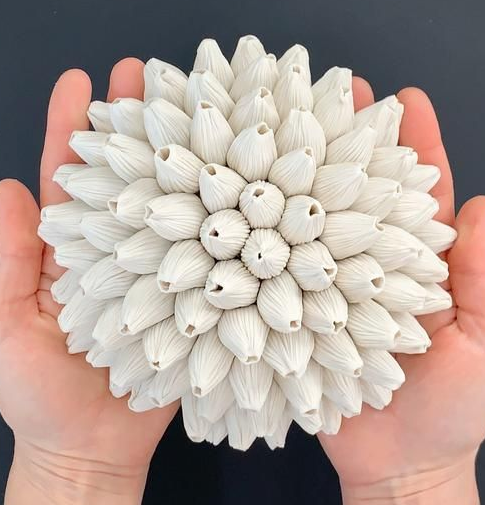}      
\caption{\emph{Left:} An artist creating volumetric decoration on a surface by adding plastic elements over a base surface. \emph{Center and right:} Examples of hand-made pottery created with such a method (Courtesy of Heather Knight).}
\label{fig:examples}
\end{figure}

In this work, we present \emph{PAVEL}, a method for decorating surfaces with \emph{pa}cked \emph{v}olumetric \emph{el}ements. Our method is inspired by hand-crafted artworks, shown in \cref{fig:examples}, that are decorated by manually packing elements onto objects. Each decoration is deformed to better cover the surfaces. These decorations are often found in hand-made ceramic, metal jewelry, and Murano's glass. \cref{fig:teaser} show examples of surfaces decorated with \emph{PAVEL} to mimic these styles.

Current methods cannot reproduce these decorations. Techniques based on surface displacement, such as procedural methods and digital sculpting, cannot reproduce these decorations since they do not model local surface topology changes. Element-based textures cannot be used either since they are generally defined in Euclidean spaces, while we model decorations on manifolds, and since they do not support deformation that maintains the elements volumes, which is required to achieve realism.

\emph{PAVEL} works in two stages. We first place decorations onto the surfaces with considerable overlap between them. In our implementation, we support both the manual placement of the elements that mimics exactly the real-world process and the automatic generation of different kinds of patterns of overlapped elements. Decorations overlap to cover the surface as much as desired by the user. We then deform the decoration to remove the overlap between them. In doing so, we preserve the volume while deforming to simulate real-world behavior. These two stages produce results that are visually similar to real-world decorations, as shown throughout the paper. In \emph{PAVEL}, we make two critical choices to make computation feasible. 

First, we use point sampling to place the decorations' centroids instead of using packing algorithms. In general, packing algorithms are just heuristics since packing is NP-hard in the Euclidean domain and just as hard on the manifold. Point sampling is significantly more efficient, by a couple of orders of magnitudes, and just as reliable in our domain. This choice favors objects having a radial symmetry but does not entirely exclude other entities, as we show in the results. 

Second, to remove the overlap, we do not use a physics engine but an implicit formulation. The use of physics engines to compute plastic deformations would seem appropriate, but it is, in fact, not the case. Physics engines evolve a valid physical configuration through time. In \emph{PAVEL}, we start with a non-physical condition in that decorations are placed with significant overlap between them. We have verified that, in these conditions, physics engines do not converge to the desired solutions, nor should they. Instead, we employ an implicit formulation inspired by \cite{ImplicitContactModeling} to remove intersections and recover the lost volume using a propagation field and fast marching approach.

The main contributions of our work are:
\begin{itemize}
    \item an overall method for decorating surfaces with packed volumetric elements supporting automatic and manual element decorations, and automatic removal of decoration overlap; to the best of our knowledge, this is the first packing method that works on 2-manifold directly; to the best of our knowledge, the decorated surfaces could not be created with any other automatic method and would require hours of manual work if sculpted manually \cite{sculptstat};
    \item an efficient procedure to remove intersections and recover elements' volume that provides a visually effective, even if not physically exact, simulation of a plastic, volume-preserving deformation;
    \item a practical demonstration of a decorative shell for an existing 3D object designed with \emph{PAVEL}, decomposed into printable parts, manufactured and attached to the base shape.
\end{itemize}

\section{Related works}
\label{sec:rela}

To the best of our knowledge, there is no prior work on packing 3D elements on surfaces. Instead, there is a vast literature on packing geometric objects, both in two- and three-dimensions. In this section, we limit our mentions to the more relevant and influential methods to our work.

\paragraph{Packing}
Packing 3D shapes in an arbitrary volume is a well-studied problem, and it is known to be NP-complete. It is relevant in graphics, engineering, operational research, and many other fields. Recent works such as \cite{Packing3D-Egelbed, Packing3D-Liu, Packing3D-Romanova} achieve an excellent volume coverage. Still, they are heavily limited in the orientation of objects, are significantly time-consuming, and can only deal with a limited number of items with few facets. Overcoming these limitations, Ma and colleagues \shortcite{PackingIrregular3D} propose an algorithmic approach well suited to complex models typically used in graphics. Their algorithm initially scales down all the objects to avoid intersections, then iteratively increases their size, optimizes rotations, and tries some object swapping to arrange local solutions. The algorithm can obtain good results but can take up to an hour to finish the computation to place less than 100 objects. Our approach differs from both since we pack elements on surfaces and since we decide element sizes upfront, which makes the problem generally harder.

The packing of 2D shapes is a relevant field of research for visualization and illustration purposes. Reinert and colleagues \shortcite{ByExamplePacking} proposes an algorithm based on Centroidal Voronoi Diagram relaxation to fill an arbitrary-shaped planar domain with multiple shapes. Saputra and colleagues \shortcite{RepulsionPack} also describe an algorithm to pack shapes in a planar domain, with a greater focus on leaving as little space as possible. To achieve their results, they allow deformations of the decorative shapes and adds small ones, in the end, to obtain better coverage and reduce space. For similar purposes, Kwan and colleagues \shortcite{CollageShapes} propose a new shape descriptor and use it to measure the touching shapes' quality. Another relevant method, by Chen and colleagues \shortcite{FabricableTileDecors}, packs tiles on top of a curved surface for fabrication and decorative purposes. They describe an \emph{attract-and-repulse} algorithm in which tiles are allowed to move and rotate, but not to deform, which can take up to fifteen minutes to place more or less 150 tiles. Two works focusing on decorations are \cite{StructuralPatterns}, and \cite{FiligreesSynth}, which investigate the possibility of transforming a surface carving holes following a pattern. In \cite{StructuralPatterns} a texture guides the building of a pattern of connected stripes taking into account the structural problems that can arise when fabricating the final object. In \cite{FiligreesSynth}, the pattern generates filigrees using the skeletal representations of simple starting elements. The described pipeline can then place them on a surface, greedily refine their position, and slightly modify them to ensure contact between close elements without overlaps. 
However, all these methods are inherently different from ours since they treat 2D shapes rather than 3D elements. Furthermore, most of these works employ specific representations that cannot be generalized to 3D.

\paragraph{Point sampling}
Placing decorations with circular symmetry over the surface of an object can be interpreted as a packing problem as well as a point sampling one. 
Blue noise point distributions and their variations can undoubtedly be used to roughly pack a surface with circular shapes. Bridson \shortcite{BridsonBlueNoise} proposes a simple and efficient algorithm to generate a blue noise distribution. Due to its greedy nature, it tends to leave significant gaps, particularly undesirable for artistic purposes. A more uniform distribution is given by a Centroidal Voronoi Diagram, which can be computed over arbitrary surfaces as described in \cite{CVDMeshing}. Knöppel and colleagues \shortcite{StripePatterns} introduce a special kind of entity distribution over a surface putting regular striped patterns that can adapt to the local characteristics of the shape, curvature included. This method is useful guidance for scattering points on the surface of the form only onto the stripes. We make use of the latter two methods to place seeds for decorative elements in our pipeline.

\paragraph{Implicit deformable models}
Algorithms to control deformable surfaces based on level set formulation are popular in Visual Computing after the seminal works by Osher, Sethian, and Malladi \shortcite{osher1988fronts,malladi1995shape}. Using implicit formulations, it is possible to model complex interface behaviors, including elastic and plastic deformations, but with huge computational complexity and stability issues \cite{barton2011conservative}.
A simplified approach is represented by the fast marching algorithm \cite{sethian1996fast}, which allows a fast estimation of the evolution of closed surfaces as a function of an underlying velocity field assuming unidirectional front propagation along the surface normals. By controlling the underlying speed, it is possible to drive the surface evolution towards the desired position.
Simplified implicit deformable models have been used in Computer Graphics to simulate soft bodies' contact, and the resulting deformation \cite{ImplicitContactModeling} and to animate deformable shapes \cite{desbrun1998active}.
A similar approach has also been recently exploited in a generalized framework to interactively control the blending of two shapes represented by signed distance fields (SDF) on regular grids \cite{SketchBasedImplicit}.
We adapt the fast marching approach to creating the effect of plastic-like volume-preserving deformation of multiple objects efficiently in our work. 

\paragraph{Surface extraction}
The most common approach to extract a mesh representation from a discrete scalar field is the Marching Cubes algorithm \cite{MarchingCubes}. When the scalar field is Boolean, an additional smoothing step is often necessary. Coeurjolly and colleagues \shortcite{VoxelRegularization} propose an optimization-based approach to extract a smooth manifold from a voxel representation to avoid this. This method works directly on the voxel field's boundary quads, moving the vertices to optimize fidelity, fairness, and smoothness. We do not use this method in our reconstruction since it is unacceptably slow for resolving our voxel grids. It produces overly smoothed results, which removes details in small decorative elements. We instead use marching cubes and a smoothing remesher to convert the deformed decorations back to a surface. 
\begin{figure}[tb]
\centering
\small
\begin{tabular}{@{}c@{\hspace{0.03in}}c@{\hspace{0.03in}}c@{}}
\includegraphics[width=0.32\linewidth]{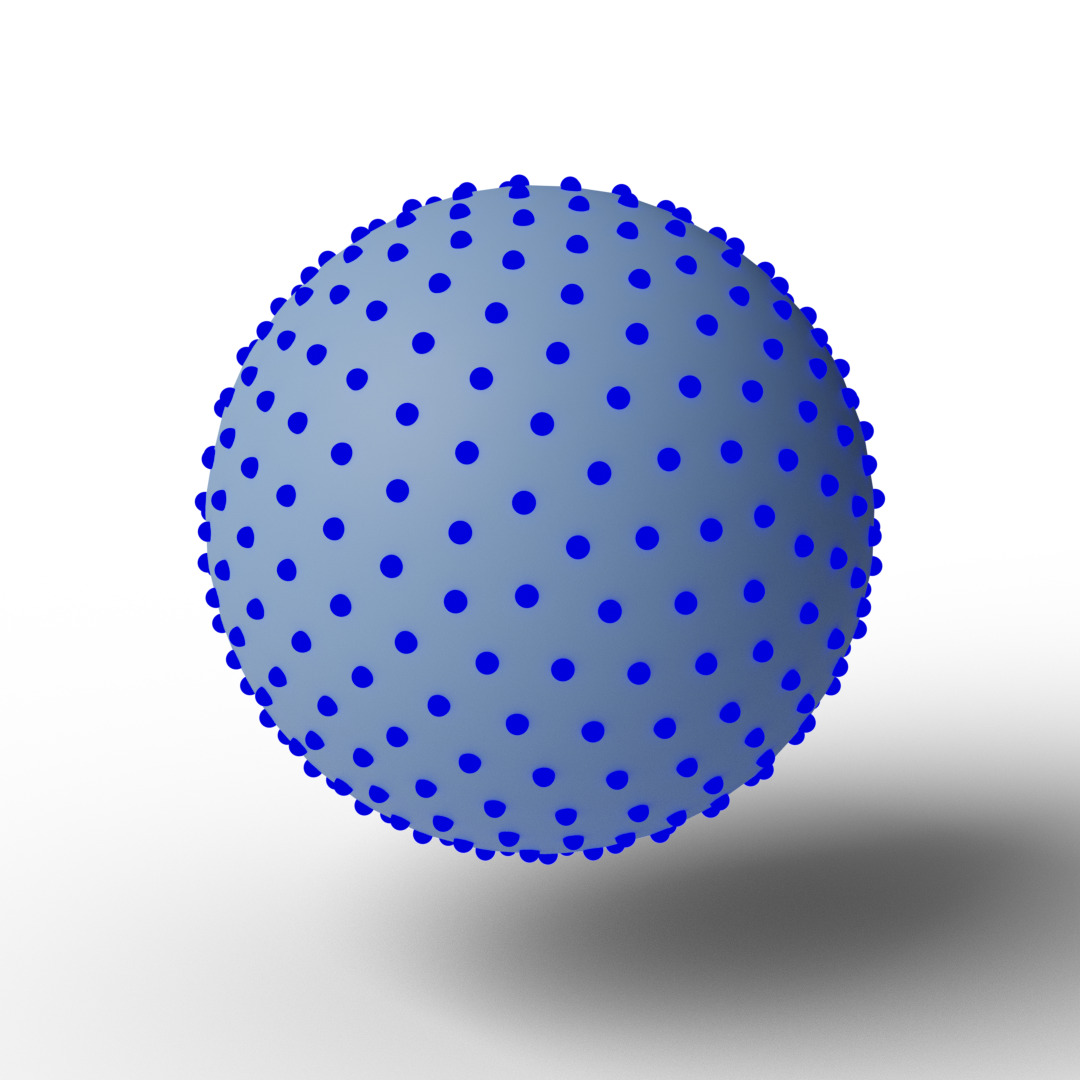} &
\includegraphics[width=0.32\linewidth]{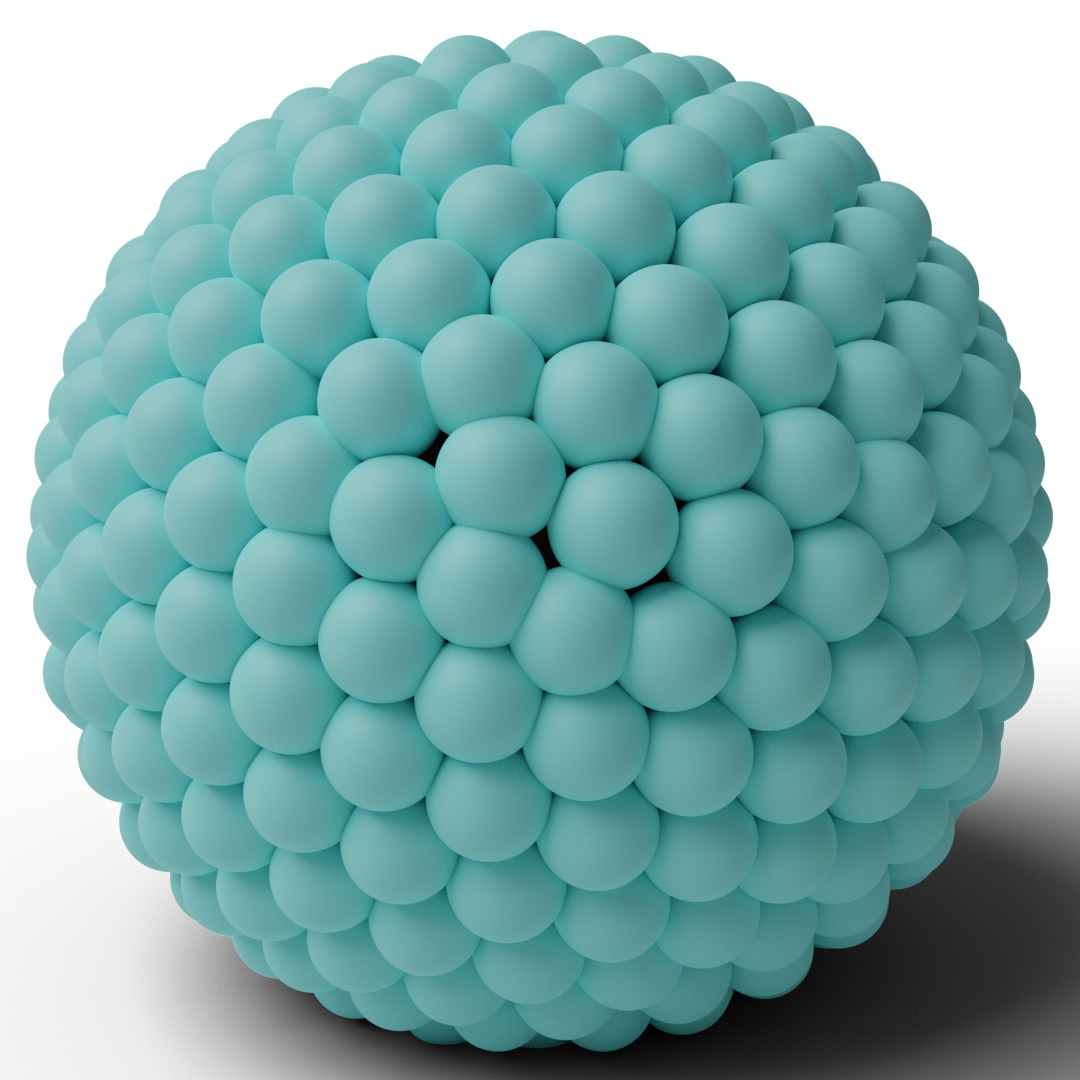} &
\includegraphics[width=0.32\linewidth]{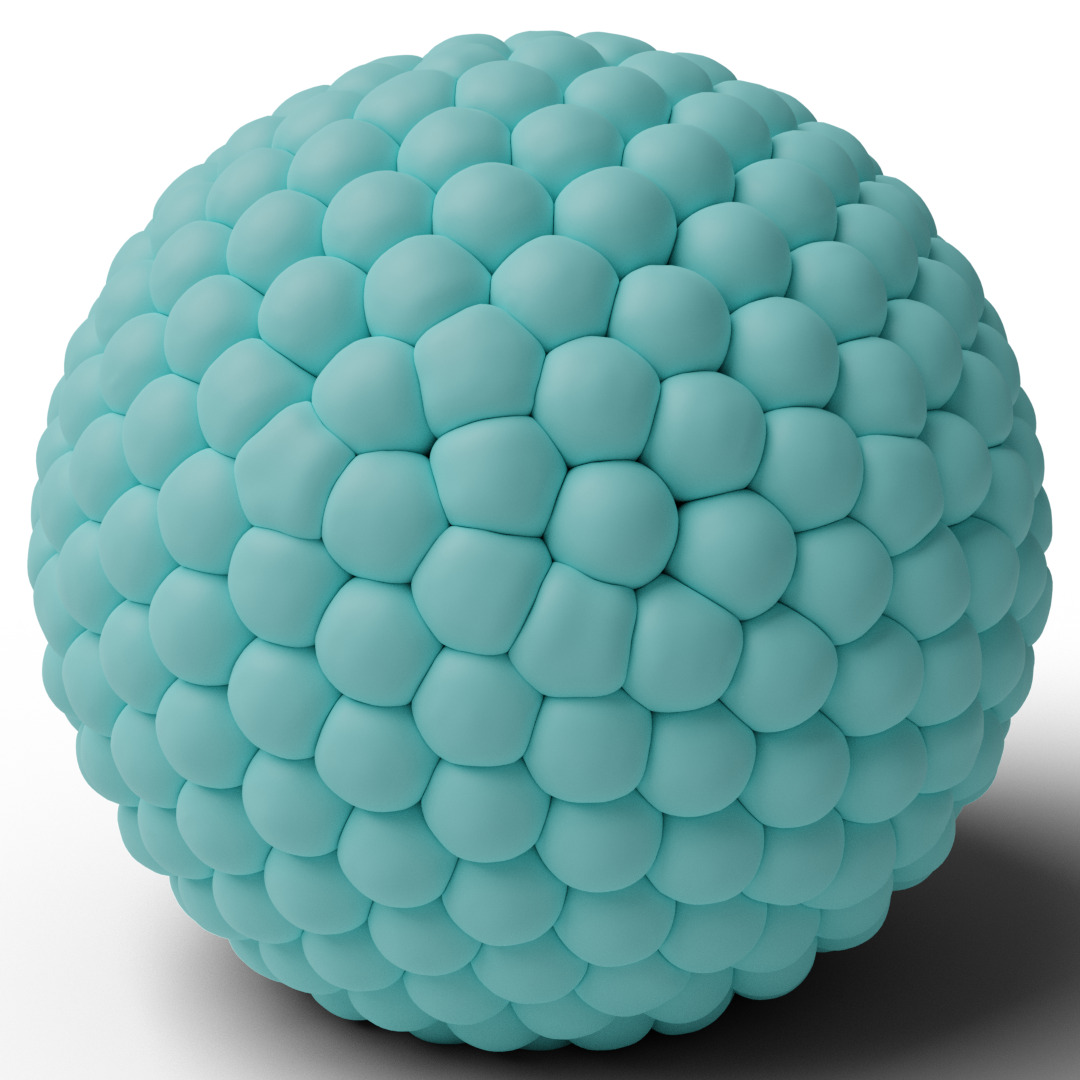} \\
\textsf{Sampled points} & \textsf{Initial placement} & \textsf{Packed decorations} \\
\end{tabular}
\caption{The proposed pipeline of \emph{PAVEL}. Automatic seed placement can create initial configuration of similar objects placed over the base shape with a partial overlap. The deformation module creates a final shape simulating packing of plastic, volume preserving elements, with some control on the material behavior. The final result can be exported as a novel shape or as a collection of printable shell elements.}
 \label{fig:decopipeline}
\end{figure}

\section{Overview}
\label{sec:over}

Our goal is to easily create element-based relief patterns over arbitrary surfaces. Our algorithm takes as input a base mesh, a mesh for each decorative element, and a small set of parameters, defined in the later Sections, that control the elements' distribution and deformations. Our method, shown in \cref{fig:decopipeline}, works in two stages. We first generate a point distribution over the base surface such that the decorative elements cover the surface well. In the second step, elements are placed in the location previously defined, and our algorithm computes the elements' deformations to eliminate the overlaps among them.

To achieve realism and mimic real-world objects, we position the decorative elements so that they are partially overlapping and, then, transform the elements' shapes. They are non-overlapping at the end while preserving their volume. This approach simulates the plastic deformations that artisans perform on their creations. While, at first sight, using a physical simulator seems the proper method to address the second step, we have found that they do not work in our case. The first concern is that simulators are very slow at our resolution and number of collisions. More importantly, though, the simulation does not necessarily converge to a meaningful state since we start with significant overlaps with a non-physical nature. Instead, we propose a new method based on an implicit representation of the elements and the Fast Marching Method (FMM)  \cite{sethian1996fast}, that can generate physically-plausible plastic deformations with bulges near the contact surfaces. We discuss this algorithm in \cref{sec:defo}.

We propose two strategies for the automatic placement phase that enable users to create both isotropic and anisotropic distributions with simple control parameters. These are respectively based on the constrained Voronoi algorithm proposed by \cite{CVDMeshing} and the equidistant stripe patterns from \cite{StripePatterns}. In case artists want direct control over the decoration placements, they may manually do it.
Additionally, in the case of base surfaces with high-curvature regions and decorations with dimensions in the same order of magnitude of the base curvature radius, we allow the user to sample an isosurface at offset $o$ to produce a more even distribution. 
We then use these points to center the decorations on them and produce intersections among adjacent elements. This step, and all its parameters, is discussed in \cref{sec:placement}.

As the pipeline aims to generate object design both for rendering and manufacturing purposes, we discuss in \cref{sec:output} how to extract smooth meshes from the voxelized elements' sets and how to generate printable shells for the decoration of real objects. A gallery of examples of decorations obtained with \emph{PAVEL} is shown in \cref{sec:results} with an analysis of the system performance, while  \cref{sec:disc} evaluates limitations and possible improvements of the approach.
\section{Deformation of packed decorations}
\label{sec:defo}

In this section, we describe how to compute the final deformations. Our algorithm takes as input a set of \emph{overlapping} element decorations placed onto an object's surface. Our goal is to remove the overlap by computing each decorative element's deformation while maintaining its volume to simulate the plastic deformations performed in real-life by artists.

\begin{figure}[tb]
\centering
\small
\begin{tabular}{@{}c@{\hspace{0.03in}}c@{\hspace{0.03in}}c@{}}
\includegraphics[width=0.32\linewidth]{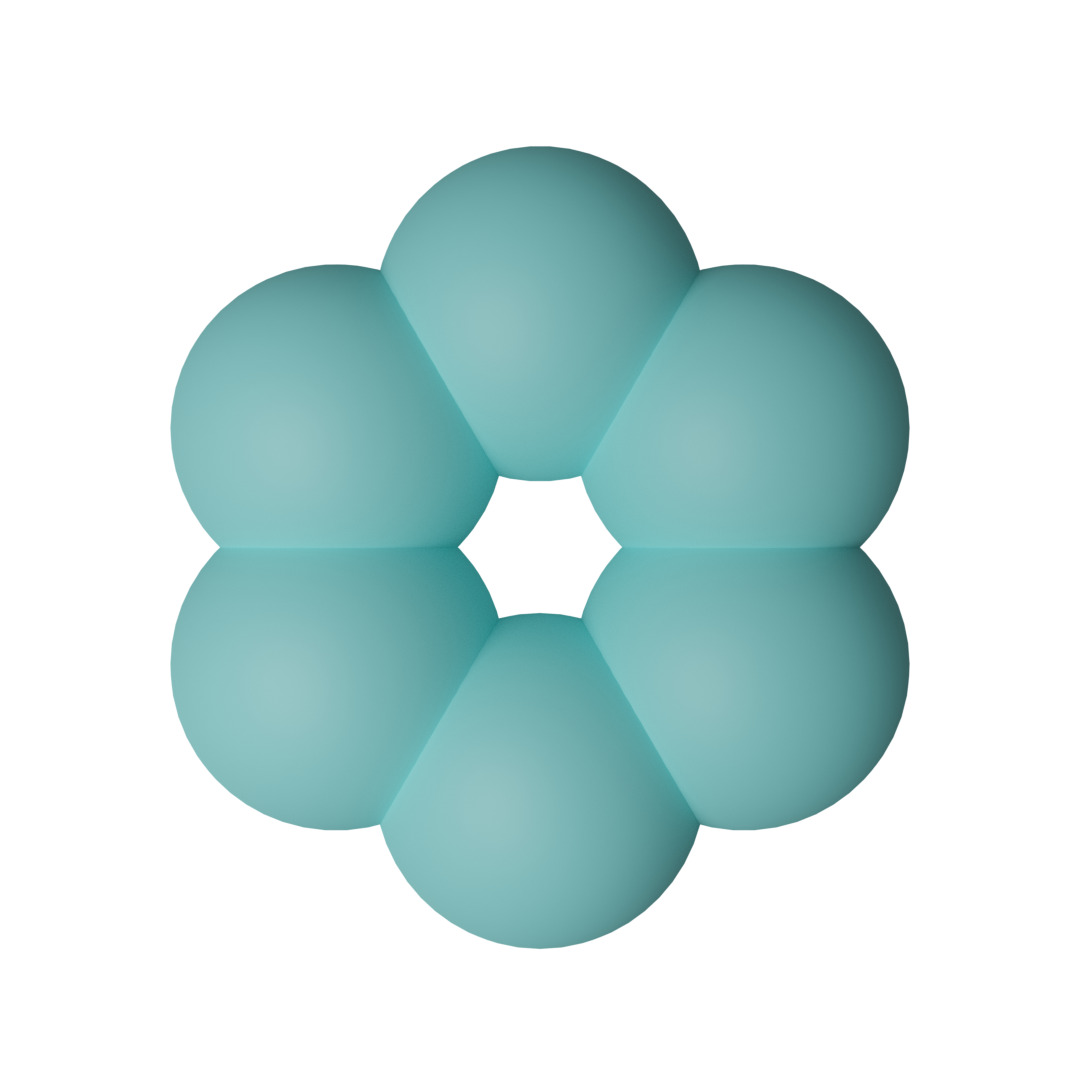} &
\includegraphics[width=0.32\linewidth]{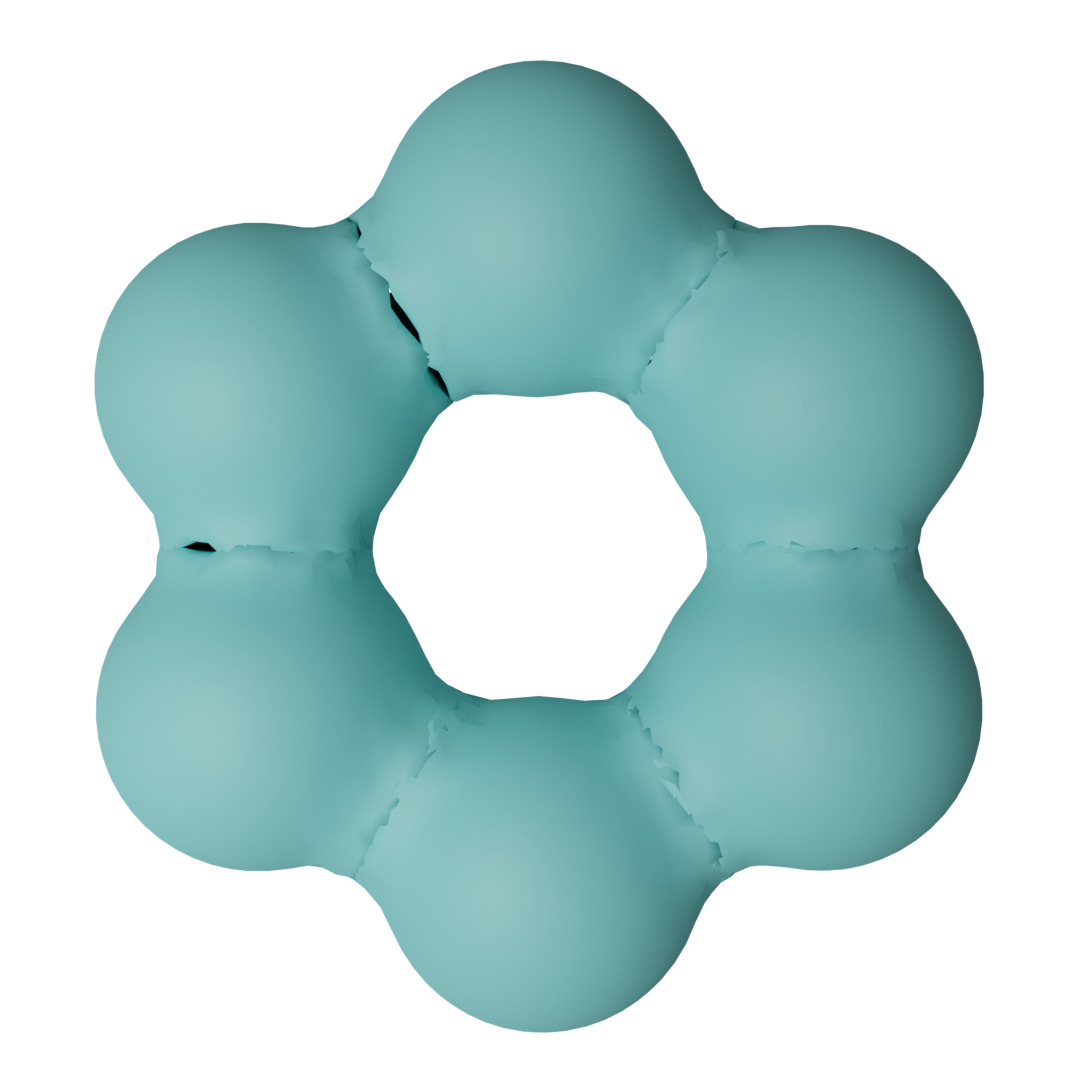} &
\includegraphics[width=0.32\linewidth]{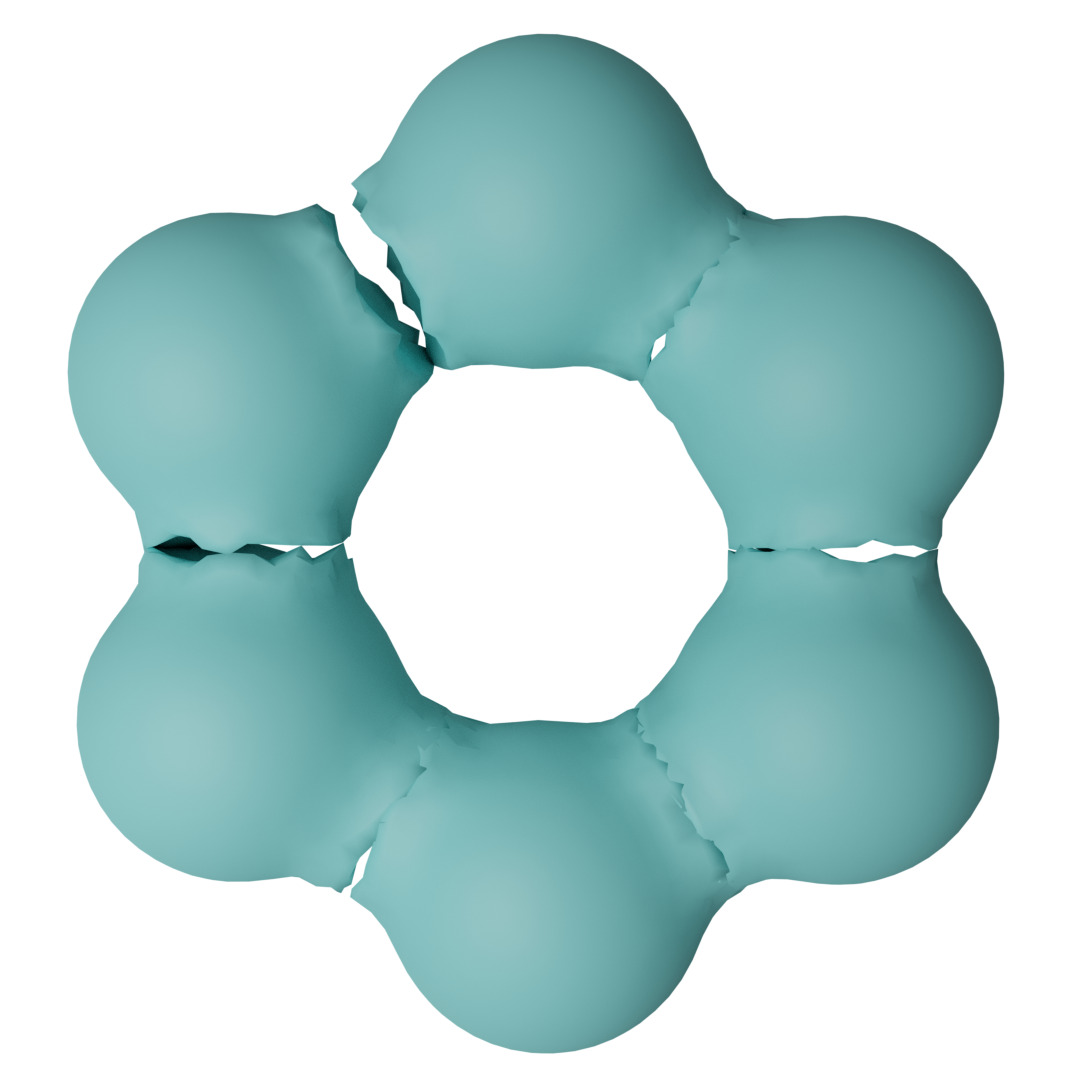} \\
\textsf{Initial overlapped} & \textsf{Artifacts} in & \textsf{More artifacts} \\
\textsf{configuration} & \textsf{packed deformations} & \textsf{if not stopped} \\
\end{tabular}
\caption{Simple example of attempting to remove overlap with a FEM simulator \cite{Houdini}. Starting from an initial condition with overlaps and forces to pin the decorations (left), the simulator attempts to resolve the overlap but breaks the decorations' shapes (middle), which becomes more noticeable if we let time increase (right). Adding more objects or manually applying correcting forces made the problems worst. We tested other simulators that exhibits artifacts that are significant, but different in nature (see text).}
\label{fig:enginesfailures}
\end{figure}

\paragraph{Comparison with Physics Engines}
To obtain the deformed elements, we cannot use existing physics engines for several reasons. First, most engines cannot accept an initial condition with large overlaps between objects, which in our case is as high as 30\% in volume. Second, complex custom forces should be defined to ensure a reasonable equilibrium, e.g., the forces that sculptors apply to get the precise deformation they desire while making sure that decorations adhere to each other.
We tested multiple physics solvers, including Houdini's FEM and XPBD solvers \cite{Houdini}, Blender's solver \cite{Blender}, and a projective dynamics solvers \cite{Vivace}. In all cases, we could not successfully remove the overlap for the reasons listed above. 
Figure \ref{fig:enginesfailures} shows a failure case.
While physics engines' behavior is reasonable, as we start the simulation in a non-physically-plausible configuration, we cannot rely on these tools for our goals.

\paragraph{\emph{PAVEL} deformations}
Since we cannot use physical based simulators, we propose a novel method based on an implicit formulation, inspired by works on interactive visual simulation of volumetric object contact \cite{ImplicitContactModeling}. We rely on fast marching \cite{sethian1996fast} using a specifically designed parametric function for the propagation field that determines the deformed shape appearance and enforce volume preservation.

We voxelize each decoration $d_i$ in its own dense grid $g_i$ that is $50\%$ larger than the decoration's dimension to ensure that the computed deformation will fit in the grid. Compared to the whole surface, this is a sparse discretization that keeps memory consumption relatively low since we are effectively discretizing a slab around the surface. While each decoration has its own representation, we ensure that the voxel dimension and the grid rotation is consistent across all voxelizations. As a consequence, we can perform Boolean operations just by keeping into account their mutual offsets. Additionally, separate voxelizations, instead of one global labeled grid, allow for simple handling of the overlaps and compatibility with the fast marching algorithm.

Given as input a set of voxelized decorations $D = \{d_0,d_1, \dotsc , d_n\}$, our first step is to resolve the overlaps. 
We do this by detecting the voxels shared by multiple decorations and removing them from all but the one with the shortest distance from the decorative element center.
Furthermore, we remove all the voxels of the decorations shared with the base objects as well.
These removals generate a new set of decorations $D'$, where each decoration has associated a measured volume loss $v_i$, equal to the sum of the voxels removed in the previous steps. We then ensure that each element recovers its original volume by expanding it in a way that simulates a plastic deformation.

\paragraph{Fast marching approach}
To compute the expansion, we use a fast marching approach \cite{sethian1996fast} to grow each element in the empty regions until they recover the original volume. To simulate a plastic deformation, which typically generates bulges near the soft bodies' contact points, we use a non-constant velocity field to drive the fast marching algorithm.
For each shape $d'_i$, we find the contact surfaces with the other shapes, and from them, we build the respective distance fields. The distance fields are then converted in velocity fields $F_i$ through a parametric formula, the \emph{volume recovery function}, that can be adapted to obtain the desired behavior. The recovery function is a piece-wise polynomial, obtained by connecting three second-degree polynomials and enforcing $C^1$ continuity, and defined by:
\begin{equation*}
f(a,b) =
     \begin{cases}
       0 & x \le 0 \\
       -\frac{x^2}{a^2}+2\frac{x}{a}& x > 0, x<a\\
       -\frac{2(x-a)^2}{(b-a)^2}-1&x > a, x\le\frac{a+b}{2}\\
       \frac{2(x-a)^2}{(b-a)^2} + b (x-a) +2 & x > \frac{a+b}{2}, x \le b\\
       0 & x > b
     \end{cases}
     \label{eq:distance2speed}
\end{equation*}

In the inset, we show an example plot of the function. We keep the velocity null in contact points to simulate the constraints related
to the attachment of surfaces. We then have two control points $a,b$ used to define the distance for maximal bulging and the maximal distance reached by the plastic wave. This allows effective control of ``material behavior'' driven by the amount and locality of the \begin{wrapfigure}[9]{l}{.4\columnwidth}
\includegraphics[width=.45\columnwidth]{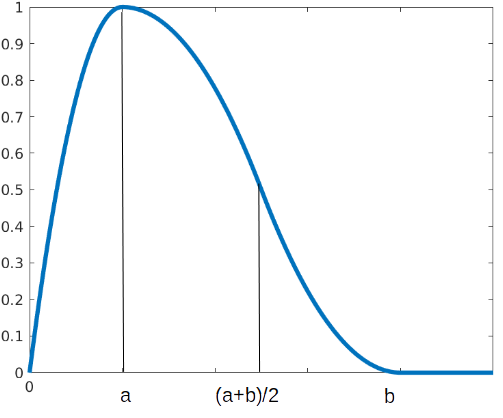}
\end{wrapfigure}
deformation.

The velocity fields are then weighted proportionally to the volume losses to simulate a faster growth of the elements more ``squeezed'' by the neighboring ones and used to compute the new arrival times for each decoration. 
To recover the volume lost by each decoration, we evolve the front of each decoration with the fast marching method. For each elements' sub-grid $g_i$, a set of active front voxels is initialized, taking those adjacent to the initial voxelization border. In contrast, we set all voxels belonging to either the initial shape, the base shape, or other elements to inactive. Each decoration can recover its volume only by occupying empty voxels.

The algorithm then iterates over all the elements. At each step, it evaluates the arrival time for all the voxels of the active front. It selects the voxel with a minimal arrival time according to the Eikonal equation's solution estimated on the base of the velocity field defined in the corresponding grid.

The selected voxel is then assigned to the current element, and it is set as inactive for the other elements' front propagation. The active front is finally updated, removing the selected voxel from it and adding to it the voxel's free neighbors.
This procedure stops when all the decorations have recovered their original volume.
It is theoretically possible that the algorithm could not find enough voxels to recover the volume when the evolution of neighboring elements occludes the front, but this is not likely to happen for a layer of small decorations over a base surface and never happened in our experiments.

The method ensures by construction that all volume is correctly maintained while deforming the decorations. The algorithm is fast since it greedily iterates over elements instead of estimating all the fronts' arrival times at once.
Fast marching is not easily parallelizable to speed up the computation. Still, as we simplify the decorations' distance fields as independent of each other, we can concurrently compute them for a significant performance gain.
We use the fast marching of \cite{scikit} and handle IO and voxelization with \cite{trimesh}.

\begin{figure}[tb]
\centering
\small
\begin{tabular}{@{}c@{\hspace{0.03in}}c@{\hspace{0.03in}}c@{}}
\includegraphics[width=0.32\linewidth]{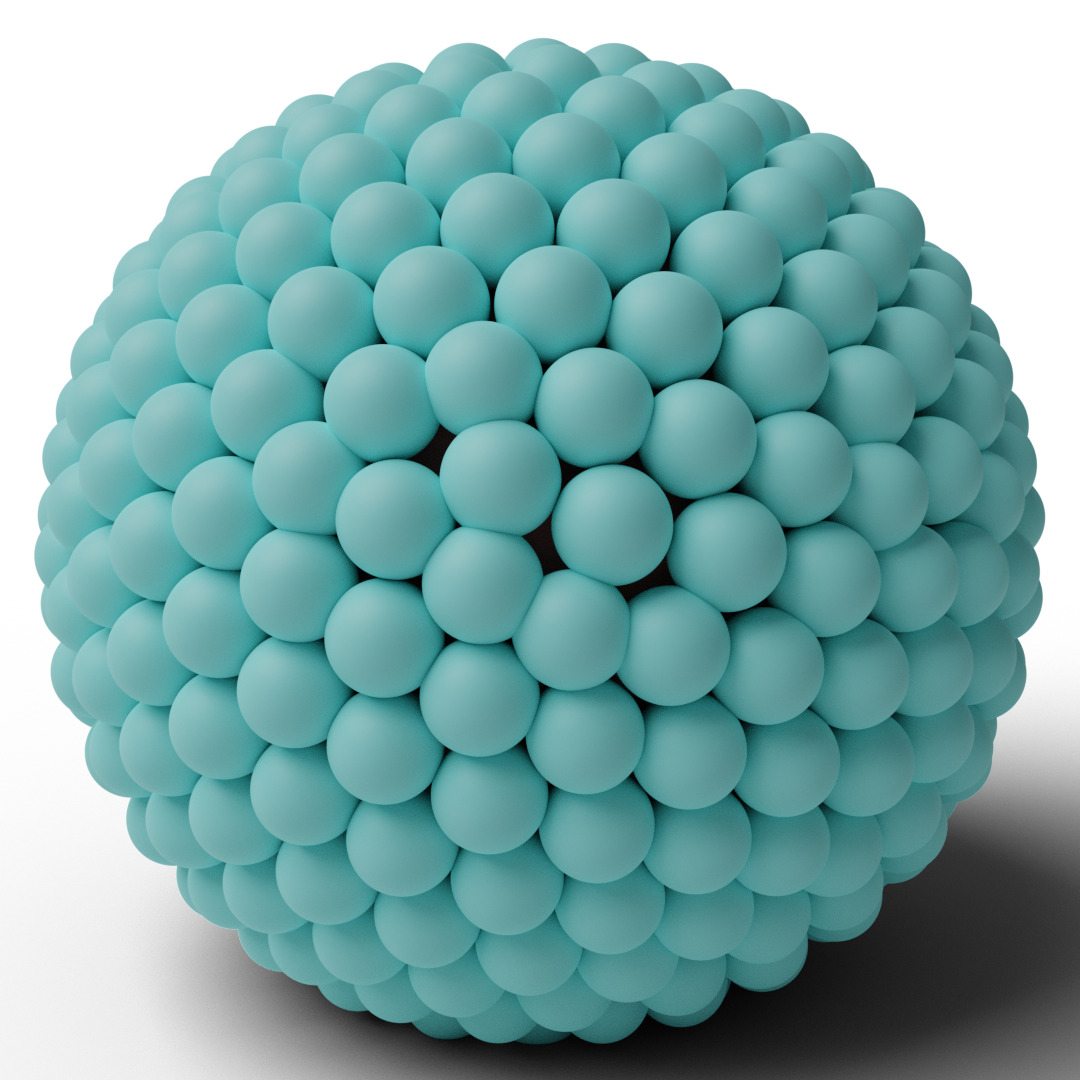} &
\includegraphics[width=0.32\linewidth]{parameters/surfcov/0003} &
\includegraphics[width=0.32\linewidth]{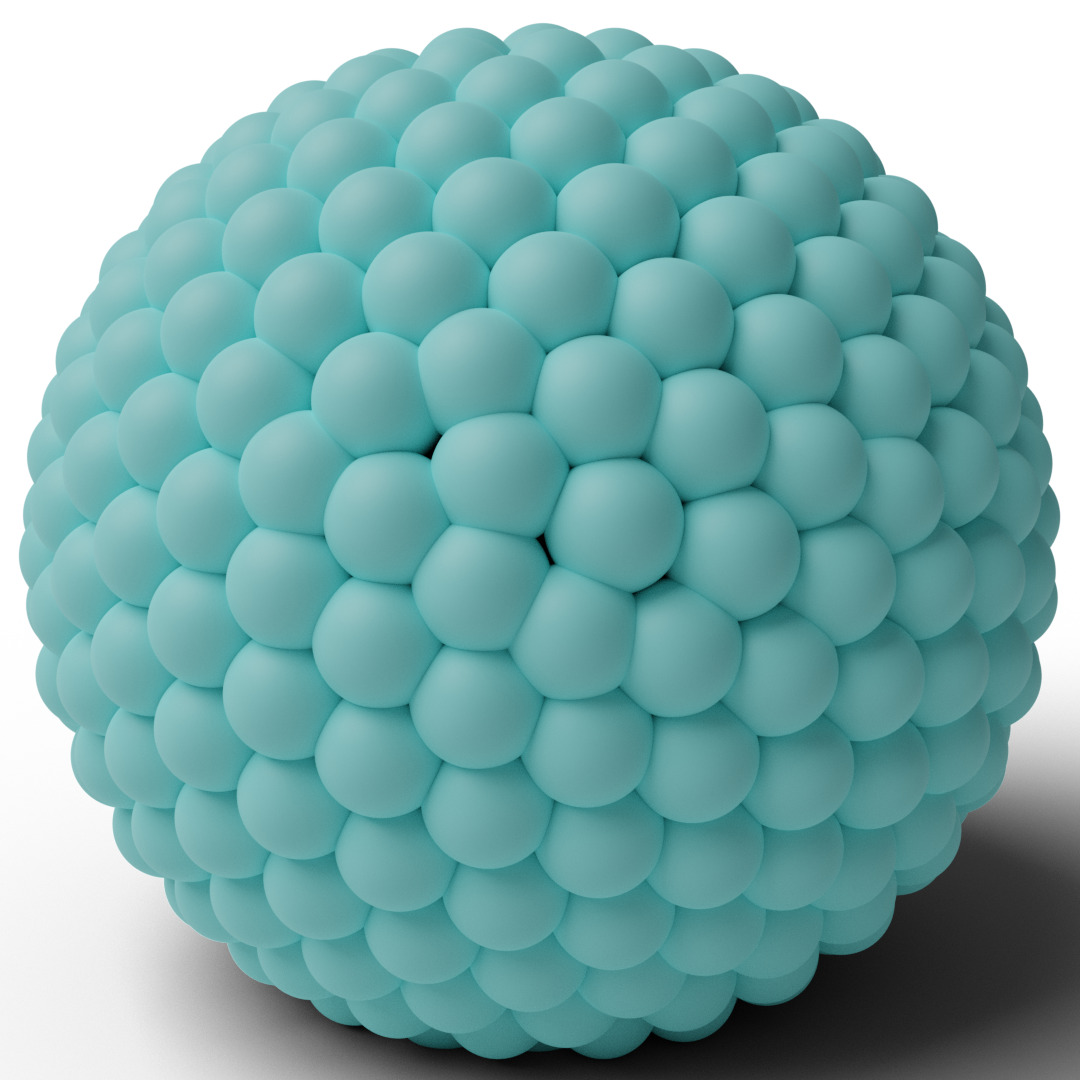} \\
\textsf{coverage} $= 1.2$ & \textsf{coverage} $= 1.4$ & \textsf{coverage} $= 1.6$ \\
\includegraphics[width=0.32\linewidth]{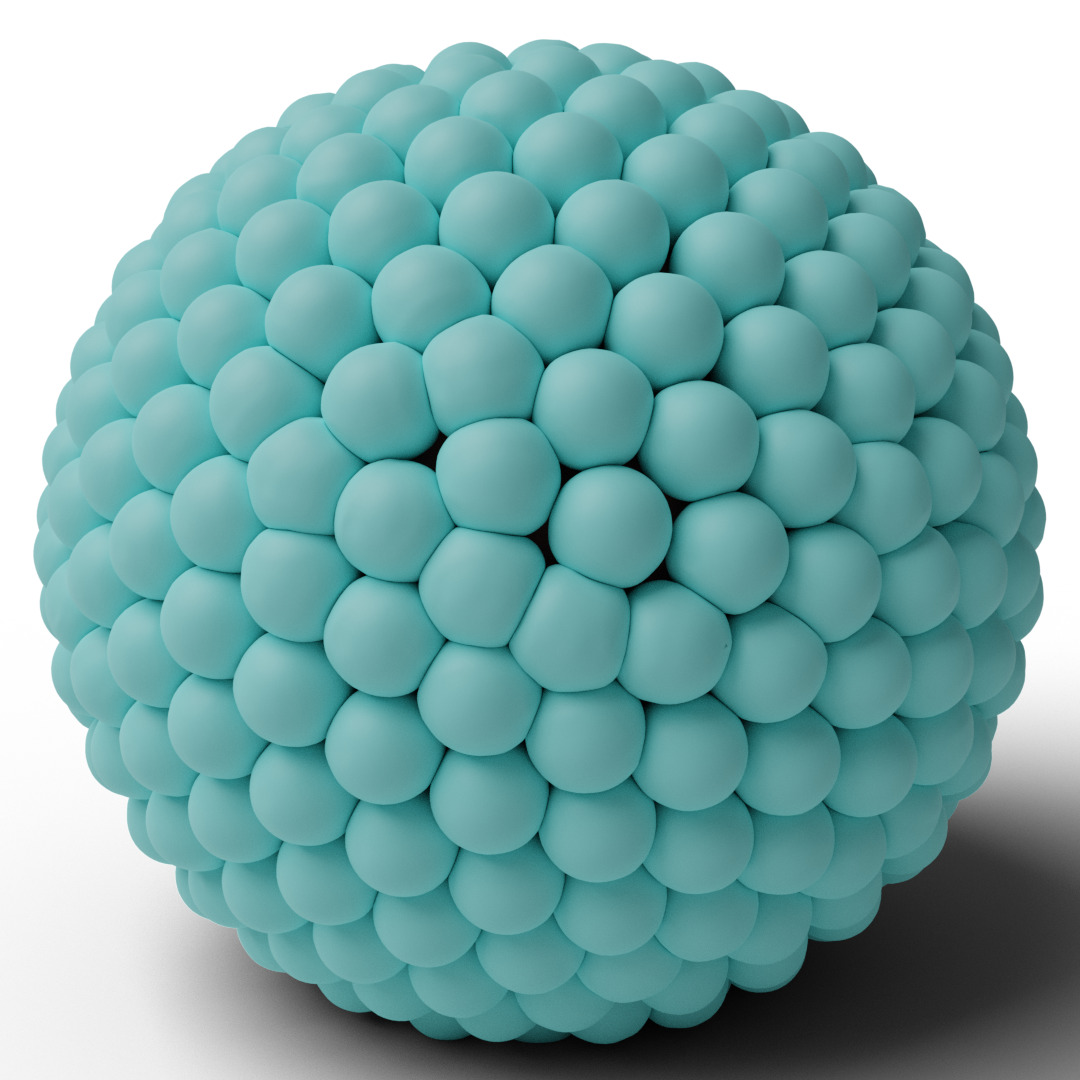} &
\includegraphics[width=0.32\linewidth]{parameters/surfcov/0004} &
\includegraphics[width=0.32\linewidth]{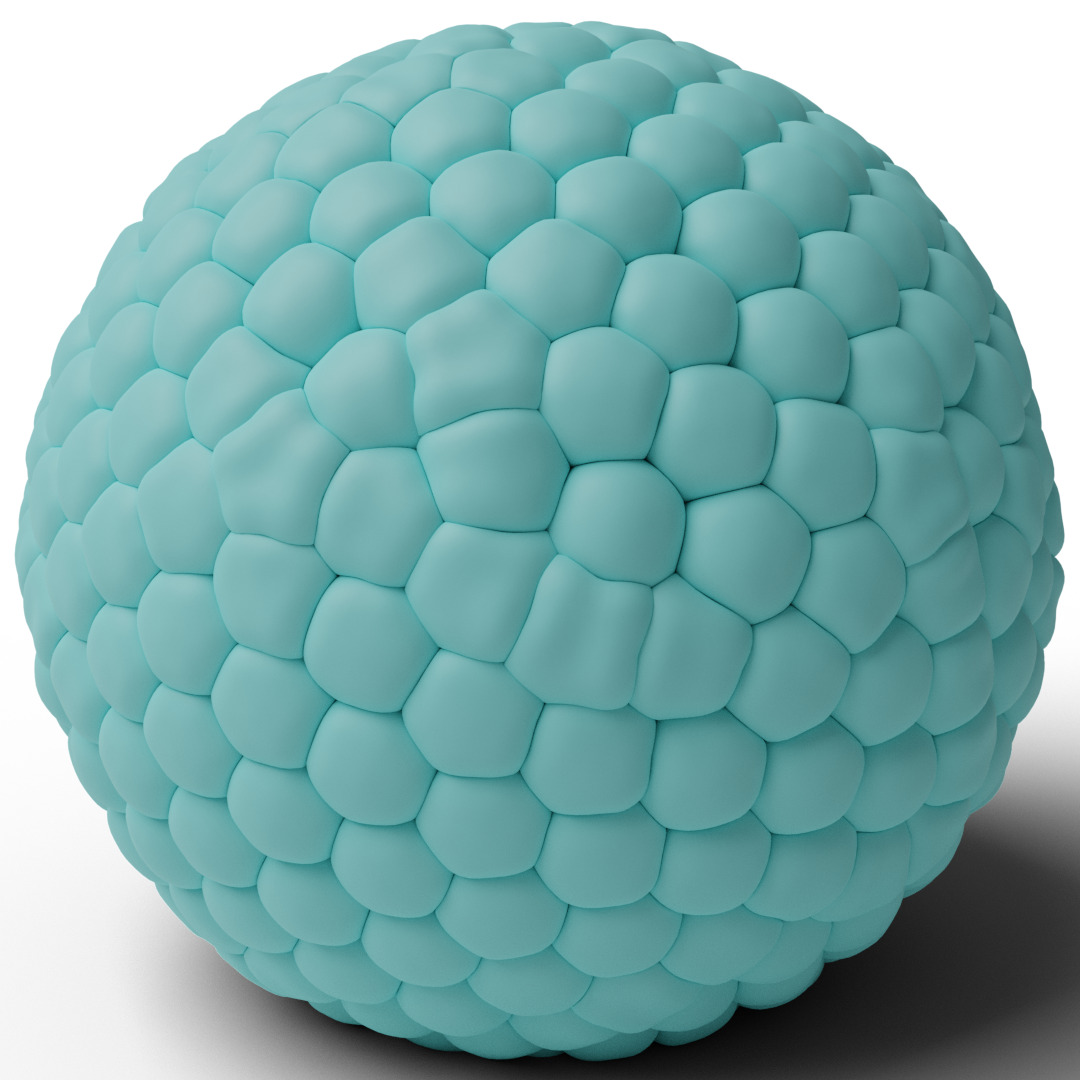} \\
\end{tabular}
\caption{The effect of different values for surface coverage on the final decoration. \emph{Top:} initial placement. \emph{Bottom:} packed elements. }
 \label{fig:surf_coverage}
\end{figure}

\paragraph{Surface coverage control}
The method described above can simulate the elements' plastic deformation if the decorative elements are initialized with a reasonable overlap. To understand what is a ``reasonable'' overlap, we performed several tests to define control parameters for the point distribution. To control the decorations' overlap, we choose to use the \emph{surface coverage}, that is, the ratio between the sum of the surface areas of the decorations' footprints and the base object area to be decorated. \cref{fig:surf_coverage} show results of packing small spheres on a base sphere with different surface coverage values.
As we can see, a ratio of 1.2 is barely enough to cause some deformation, while a ratio of 1.6 produces excessive and unrealistic distortions of the elements' shape.

\begin{figure}[b]
\centering
\small
\begin{tabular}{@{}c@{\hspace{0.03in}}c@{\hspace{0.03in}}c@{}}
\includegraphics[width=0.3\linewidth]{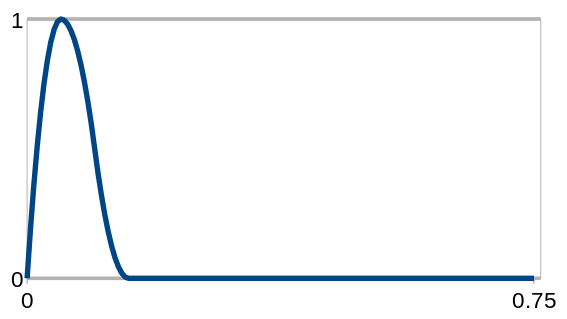} & 
\includegraphics[width=0.3\linewidth]{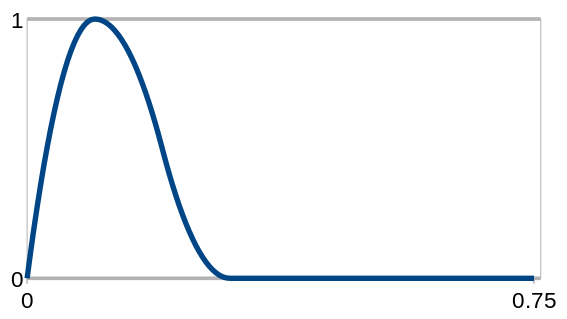} &
\includegraphics[width=0.3\linewidth]{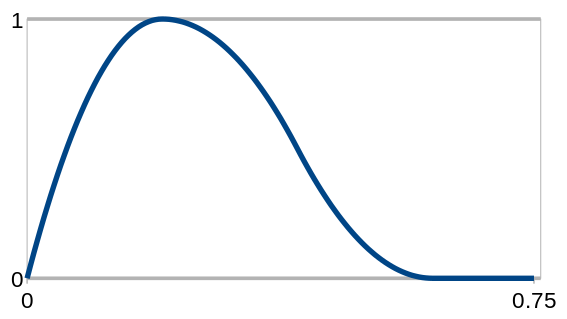} \\
$a=0.05, b=0.15$ & $a=0.1, b=0.3$ & $a=0.2, b=0.6$ \\
\includegraphics[width=0.3\linewidth]{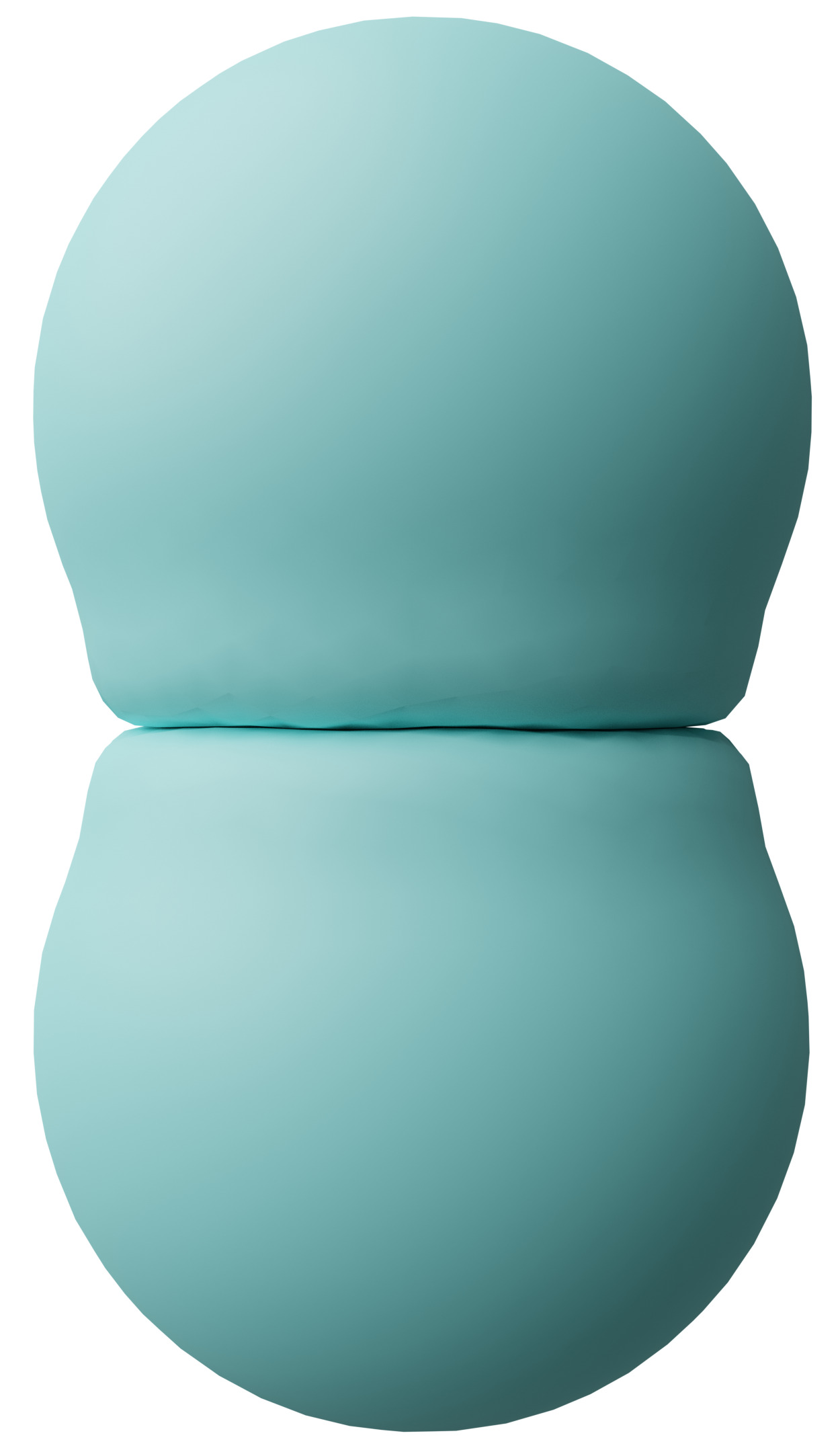} &
\includegraphics[width=0.3\linewidth]{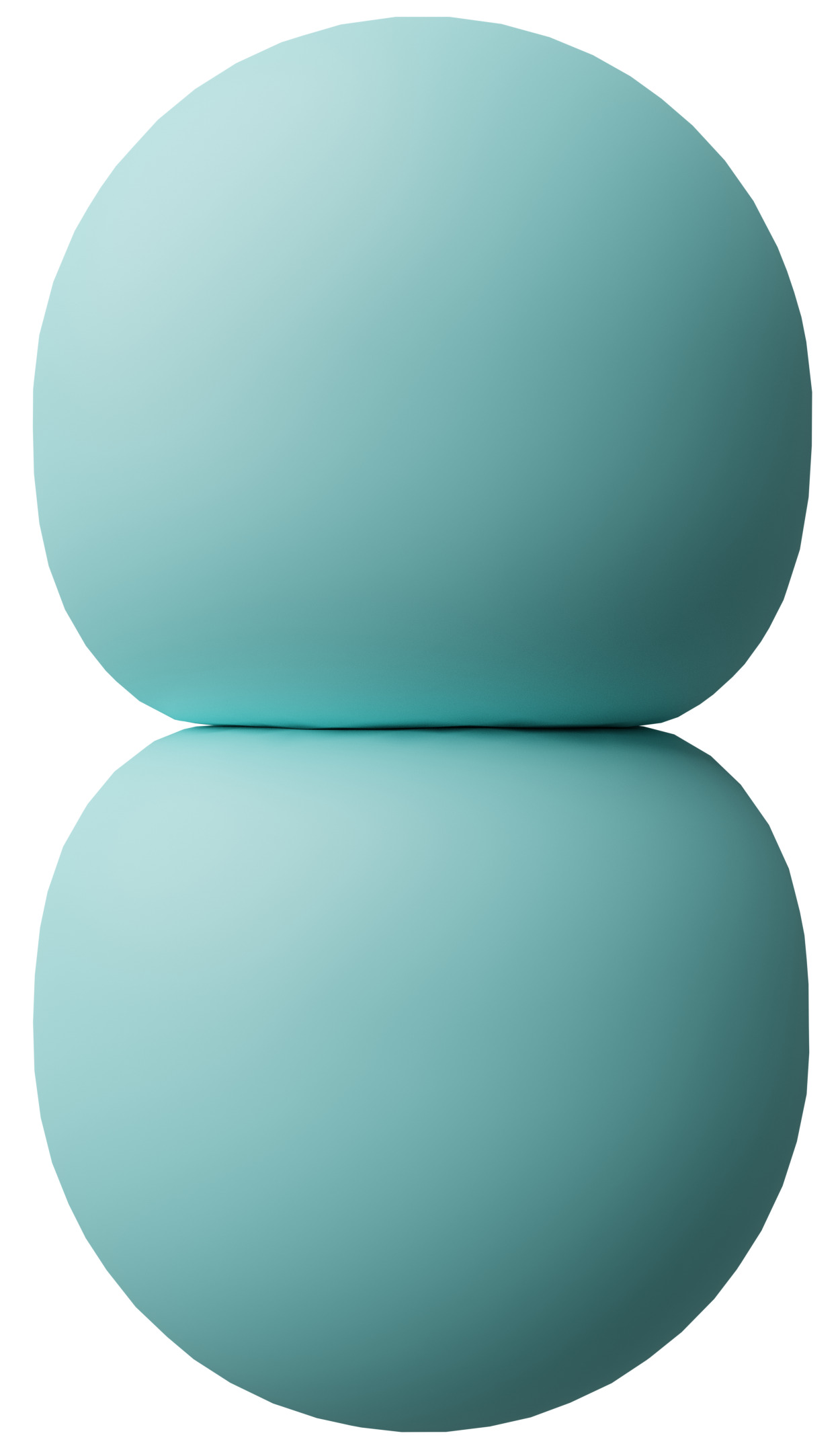} &
\includegraphics[width=0.3\linewidth]{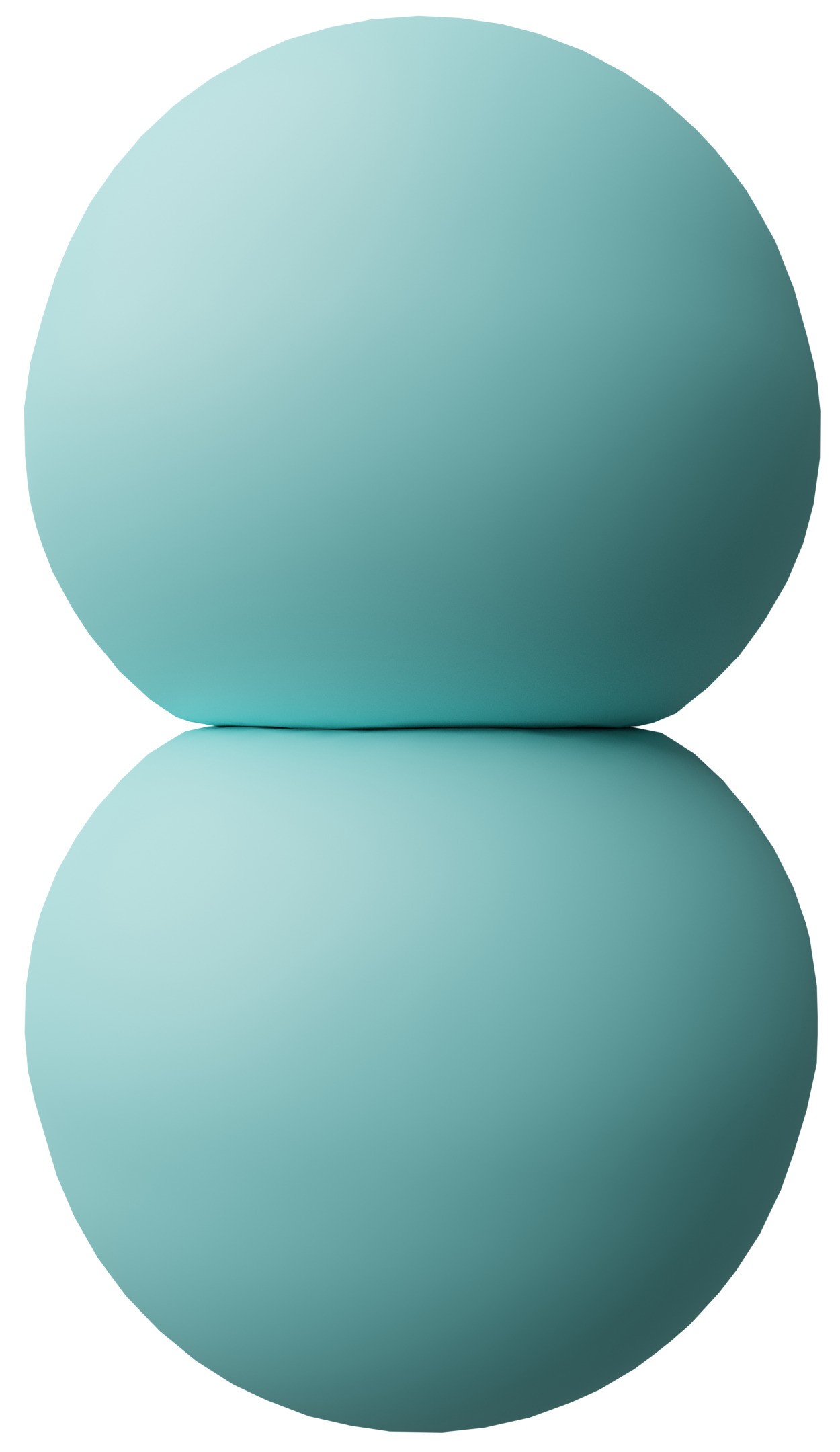} \\
\end{tabular}
 \caption{The effect of different choices for the volume recovery function parameters on the deformation of a couple of spheres isolated from Figure~\ref{fig:surf_coverage}.}
 \label{fig:function_parameters}
\end{figure}

\paragraph{Volume recovery parameters}
We also performed extensive tests to choose the parameters of the volume recovery function (Eqn.~\ref{eq:distance2speed}). \Cref{fig:function_parameters} shows the results obtained with the same overlapping spheres but different choices for the values of the $a$ and $b$ parameters. By shifting the control parameters, it is possible to simulate different plastic behaviors. 
All the results shown in the paper, but \Cref{fig:function_parameters}, are obtained with the same settings $(a=0.1,b=0.3)$.

\begin{figure}[tb]
\centering
\small
\begin{tabular}{@{}c@{\hspace{0.03in}}c@{}}
\includegraphics[width=0.46\linewidth]{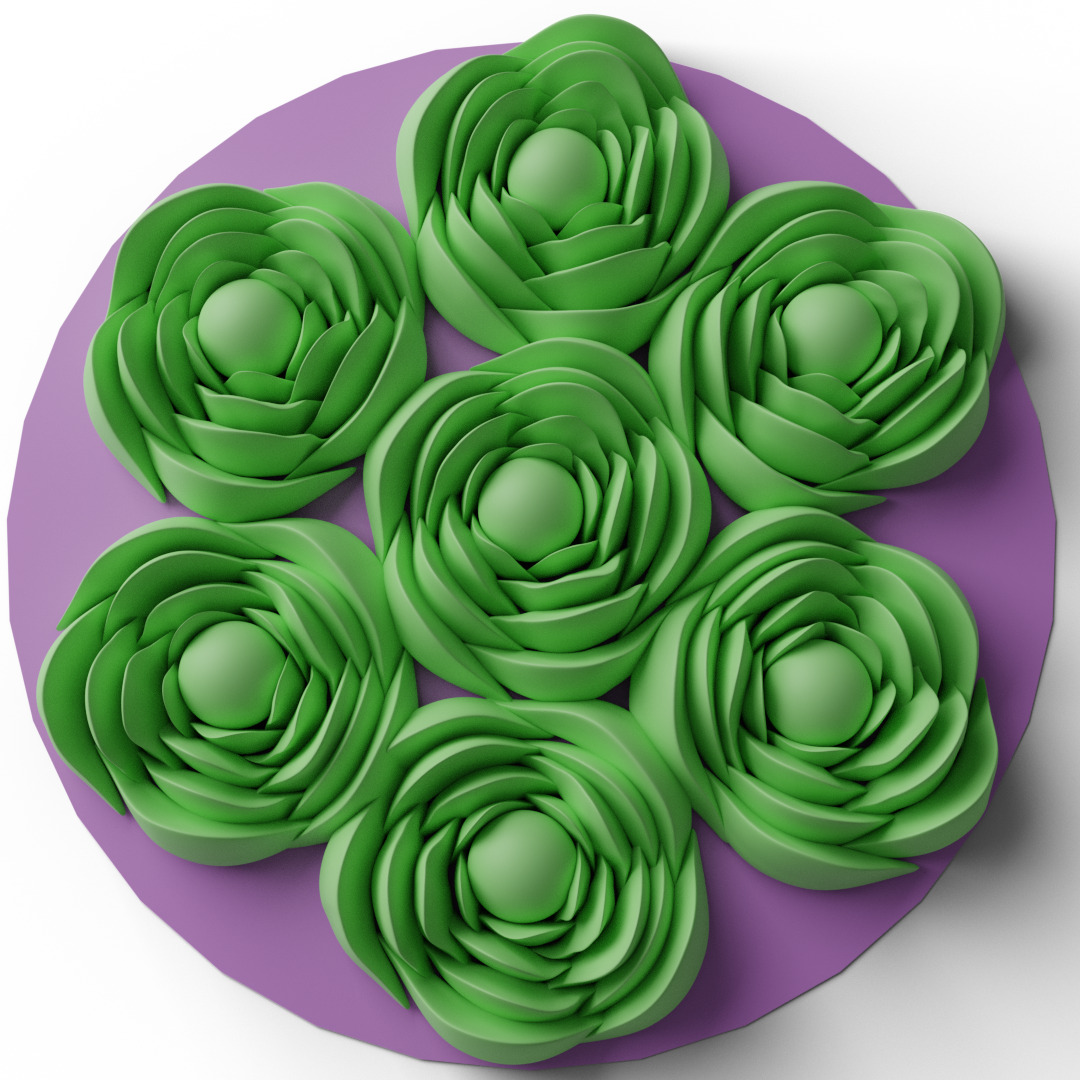} &
\includegraphics[width=0.46\linewidth]{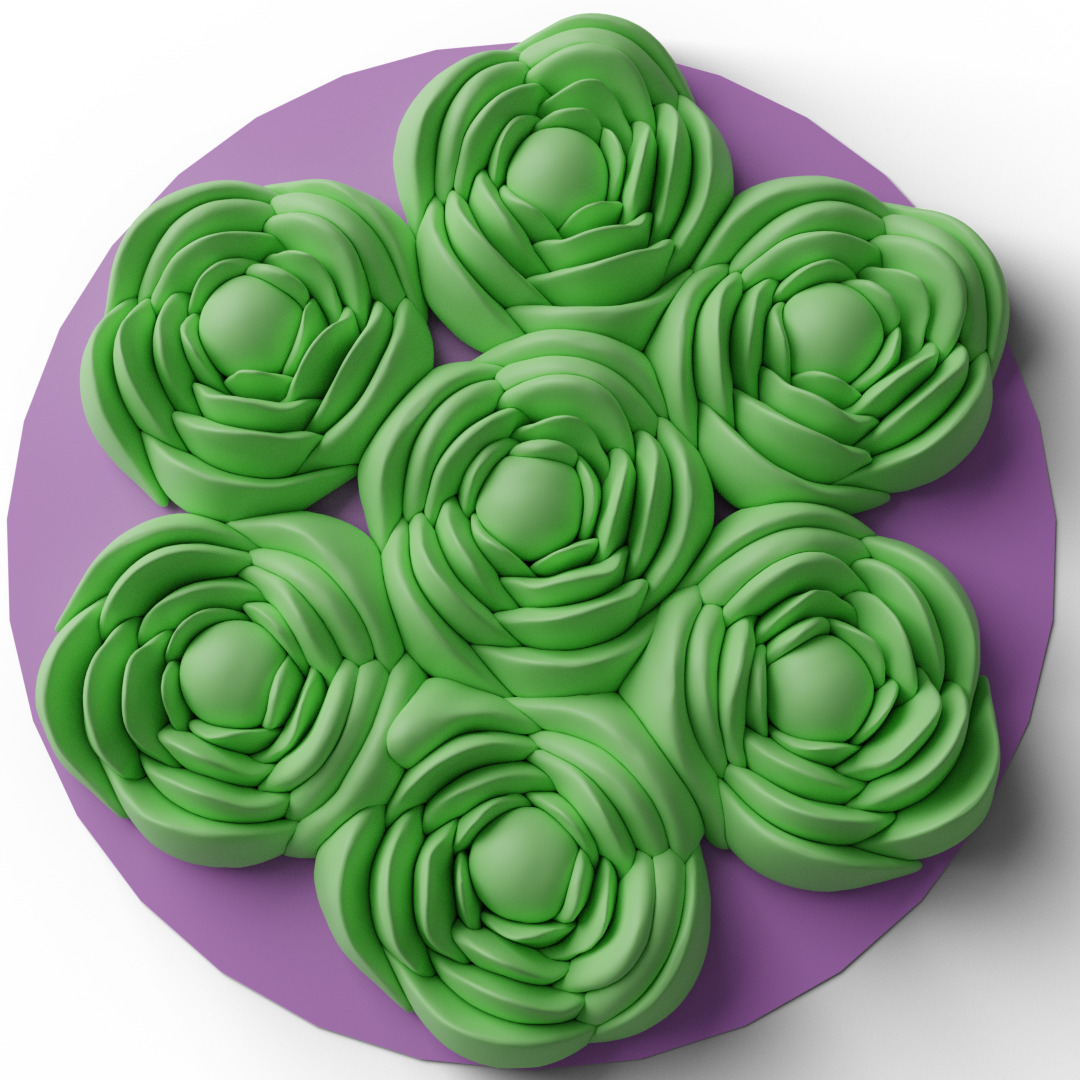} \\
\textsf{Initial placement} & \textsf{Packed decorations} \\
\includegraphics[width=0.46\linewidth]{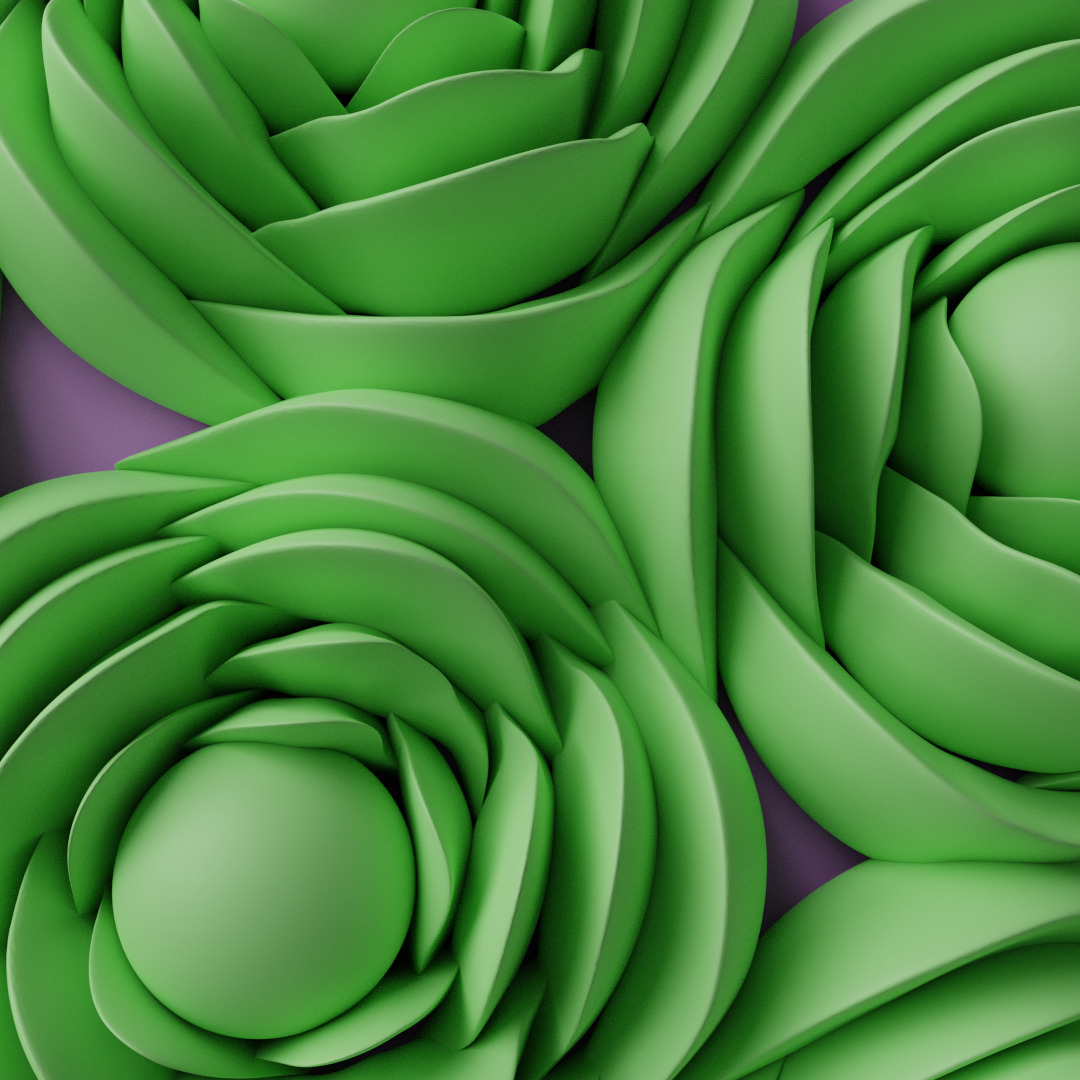} &
\includegraphics[width=0.46\linewidth]{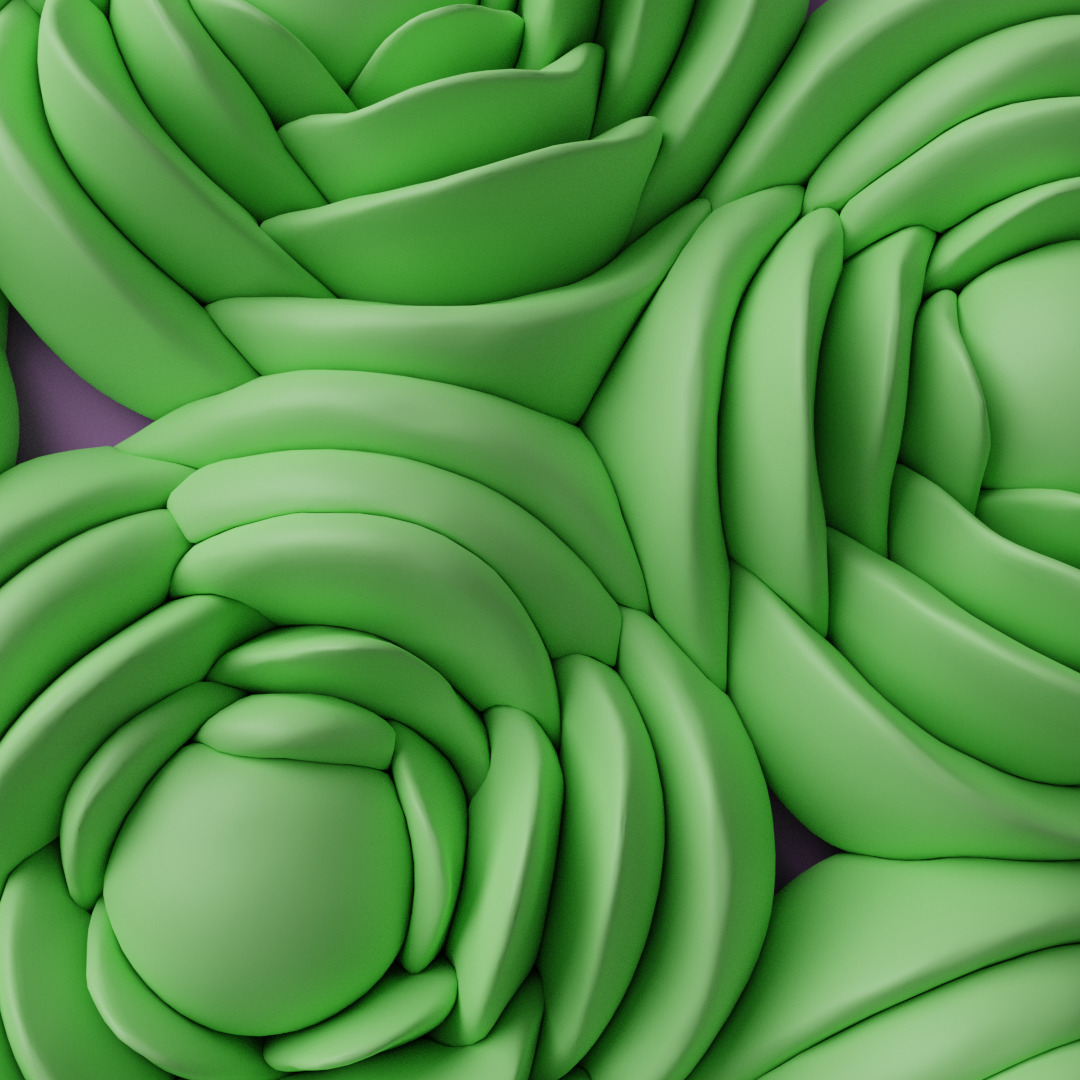} \\
\includegraphics[width=0.46\linewidth]{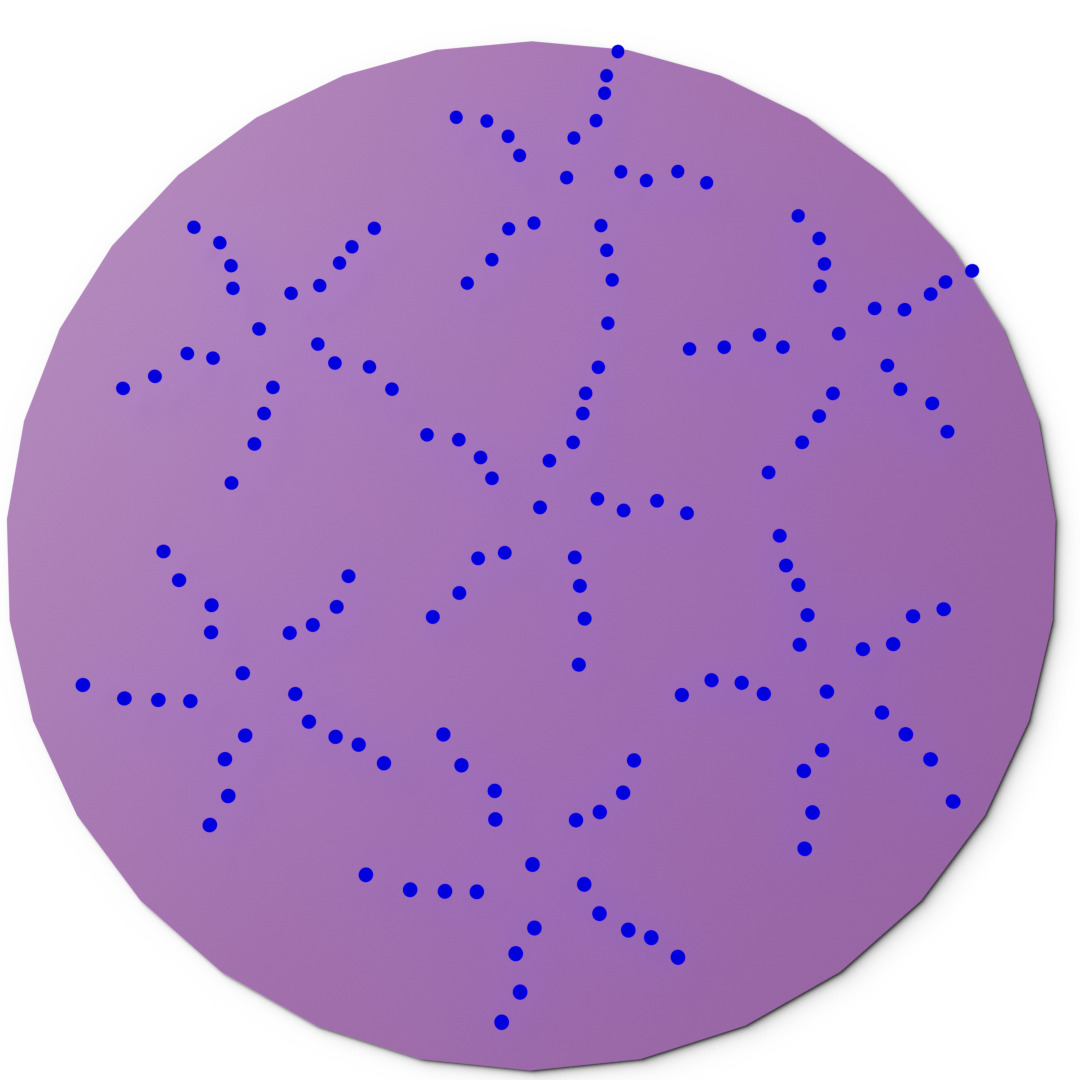} &
\includegraphics[width=0.46\linewidth]{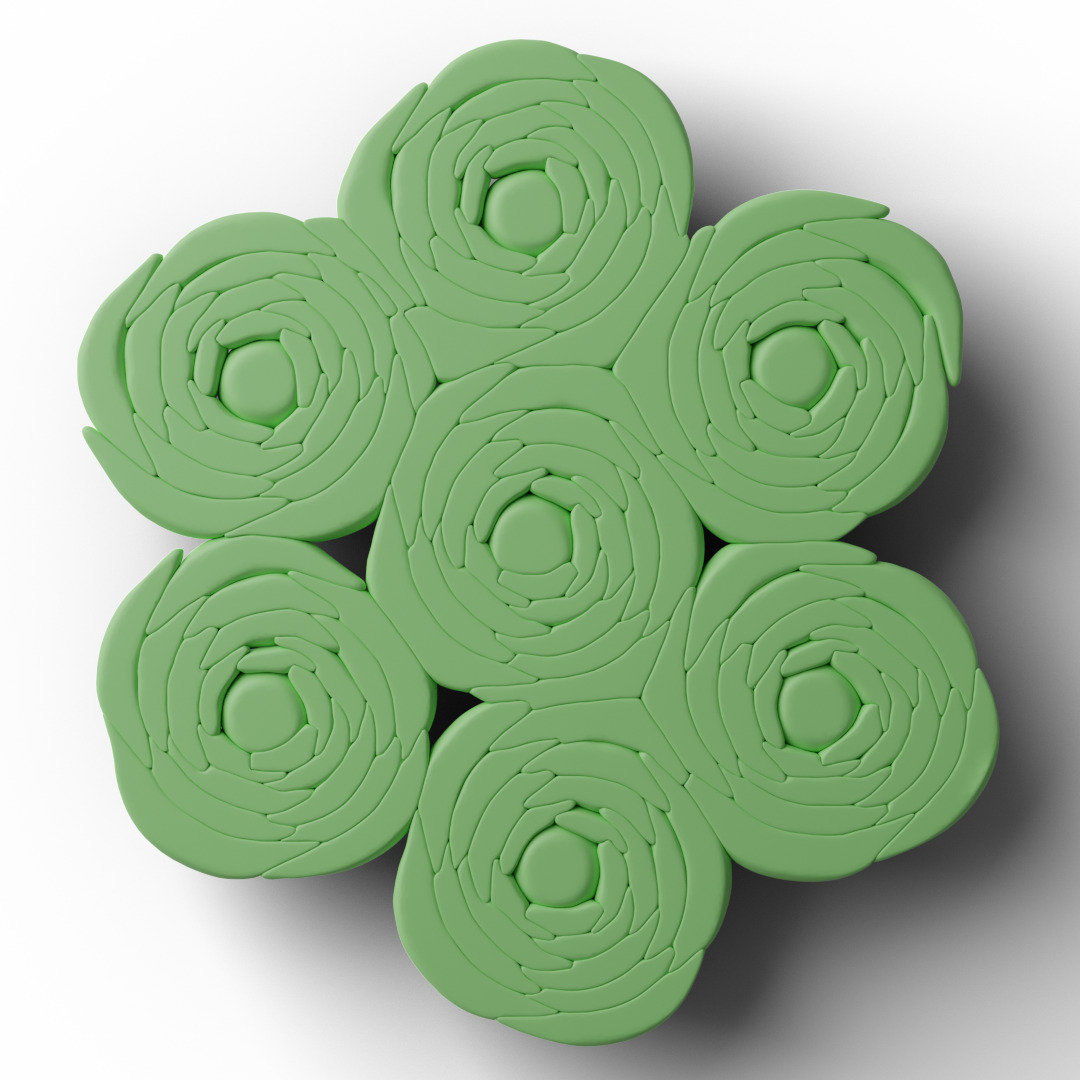} \\
\textsf{Manually-placed decorations} & \textsf{Packed decorations (back view)} \\
\end{tabular}
\caption{Example of decoration obtained by manually placing elements and having \emph{PAVEL} resolve the overlap. Here each petal is a separate decorative element and the deformation is computed to simulate sculpting clay. Note from the back view, how \emph{PAVEL} can resolve the overlap in a very complex case with many self-intersecting, non-symmetric, elements.}
\label{fig:manual}
\end{figure}

As it is possible to see, for example, in \Cref{fig:manual}, this choice provides results that are visually quite close to real plastic deformation.

\section{Placement of decorations}
\label{sec:placement}

The deformation step is enough to obtain the final result if the user places the decorations manually on the base object's surface, as shown in \cref{fig:manual}. In this manner, the user directly controls the decorations' location and the surface coverage, while \emph{PAVEL} controls the material behavior using the deformation parameters described above.
In the system, we supply the automatic placement methods described in the following.

\paragraph{Automatic placement by Point Sampling}
While manual placement provides complete artistic control, the decoration of complex surfaces with small decorative elements would be very time consuming, just like in real-world crafts. To simplify creating complex decorations, we also support the automatic placement of decorative elements using two procedural methods.

Our placement methods are based on point sampling algorithms. We support the placement of objects with approximate 2D radial symmetry, with the symmetry axis aligned to the base surface's local normal. We also assume that the decorations are small compared to the base surface's curvature to ignore the variations in the local volume extruded from the base surface. Both of these are reasonable assumptions found in the handcrafted models that inspired our work.

Instead of a packing algorithm, our use of point sampling is motivated by a trade-off between speed and packing quality. It's worth reminding that optimal packing is NP-hard in the Euclidean domain and just as hard in the intrinsic manifold metric. This means that all published packing algorithms are approximations that trade-off element shapes, packing quality, and computation time. In general, it is well known in both the literature and the practice that most packing algorithms fail or are not optimal for all but the simplest shapes. Note also that we do not allow to change the decoration size, which is the most used approximation for soft packing in graphics. For all these reasons, our point sampling approaches are just another heuristic within their stated constraints. If better placement heuristics are available, we could integrate them into our pipeline with ease. 

\paragraph{Isotropic seeding}
Our first automatic placement method produces isotropic distributions that work well for decorations with cylindrical symmetry. The user controls the density of decorations by setting the desired surface coverage, and the algorithm instantiates the correct number of seeds to guarantee the target density.

\begin{figure}[tb]
\centering
\small
\begin{tabular}{@{}c@{\hspace{0.03in}}c@{\hspace{0.03in}}c@{}}
\includegraphics[width=0.32\linewidth]{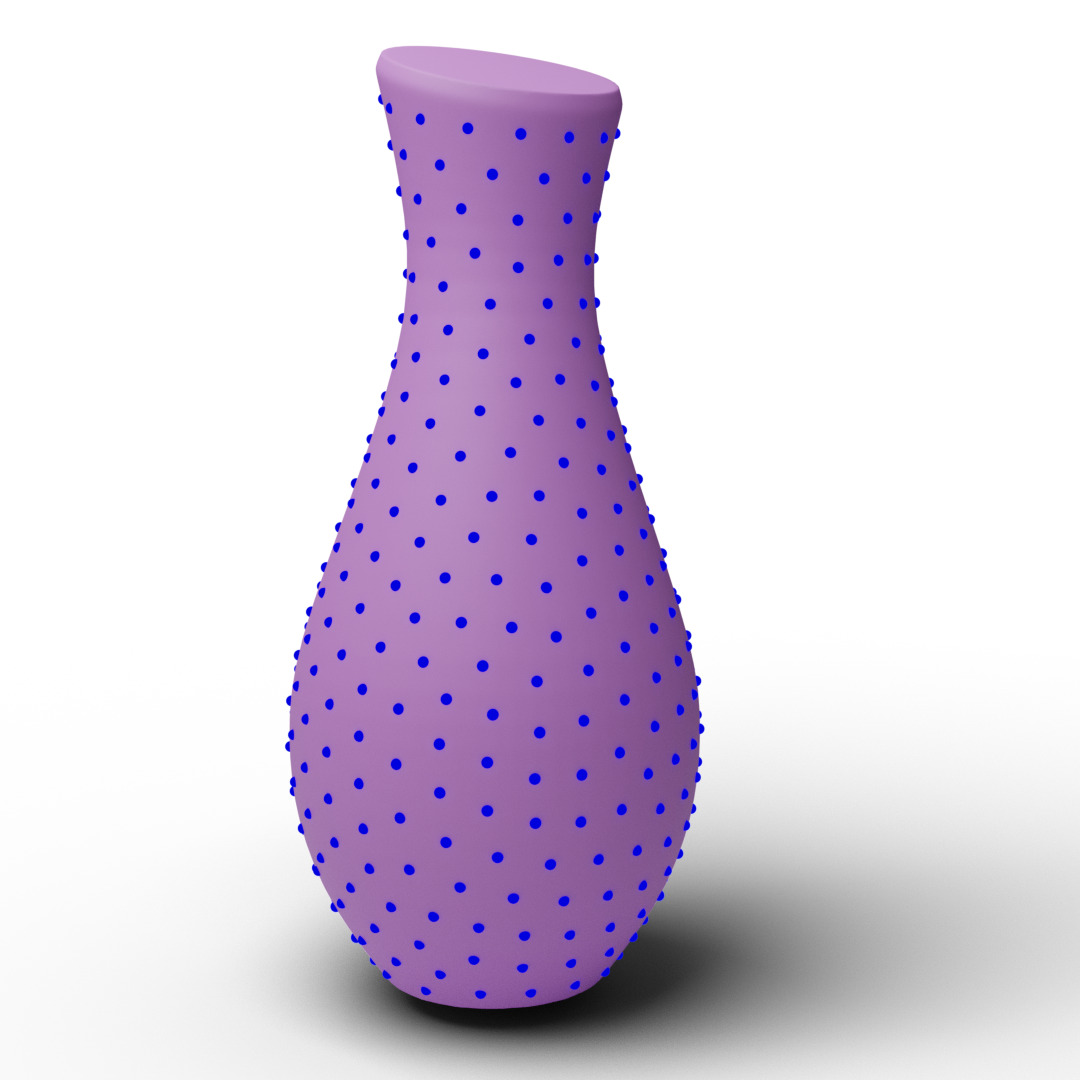} &
\includegraphics[width=0.32\linewidth]{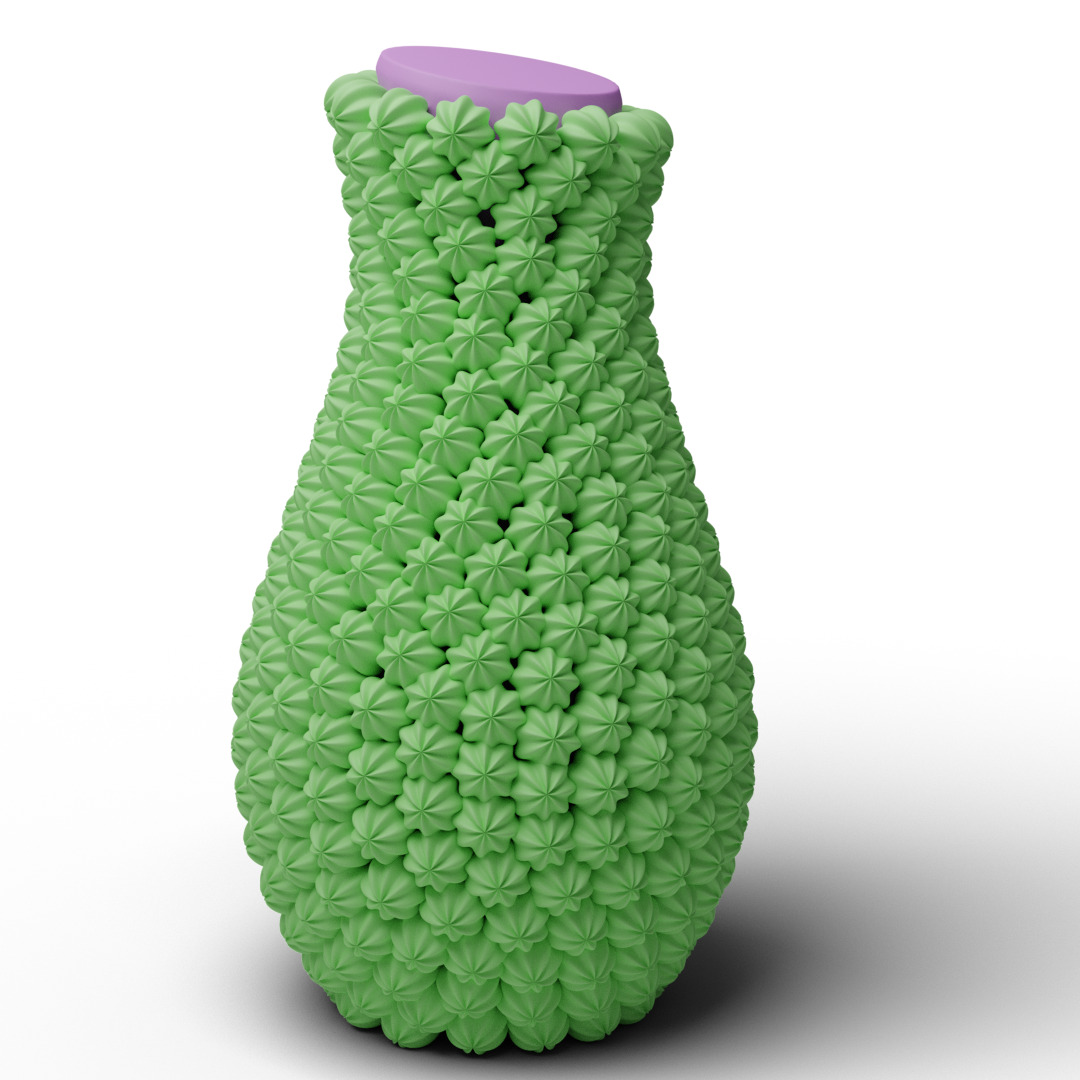} &
\includegraphics[width=0.32\linewidth]{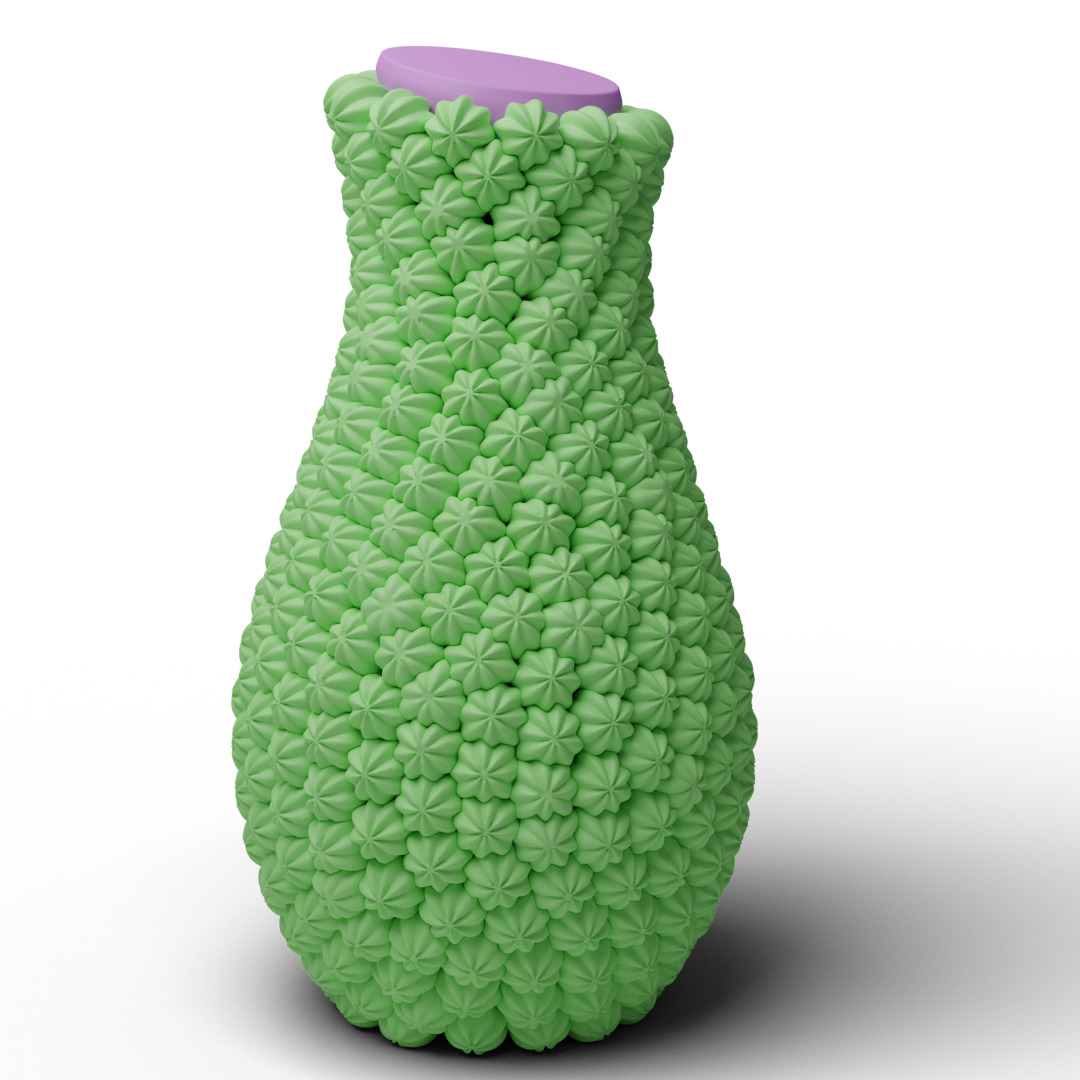} \\
\textsf{Sampled points} & \textsf{Initial placement} & \textsf{Packed decorations} \\
%\raisebox{0.08\linewidth}{\includegraphics[width=0.16\linewidth]{results/vase/deco2}} &
\includegraphics[width=0.32\linewidth]{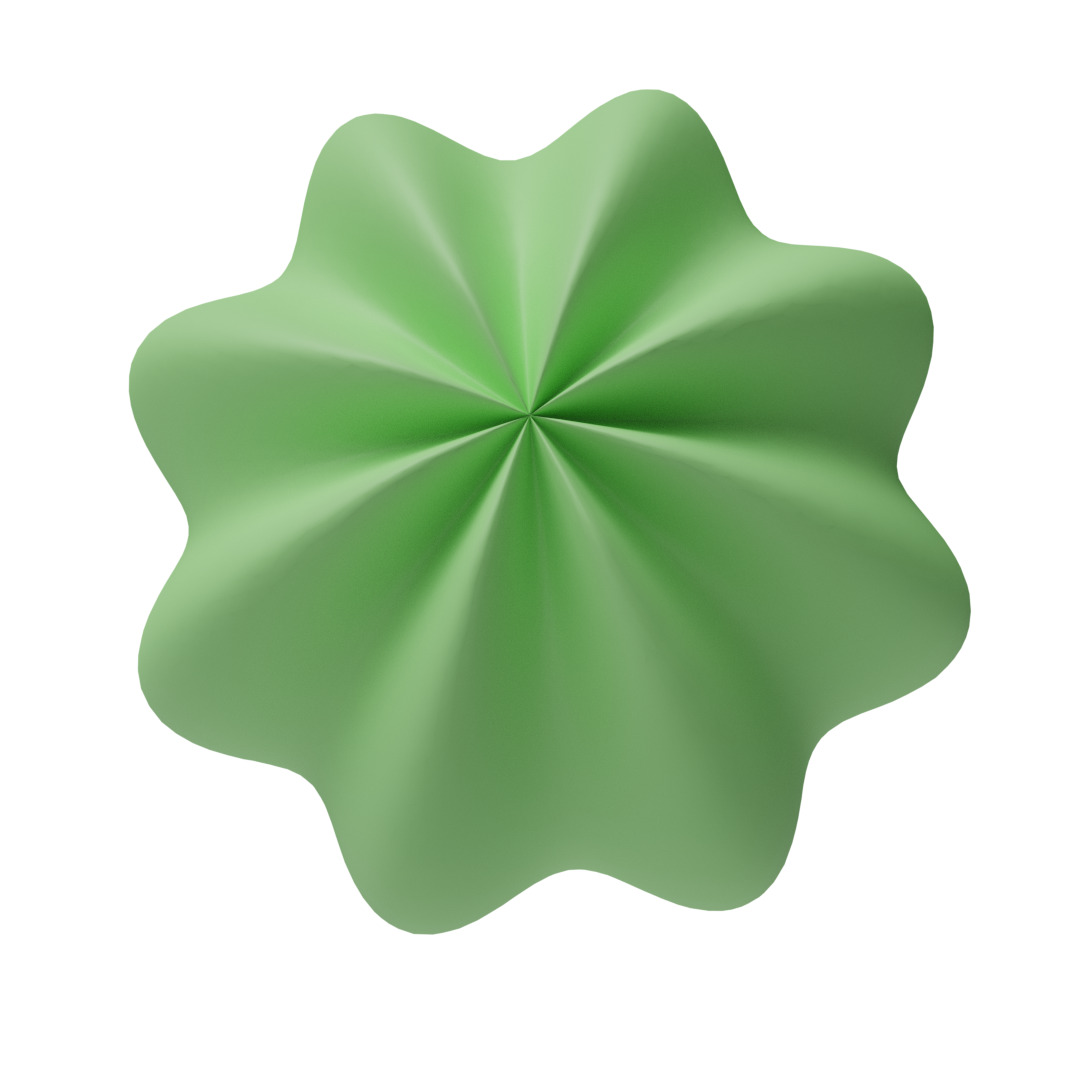} &
\includegraphics[width=0.32\linewidth]{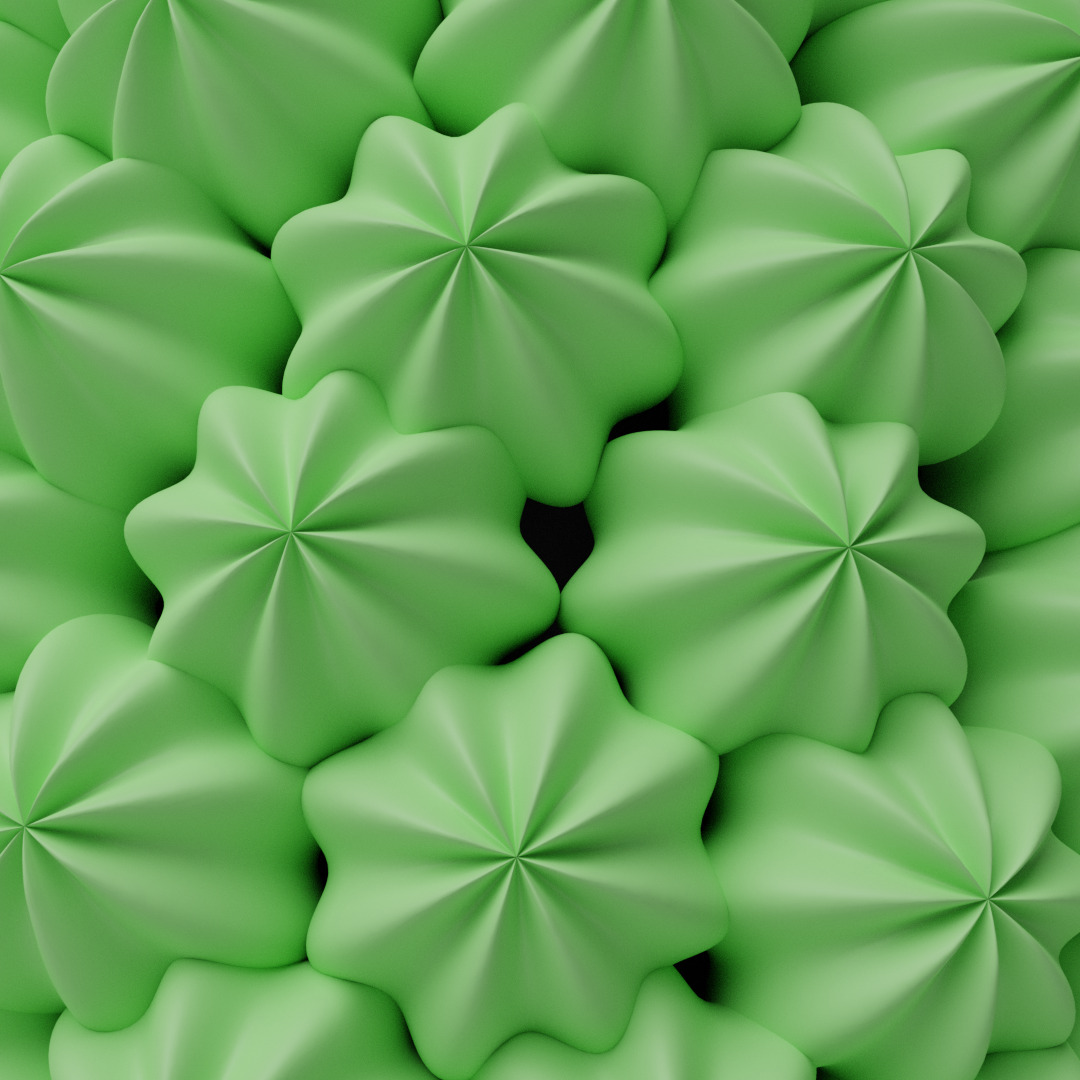} &
\includegraphics[width=0.32\linewidth]{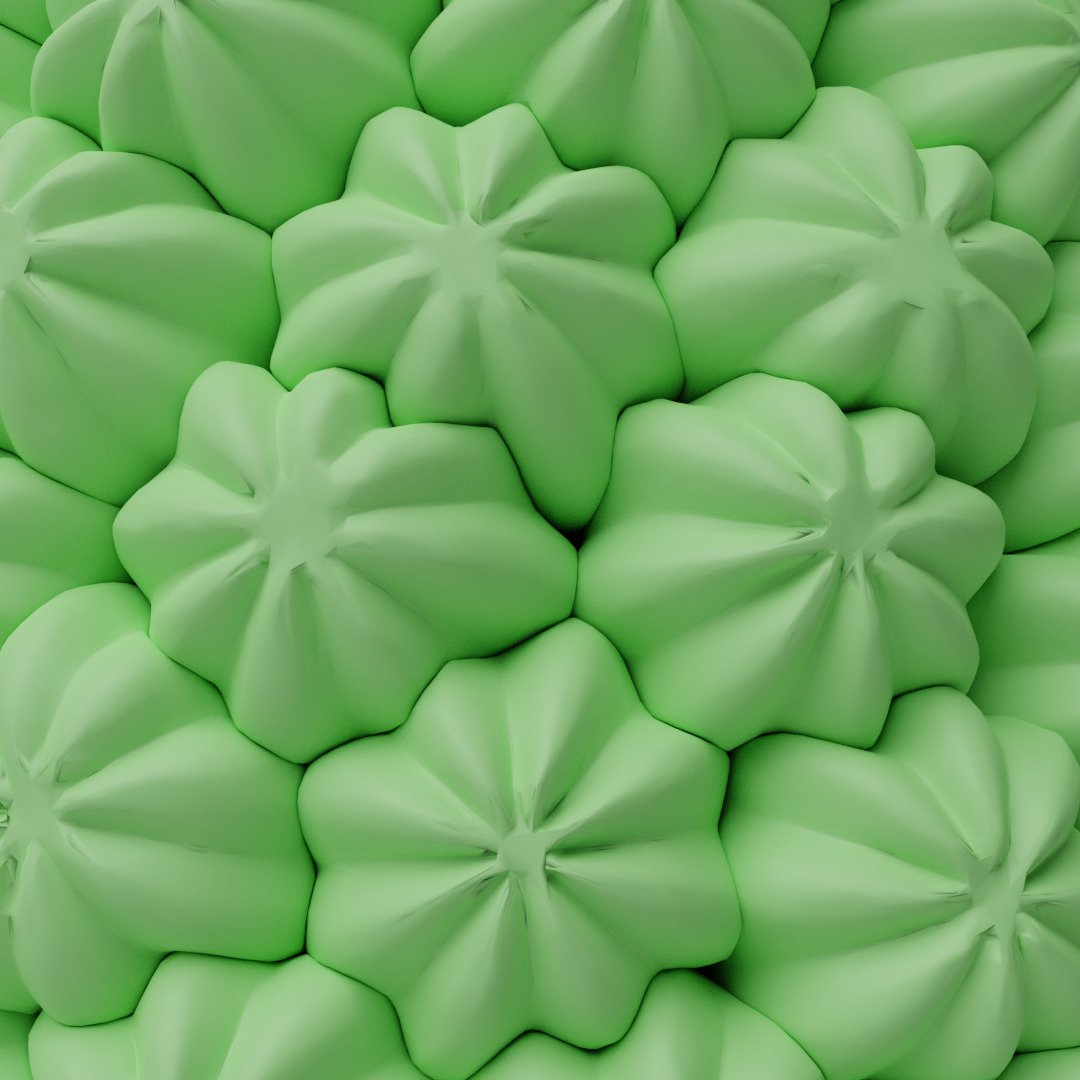} \\
\textsf{Decorative element} & \textsf{Close view (initial)} & \textsf{Close view (packed)} \\
\end{tabular}
\caption{Isotropic seeding with CVT on a vase.}
\label{fig:cvd}
\end{figure}

To ensure a high-quality distribution, we rely on optimization using a method based on the Centroidal Voronoi Tessellation (CVT) and its efficient implementation described in \cite{YanLLSW09} and available in the Geogram library \cite{Geogram}. We generate the base surface's tessellation with the desired number of vertexes and use them as the centers of the decorations.
When converged, this iterative approach produces a point distribution with good statistical properties, resulting in a high-quality decorated surface. To obtain a more ``hand-made'' look, we allow users to reduce the number of iterations. \Cref{fig:cvd} shows an example results obtained with this method that closely resembles the hand-made objects shown in \cref{fig:examples}.

\begin{figure}[b]
\centering
\small
\begin{tabular}{@{}c@{\hspace{0.03in}}c@{\hspace{0.03in}}c@{}}
\includegraphics[width=0.32\linewidth]{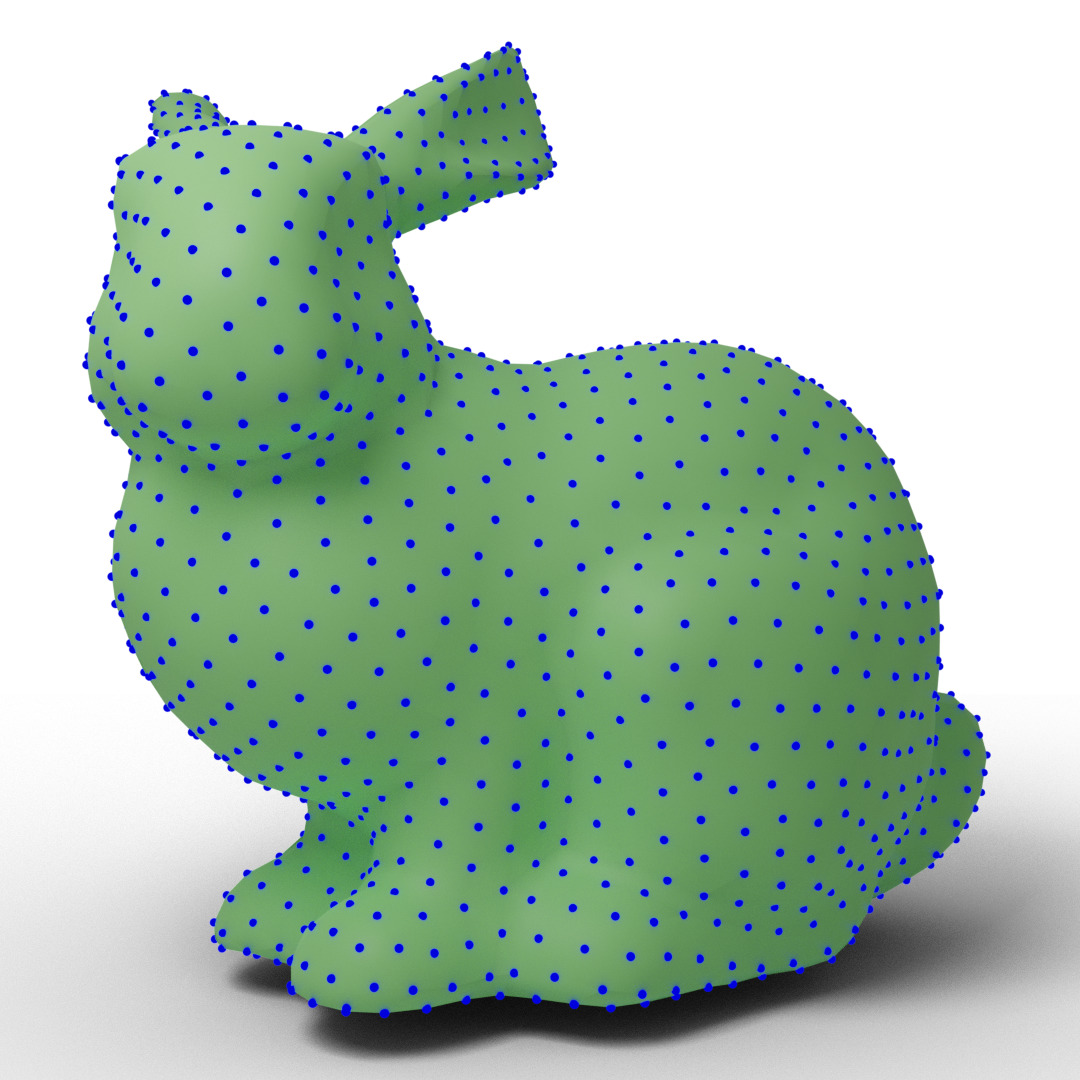} &
\includegraphics[width=0.32\linewidth]{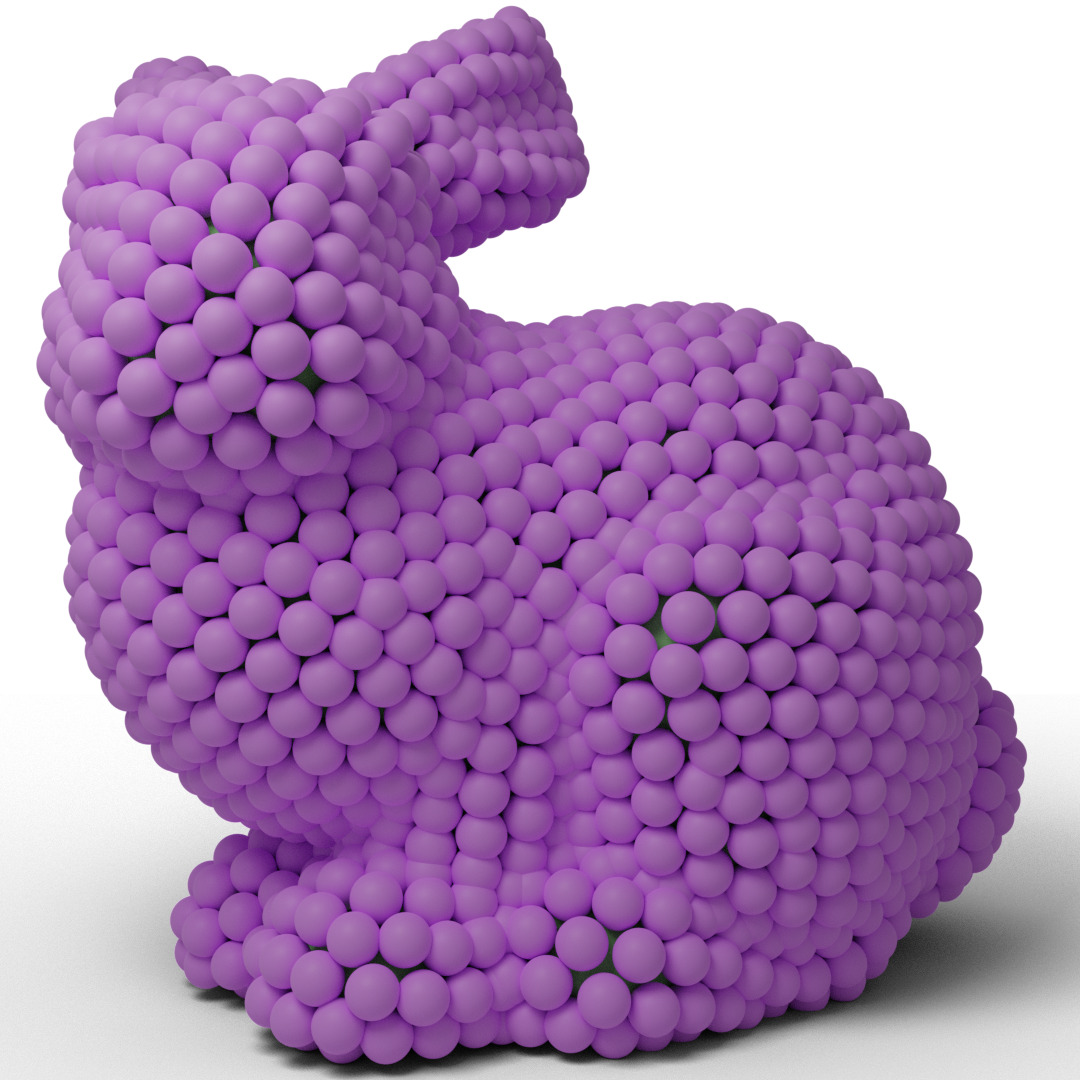} &
\includegraphics[width=0.32\linewidth]{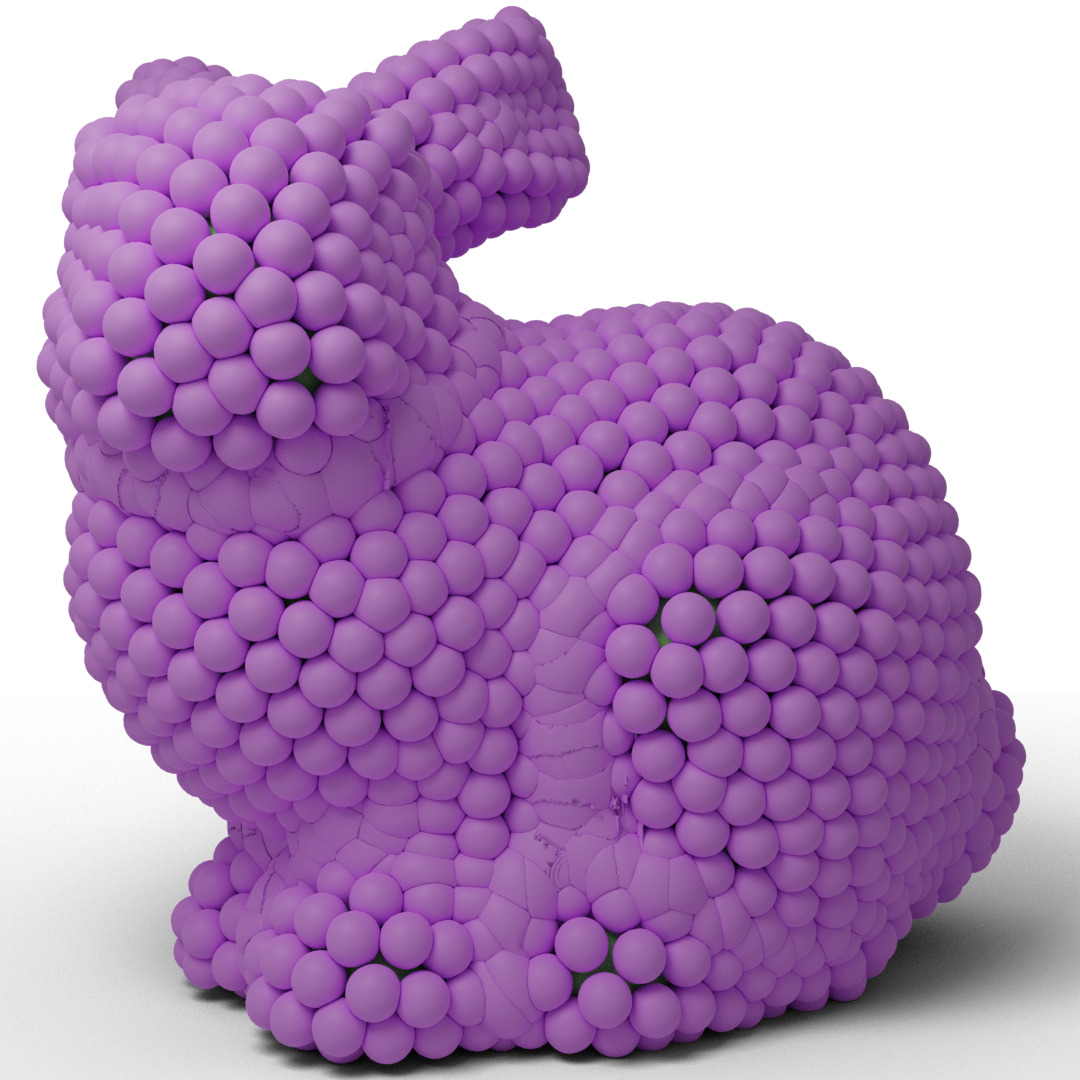} \\
\multicolumn{3}{c}{\textsf{On-surface sampling}} \\
\includegraphics[width=0.32\linewidth]{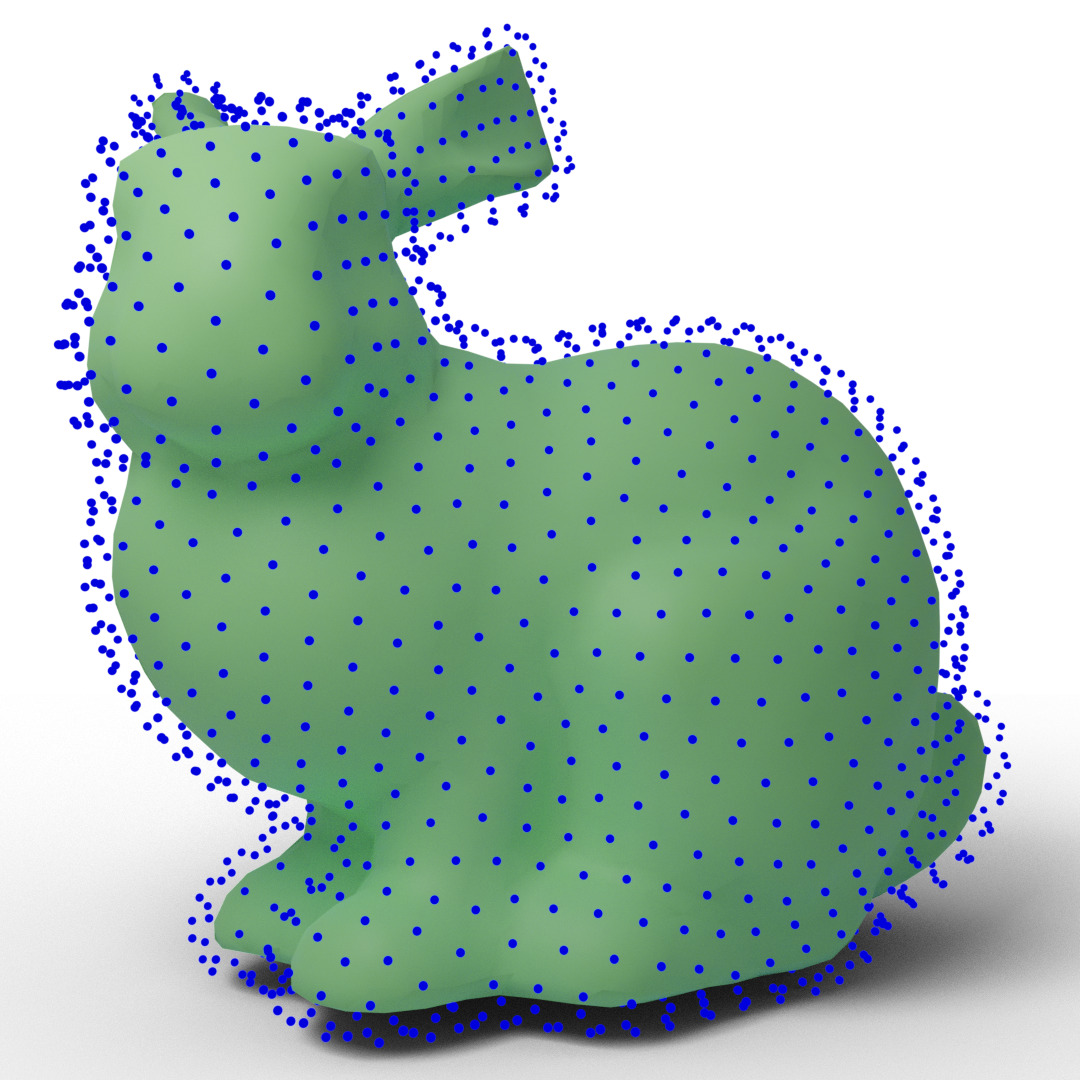} &
\includegraphics[width=0.32\linewidth]{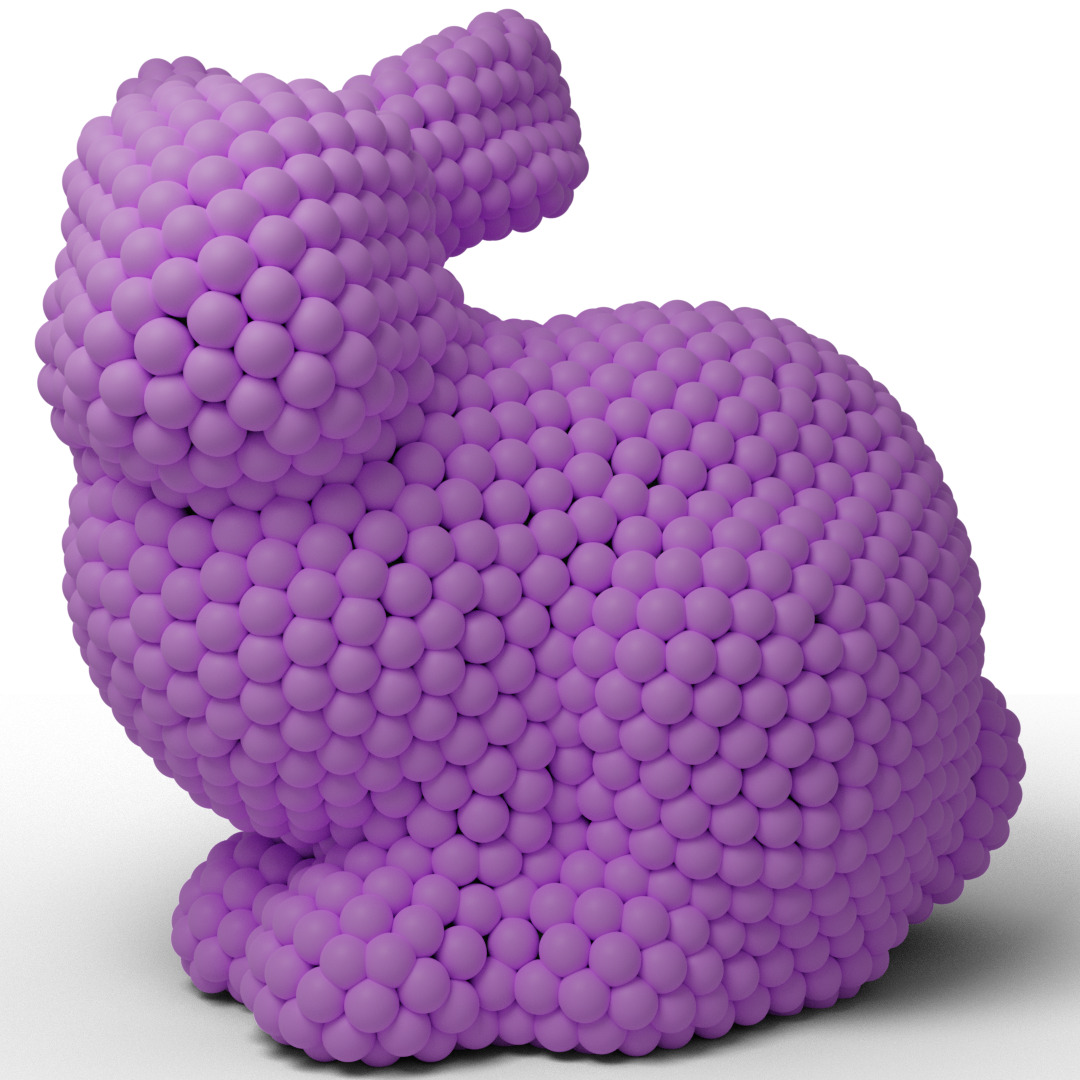} &
\includegraphics[width=0.32\linewidth]{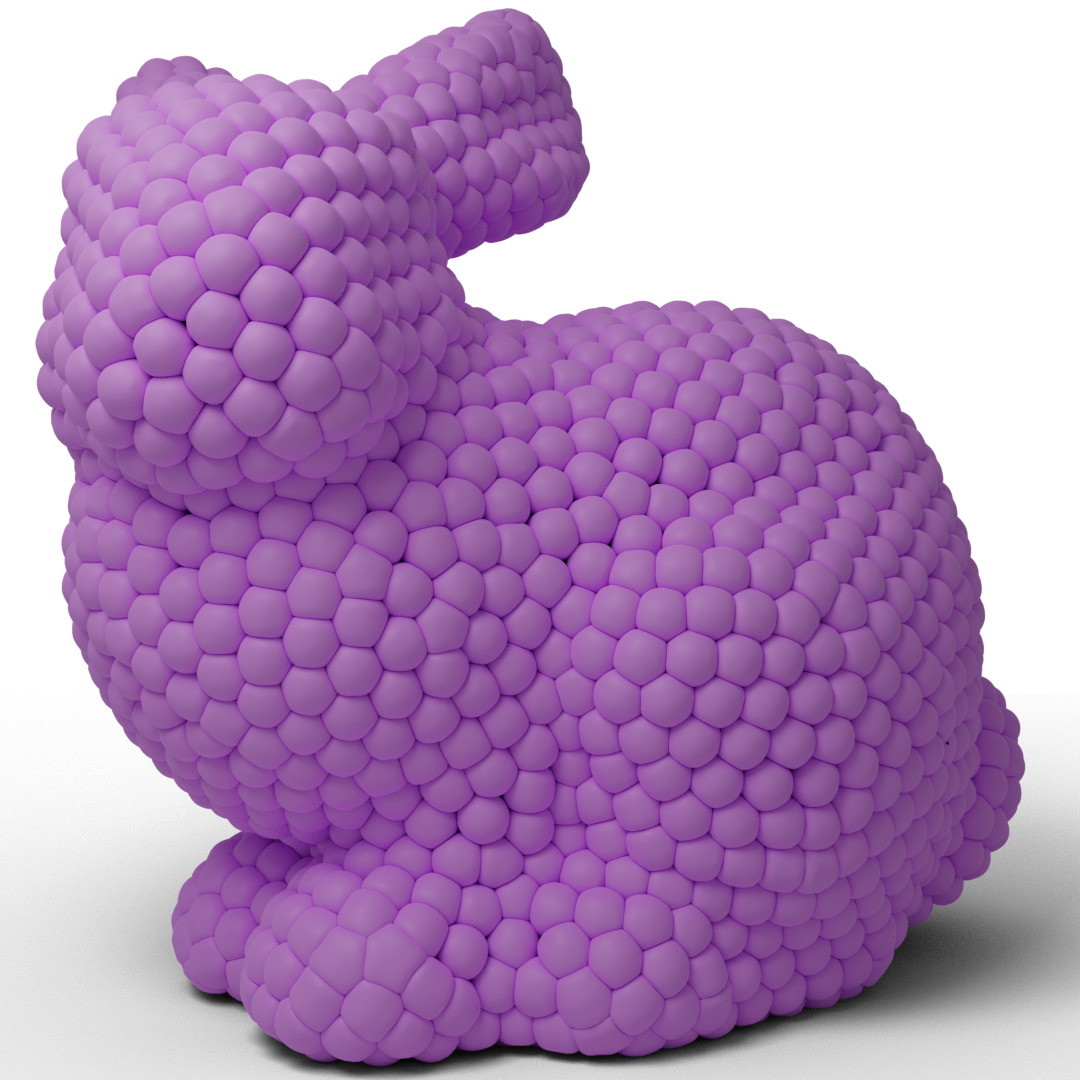} \\
\multicolumn{3}{c}{\textsf{Offset surface sampling}} \\
\end{tabular}
\begin{tabular}{@{}c@{\hspace{0.03in}}cc@{\hspace{0.03in}}c@{}}
\includegraphics[width=0.22\linewidth]{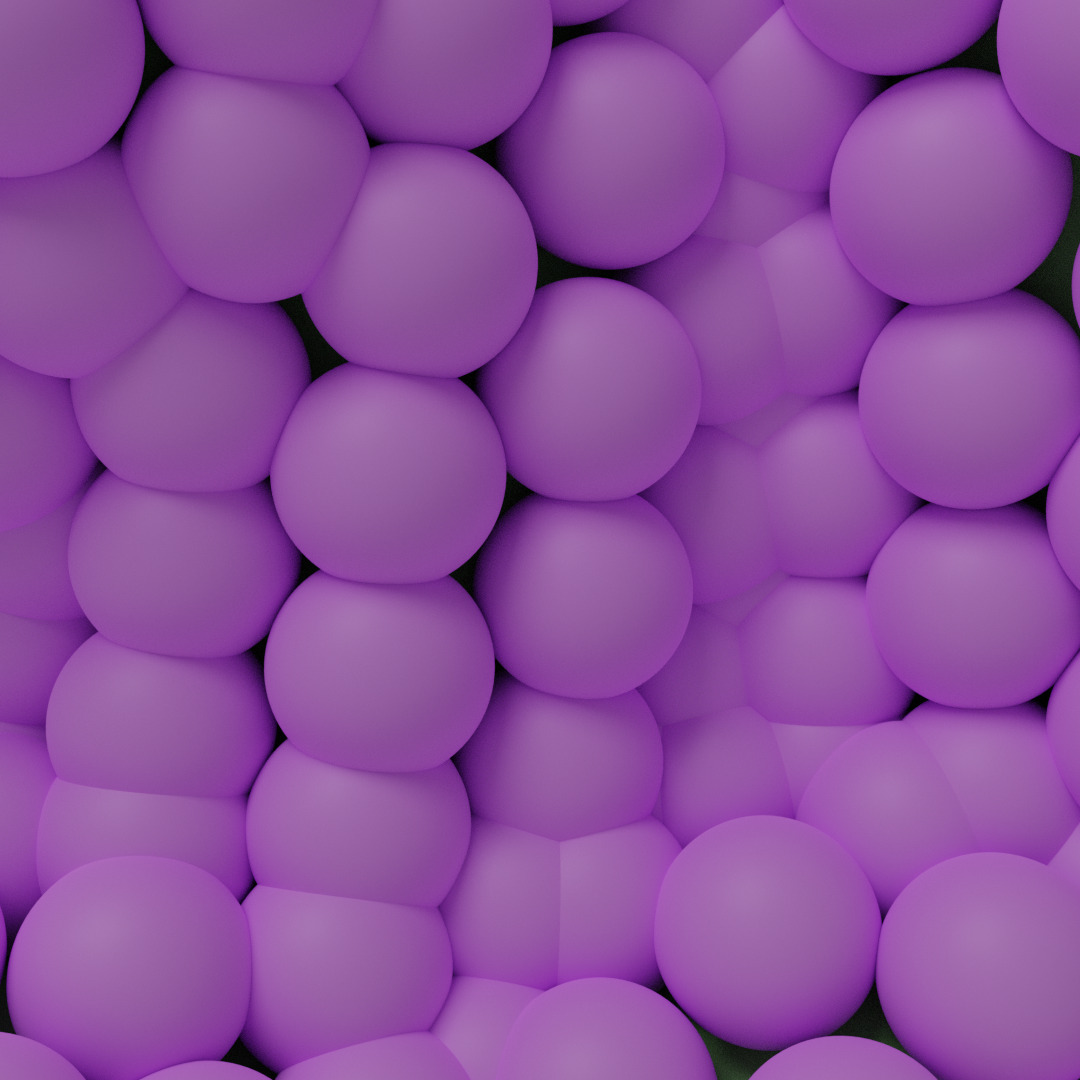} &
\includegraphics[width=0.22\linewidth]{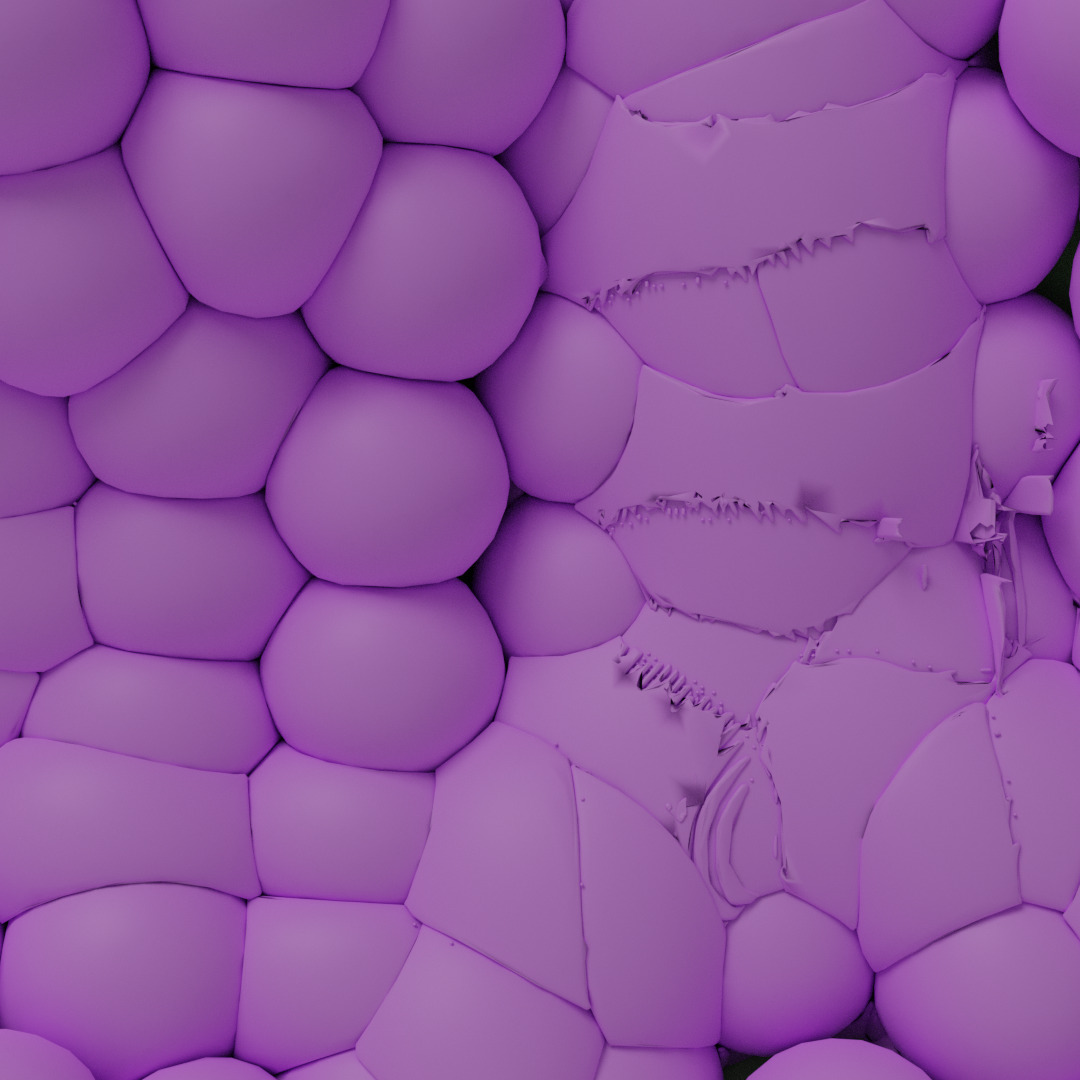} &
\includegraphics[width=0.22\linewidth]{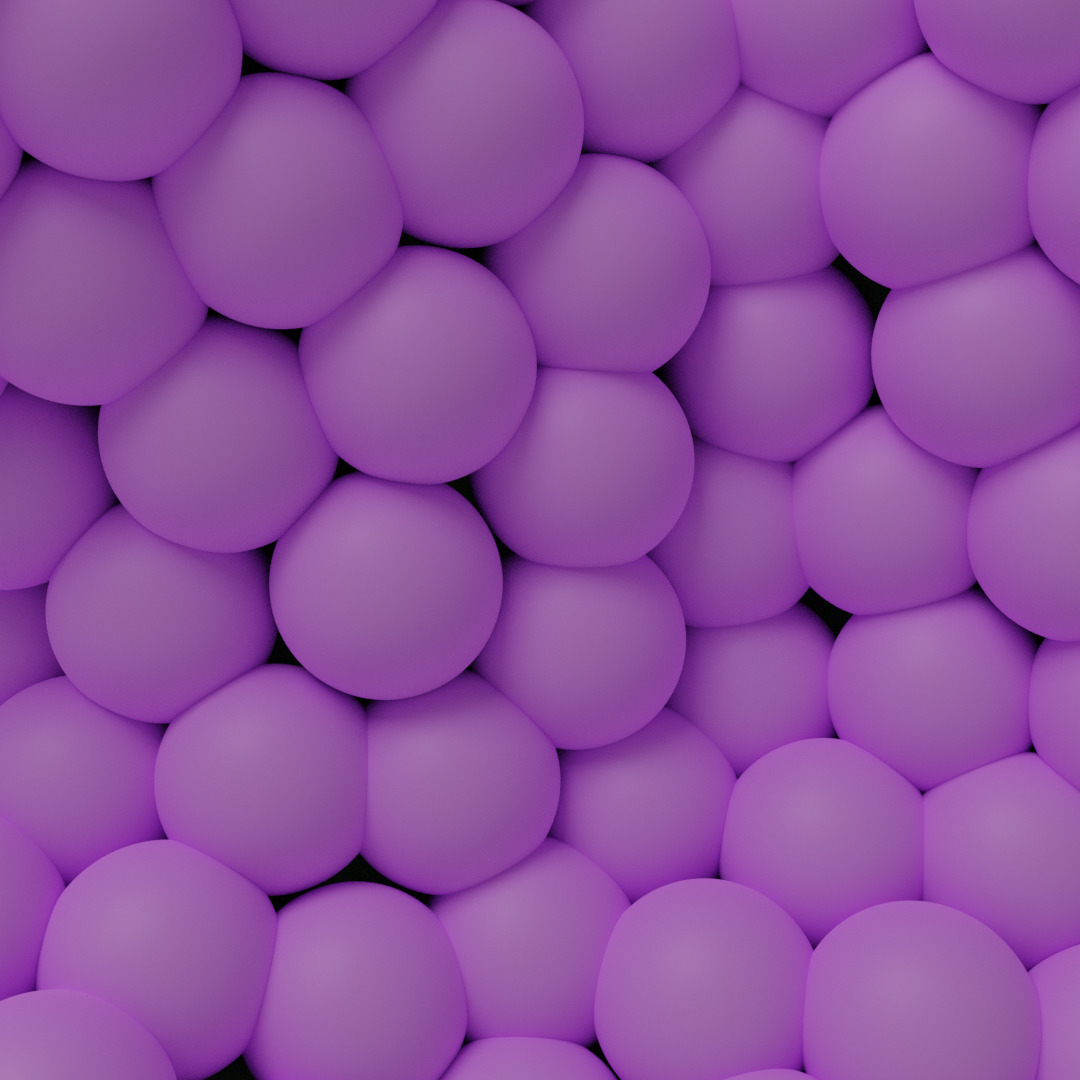} &
\includegraphics[width=0.22\linewidth]{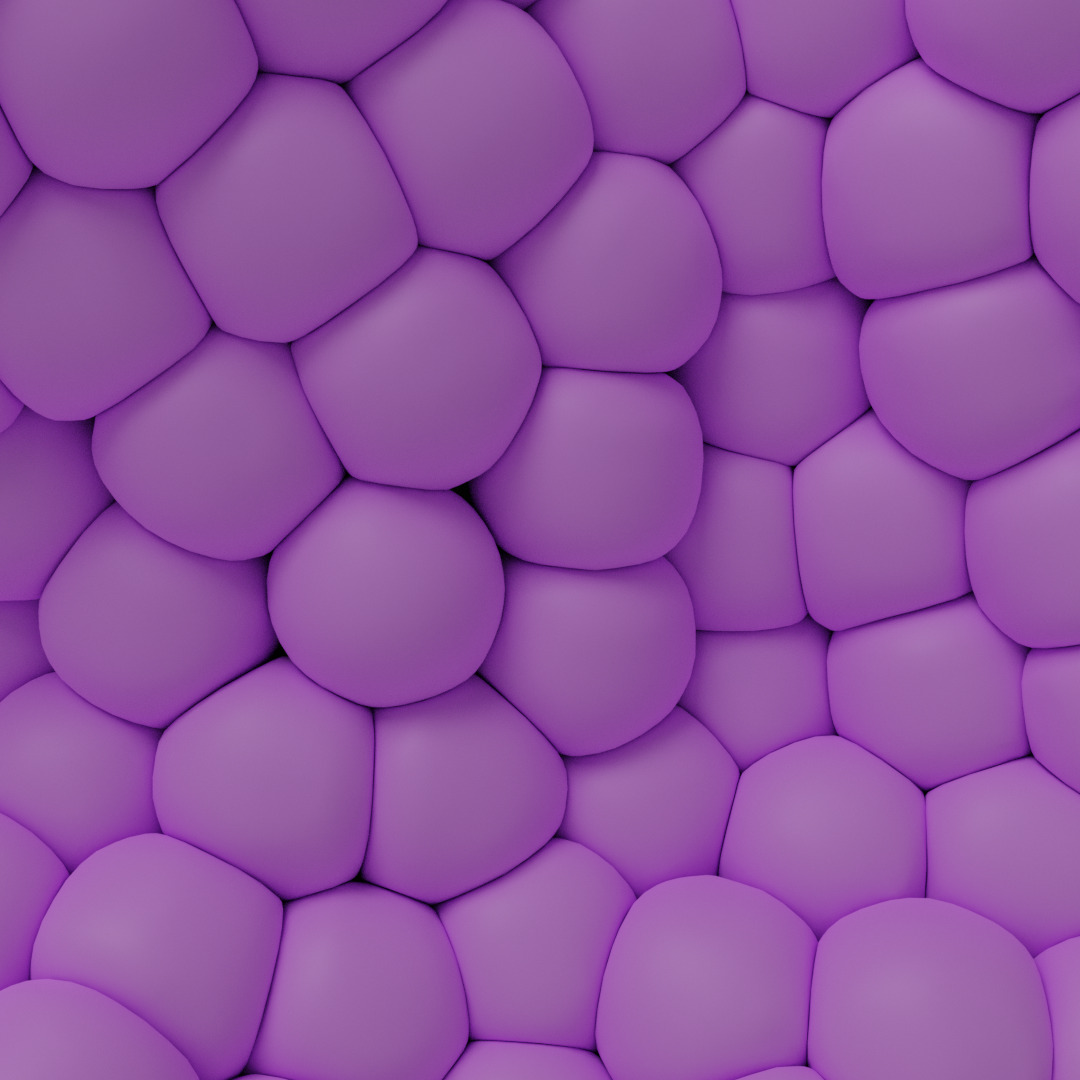} \\
\multicolumn{2}{c}{\textsf{On-surface sampling}} &
\multicolumn{2}{c}{\textsf{Offset surface sampling}} \\
\end{tabular}
\caption{Comparison between sampling on the surface (top) or on an isosurface at distance $d$ (middle). Using the isosurface allows for more even packing and deformations.} \label{fig:samplingcomparison}
\end{figure}

\paragraph{Offset surface}
The previous method works well if the decorative elements are small compared to the surface curvature. If that assumption is not valid, the decorated surface has visible artifacts, shown in \cref{fig:samplingcomparison}. 
The main problem is that our method generates equally spaced seeds that represent the bottom of the decorations but does not take into account decorations' volume and surface curvature. 
Decorations overlap more in concave regions than convex ones. The volume above the surface is smaller in the former case and can result in heavily compressed areas in concave zones and non-contacting in convex regions. When decorations are small compared to the surface curvature, this effect is negligible. But it becomes more noticeable as the decorations' size increases.

For this reason, we also provide a method that considers the volume of the elements when point sampling their locations. The basic idea is to extract an isosurface at a signed distance $d$ from the base mesh, perform the sampling on it, and then map back the seed points on the original base surface, projecting them along the local normal direction.

An appropriate choice of $d$ is the height of the element's z-section with the maximal area.
In this way, we can handle large decoration sizes even on surfaces with varying curvature. We show a clear example of this in \cref{fig:samplingcomparison}. Here the bunny features gulfs and bulges. They are handled without problems by sampling the offset surface. Conversely, sampling on the surface reveals significant artifacts. 
Since decorations are small in general, and the modified sampling has a non-negligible computational cost (see \cref{sec:results}), we leave the choice of whether to use the offset surface to the user.

\paragraph{Stripe-based, anisotropic seeding}
To support the automatic generation of anisotropic element patterns, we also provide a stripe-based sampling approach. This method can create gridded patterns of decorations with approximate cylindrical symmetry and even with elongated shapes.

\begin{figure}[tb]
\centering
\small
\begin{tabular}{@{}c@{\hspace{0.03in}}c@{\hspace{0.03in}}c@{}}
\includegraphics[width=0.32\linewidth]{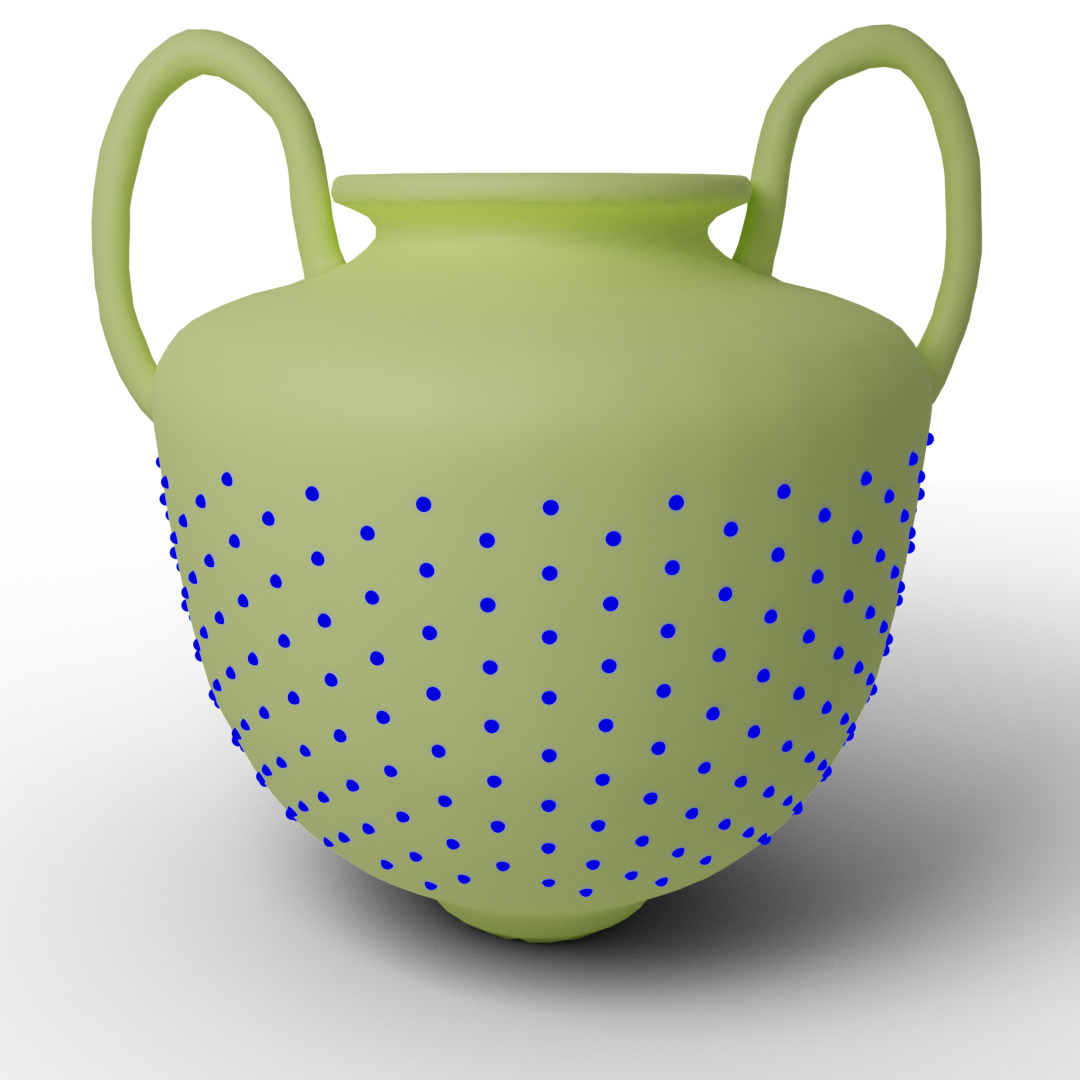} &
\includegraphics[width=0.32\linewidth]{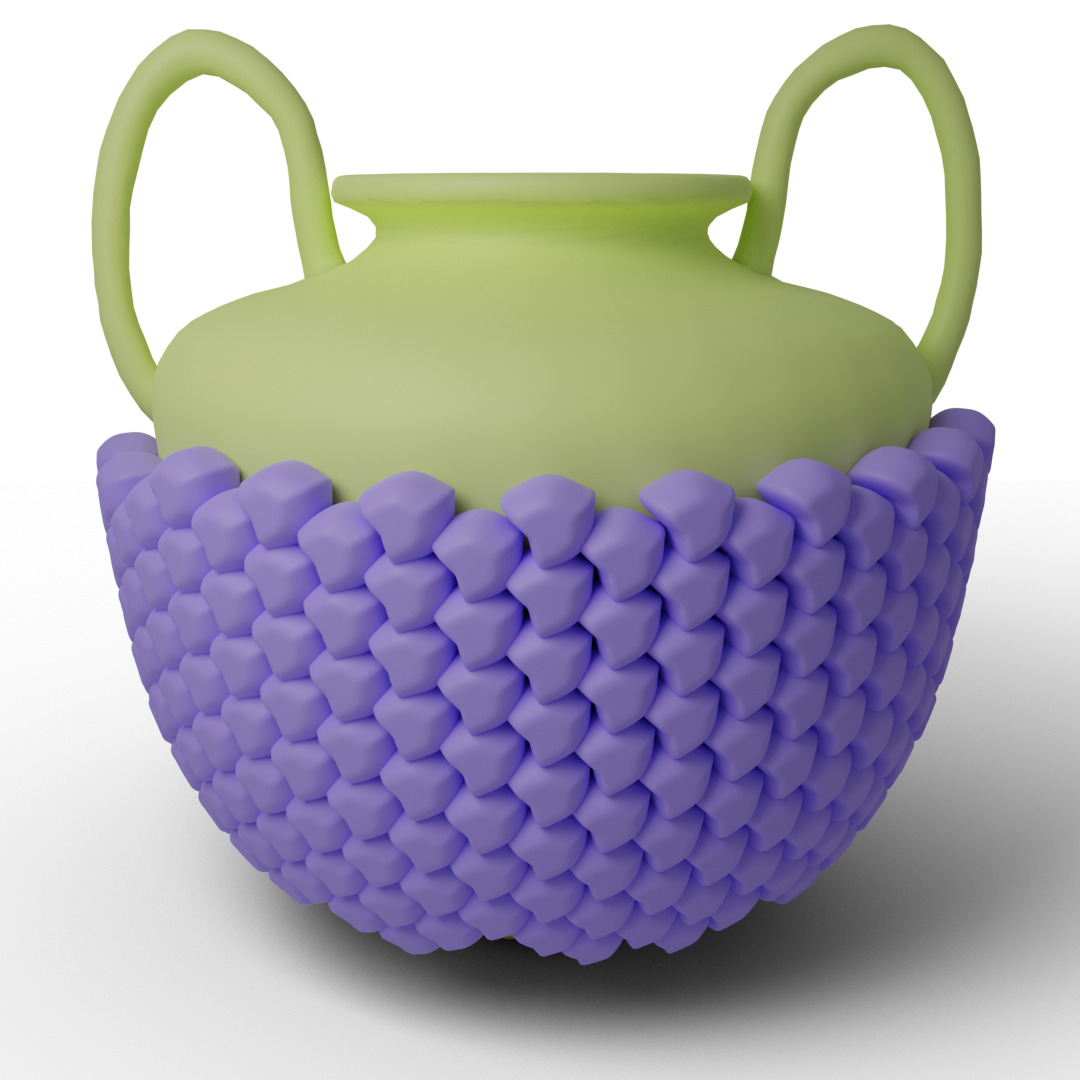} &
\includegraphics[width=0.32\linewidth]{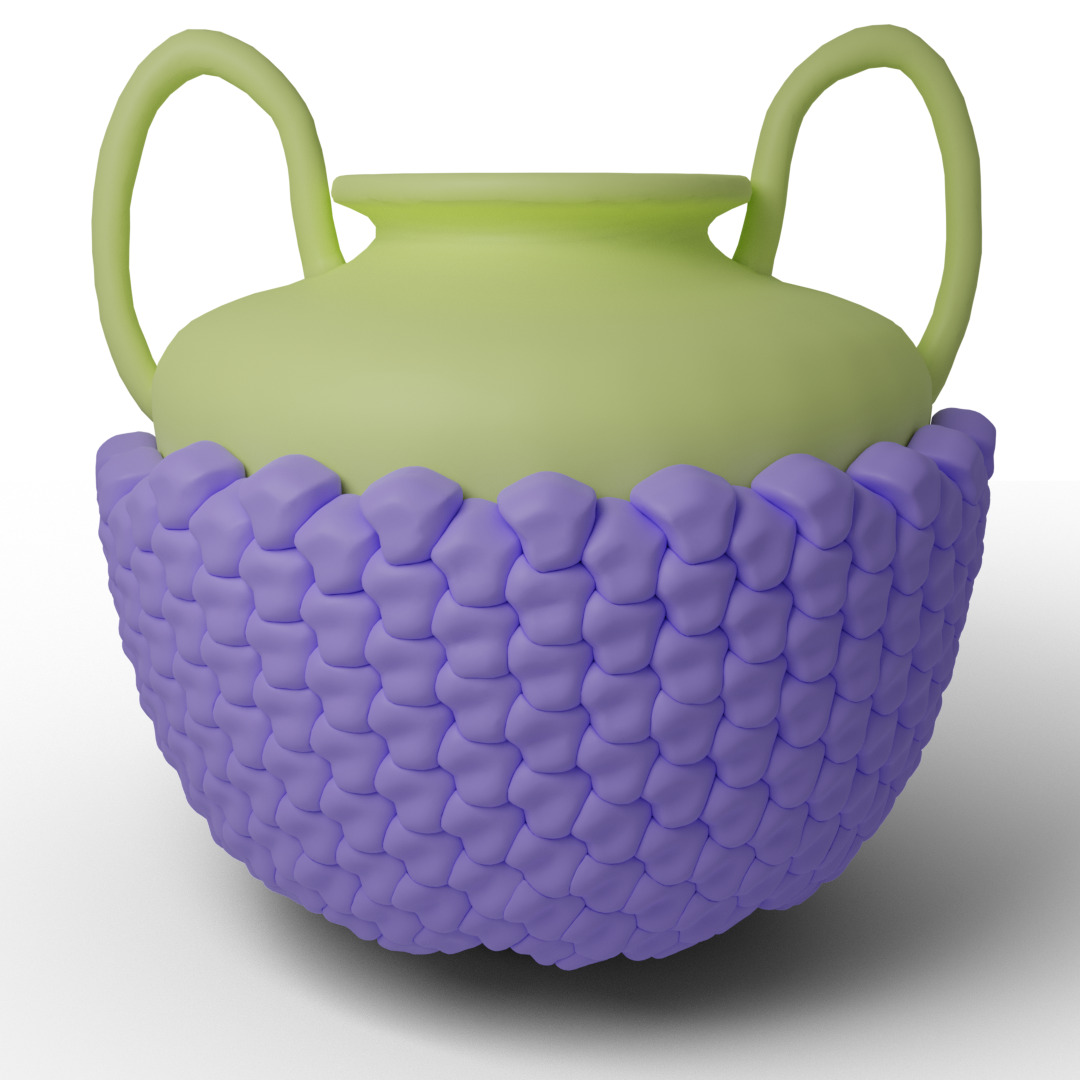} \\
\textsf{Sampled points} & \textsf{Initial placement} & \textsf{Packed decorations} \\
\includegraphics[width=0.30\linewidth]{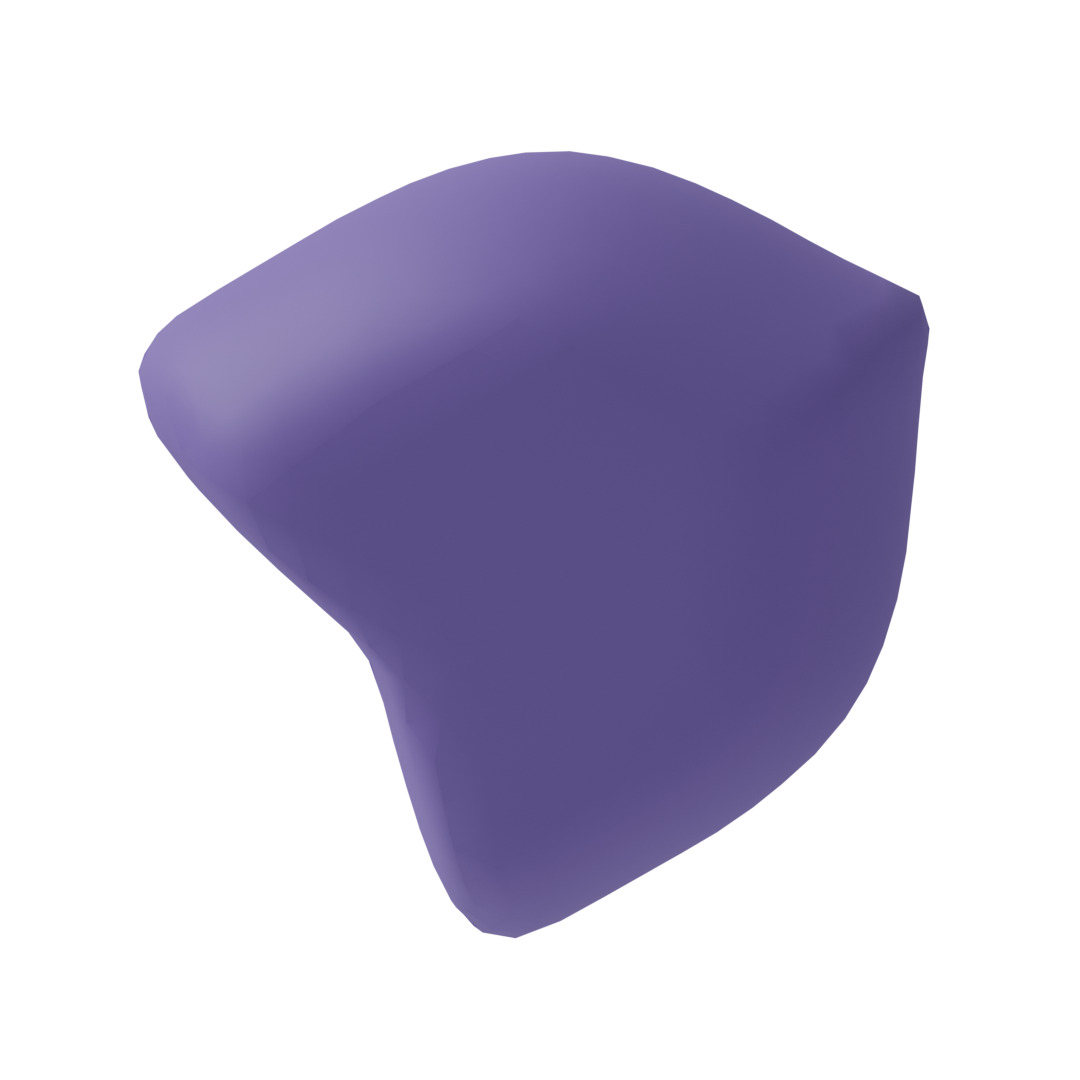} &
\includegraphics[width=0.30\linewidth]{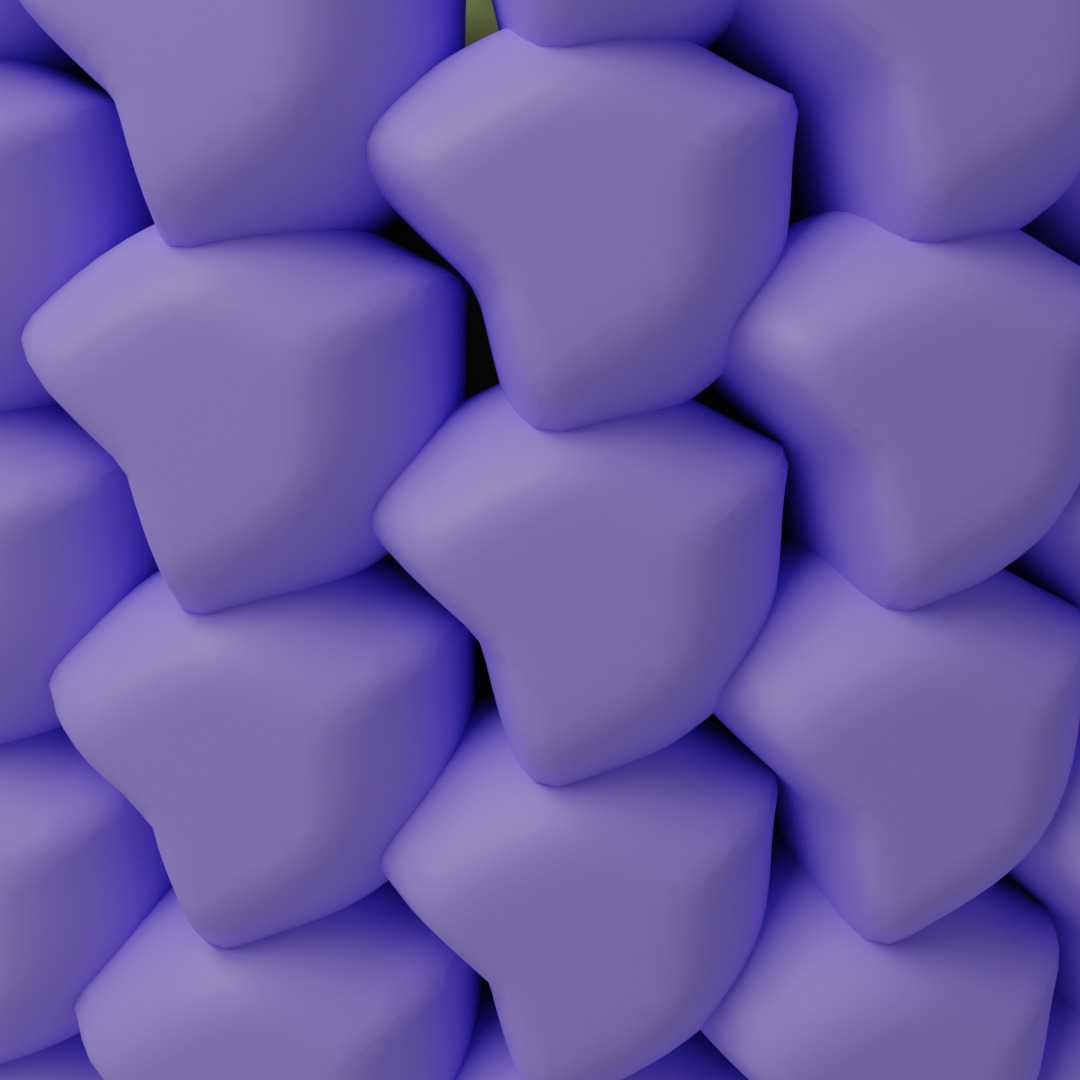} &
\includegraphics[width=0.30\linewidth]{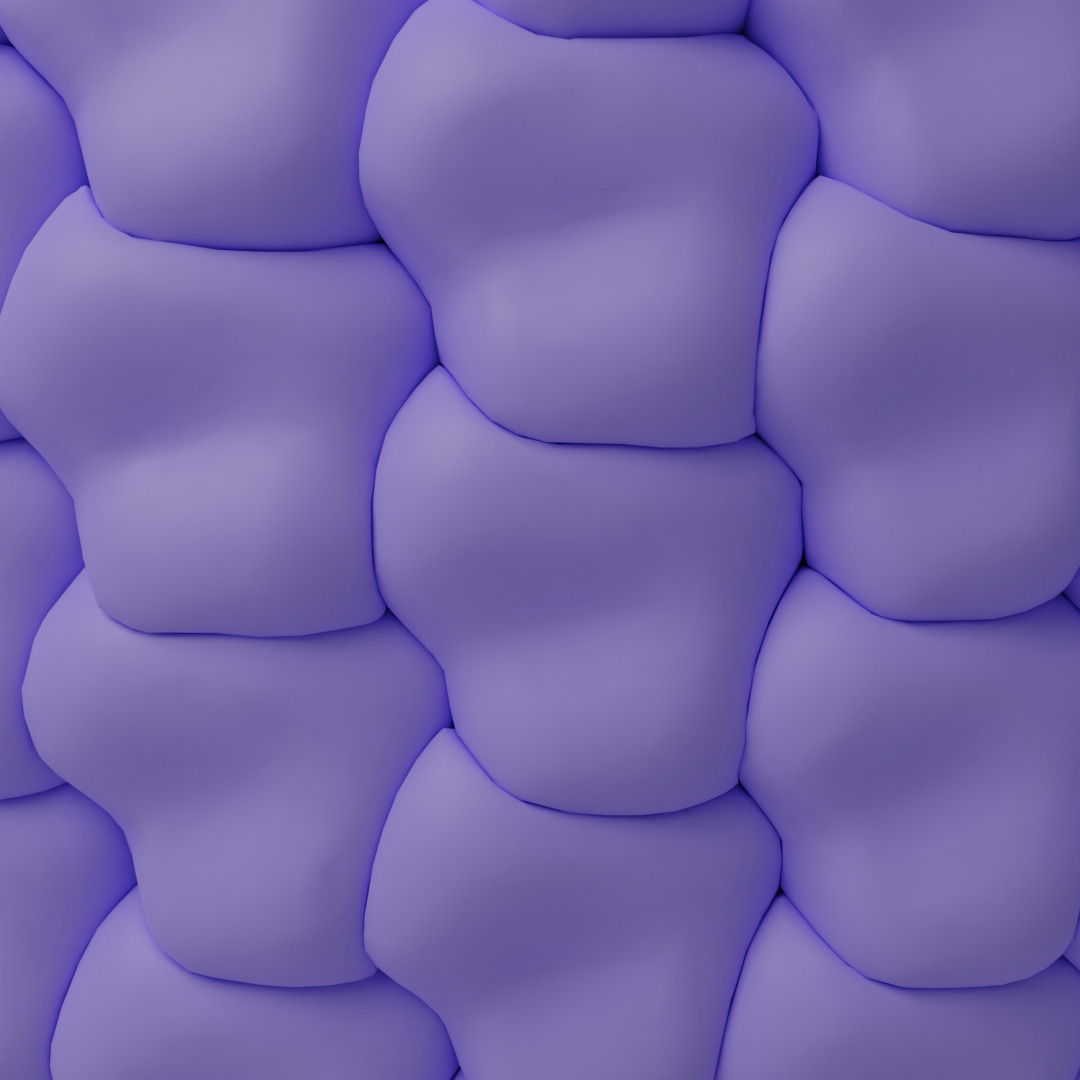} \\
\textsf{Decorative element} & \textsf{Close view (initial)} & \textsf{Close view (packed)} \\
\end{tabular}
\caption{Non-isotropic elements' sampling on non perpendicular stripes' directions.}
\label{fig:anisotropicDecos2}
\end{figure}

Our method is based on \cite{StripePatterns} that allows us to generate patterns of evenly spaced stripes over a generic base surface.
The method is based on a simple optimization problem and approximately maintains a user-specified orientation and spacing of the lines by
automatically inserting branch points where necessary. Furthermore, the generated patterns are globally continuous, which avoids visible seams after the applications of our decorative elements, and with frequencies not depending on the mesh resolution.
We exploit the stripe-based method by generating two crossing stripe patterns with a user-defined crossing angle and sampling seed points at the line crossings.
In this manner, we obtain lines with points sampled at constant distances and with a controlled offset among the decorative elements of two adjacent stripes.
\Cref{fig:anisotropicDecos2} shows an example with crossing stripes creating an angle of 60 degrees and elements placed with a constant orientation concerning the main stripe.

\begin{figure}[tb]
    \centering
    \small
\begin{tabular}{@{}c@{\hspace{0.03in}}c@{\hspace{0.03in}}c@{}}
\includegraphics[width=0.32\linewidth]{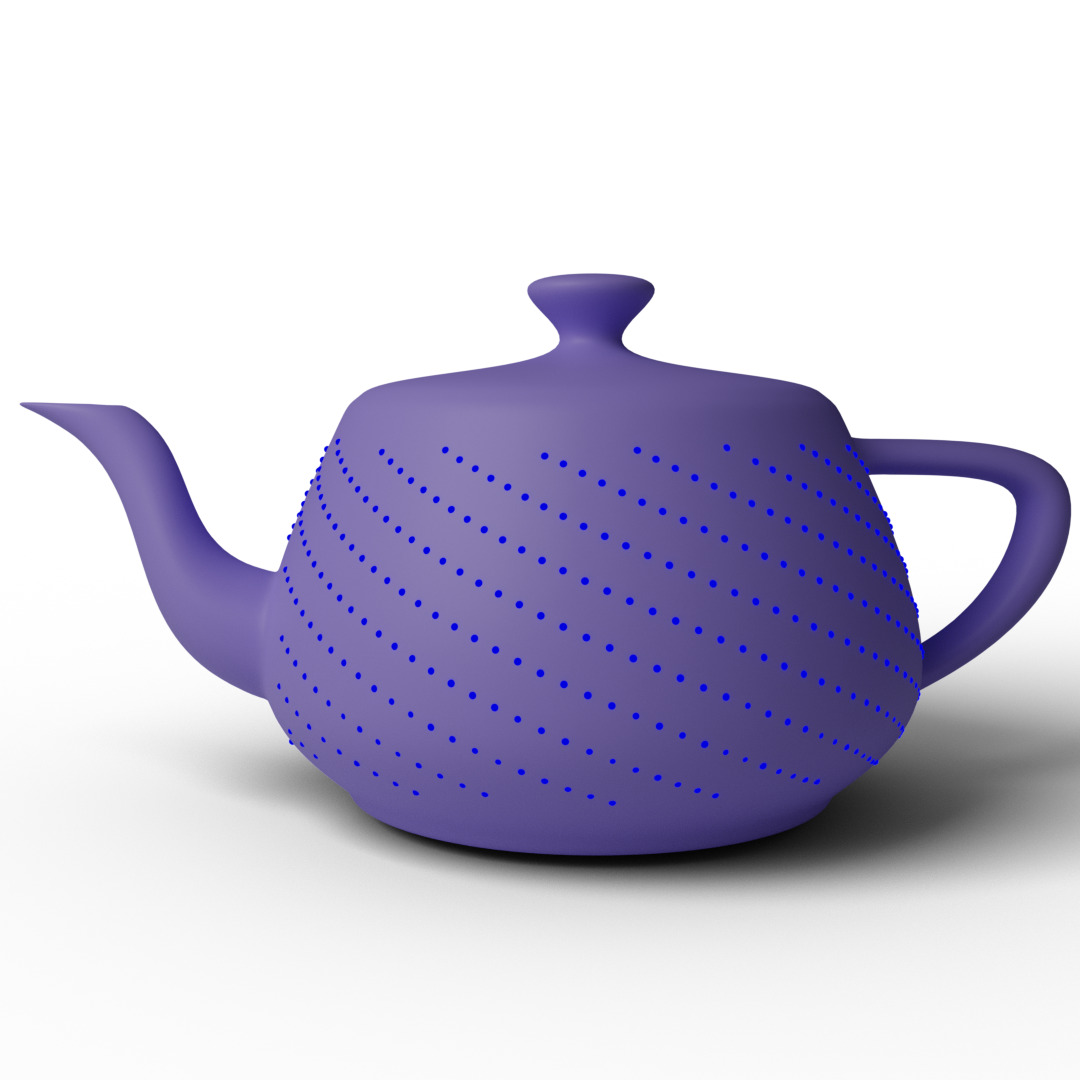} &
\includegraphics[width=0.32\linewidth]{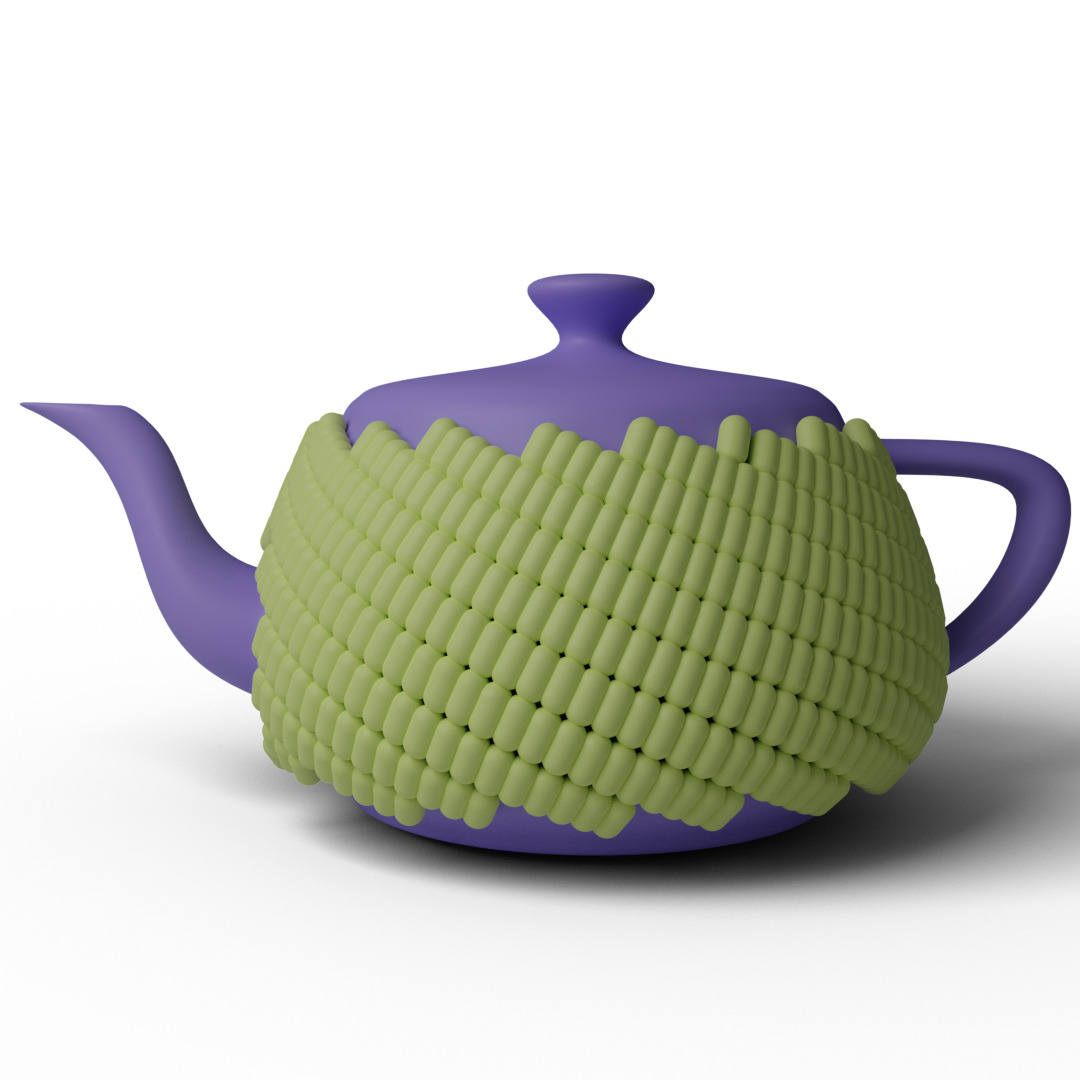} &
\includegraphics[width=0.32\linewidth]{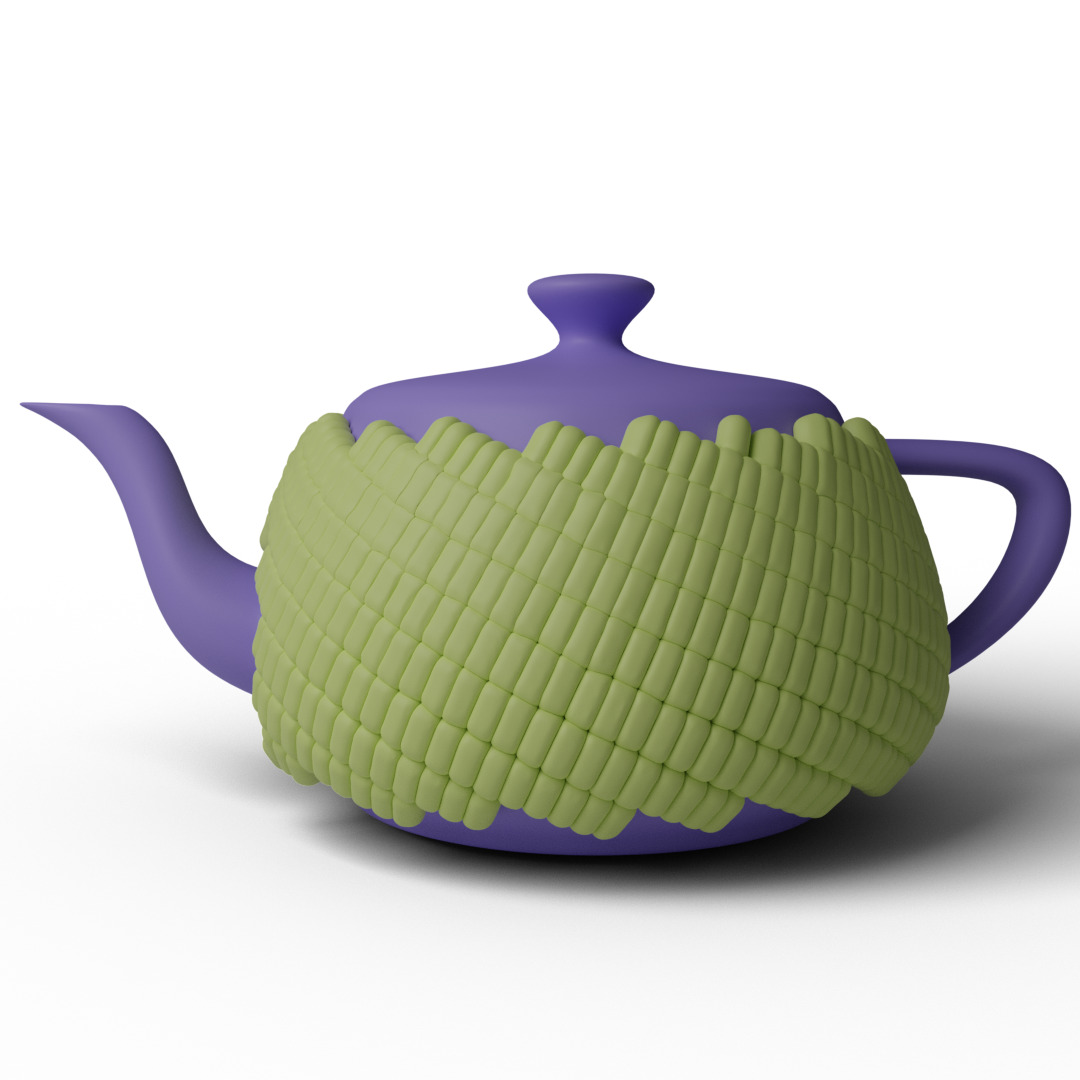} \\
\textsf{Sampled points} & \textsf{Initial placement} & \textsf{Packed decorations} \\
\includegraphics[width=0.30\linewidth]{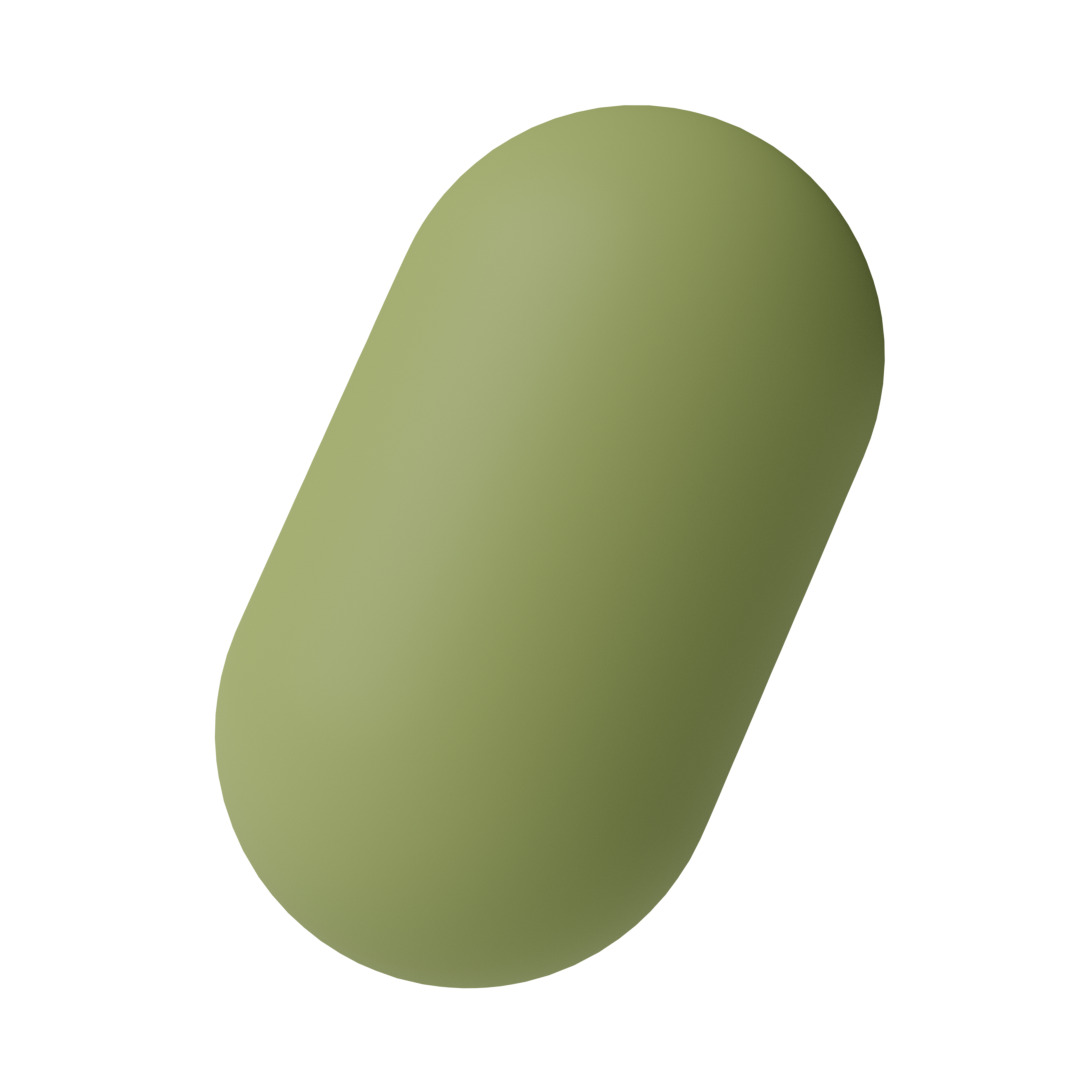} &
\includegraphics[width=0.30\linewidth]{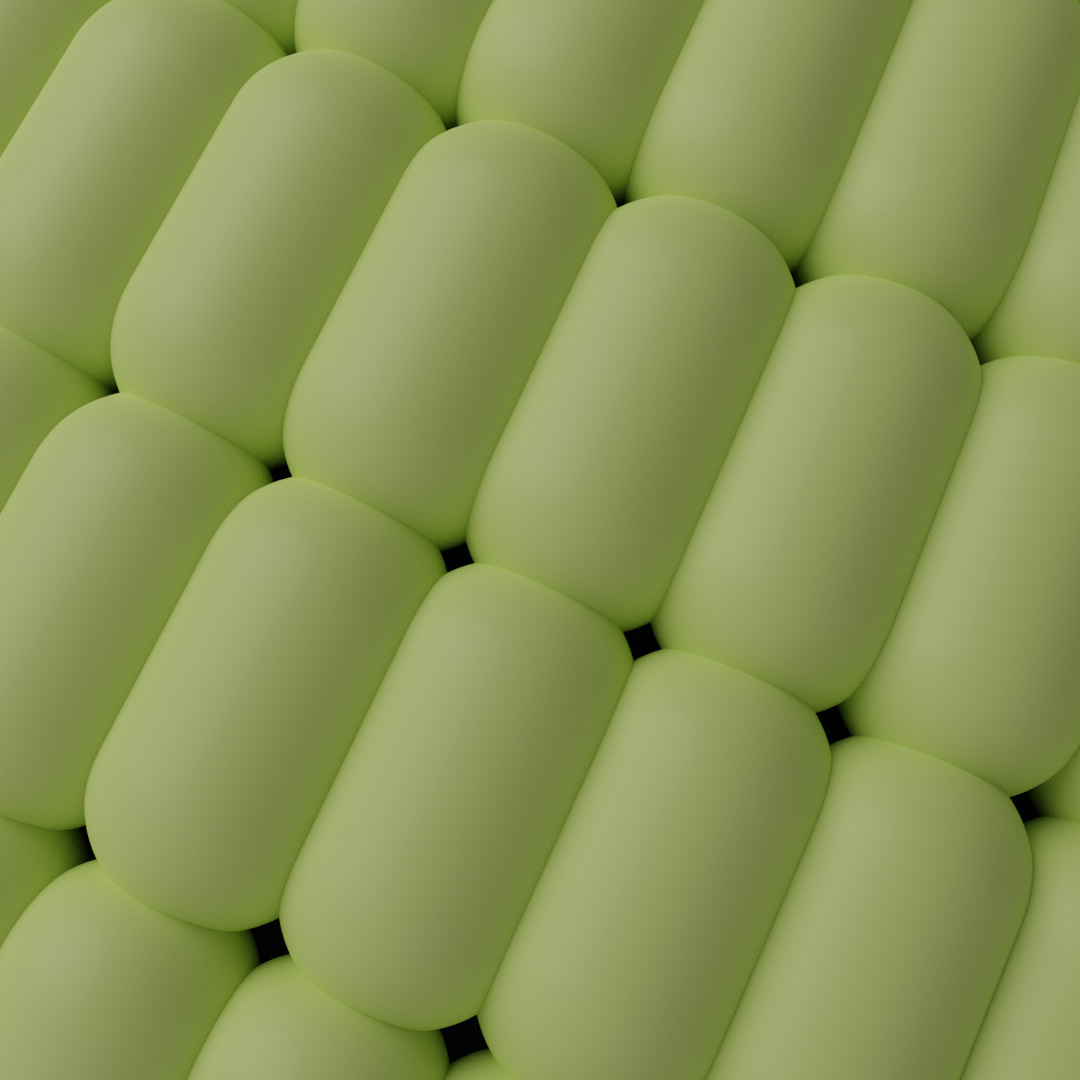} &
\includegraphics[width=0.30\linewidth]{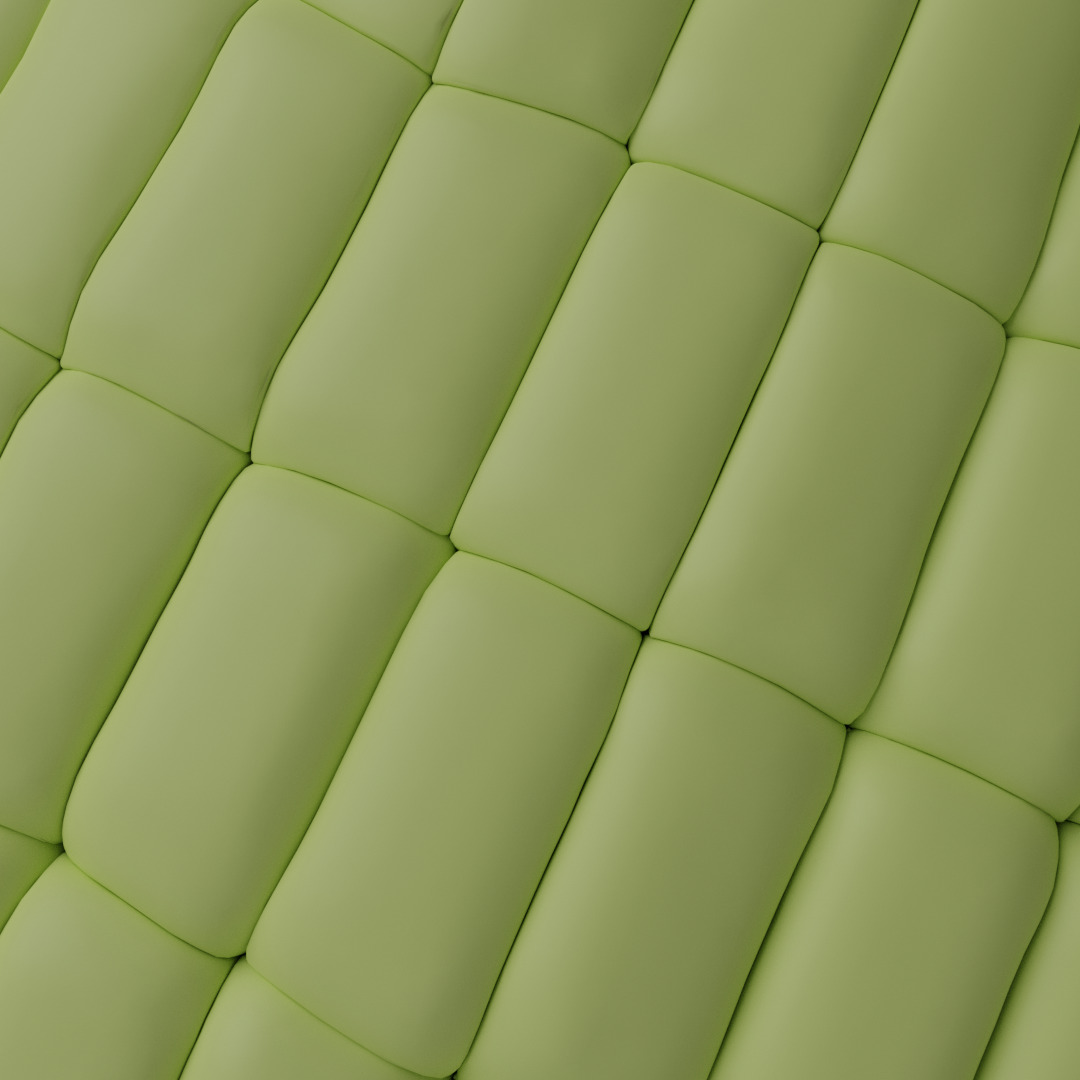} \\
\textsf{Decorative element} & \textsf{Close view (initial)} & \textsf{Close view (packed)} \\
\end{tabular}
   \caption{Using the stripe-based sampling, we can pack on a generic shape like the teapot non-isotropic decorations aligned with a directional field.}
    \label{fig:anisotropicDecos}
\end{figure}

Since we can independently set the sample spacing on the main stripes and the distance between adjacent stripes, we are no longer forced to pack objects with approximate radial symmetry, but we can insert elements with arbitrary elongation. Furthermore, since the stripes encode a local directional field, we can use non-isotropic decorative elements oriented in a controlled way with respect to the stripes' directions. \Cref{fig:anisotropicDecos} shows an example with perpendicular stripes-based sampling and non-isotropic elements with main directions aligned with the stripes fields.

\begin{figure}[b]
\centering
\small
\begin{tabular}{@{}c@{\hspace{0.03in}}c@{\hspace{0.03in}}c@{}}
\includegraphics[width=0.32\linewidth]{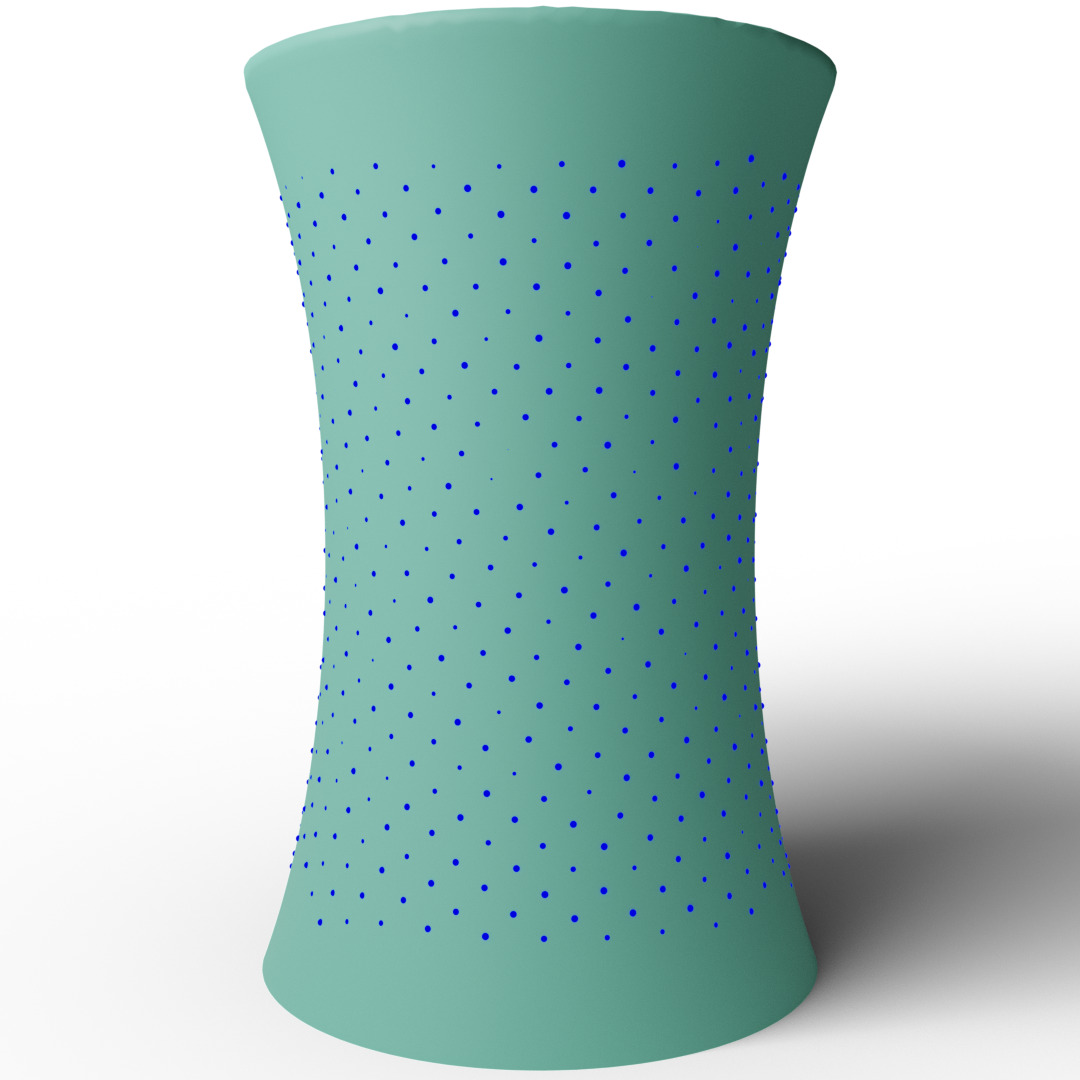} &
\includegraphics[width=0.32\linewidth]{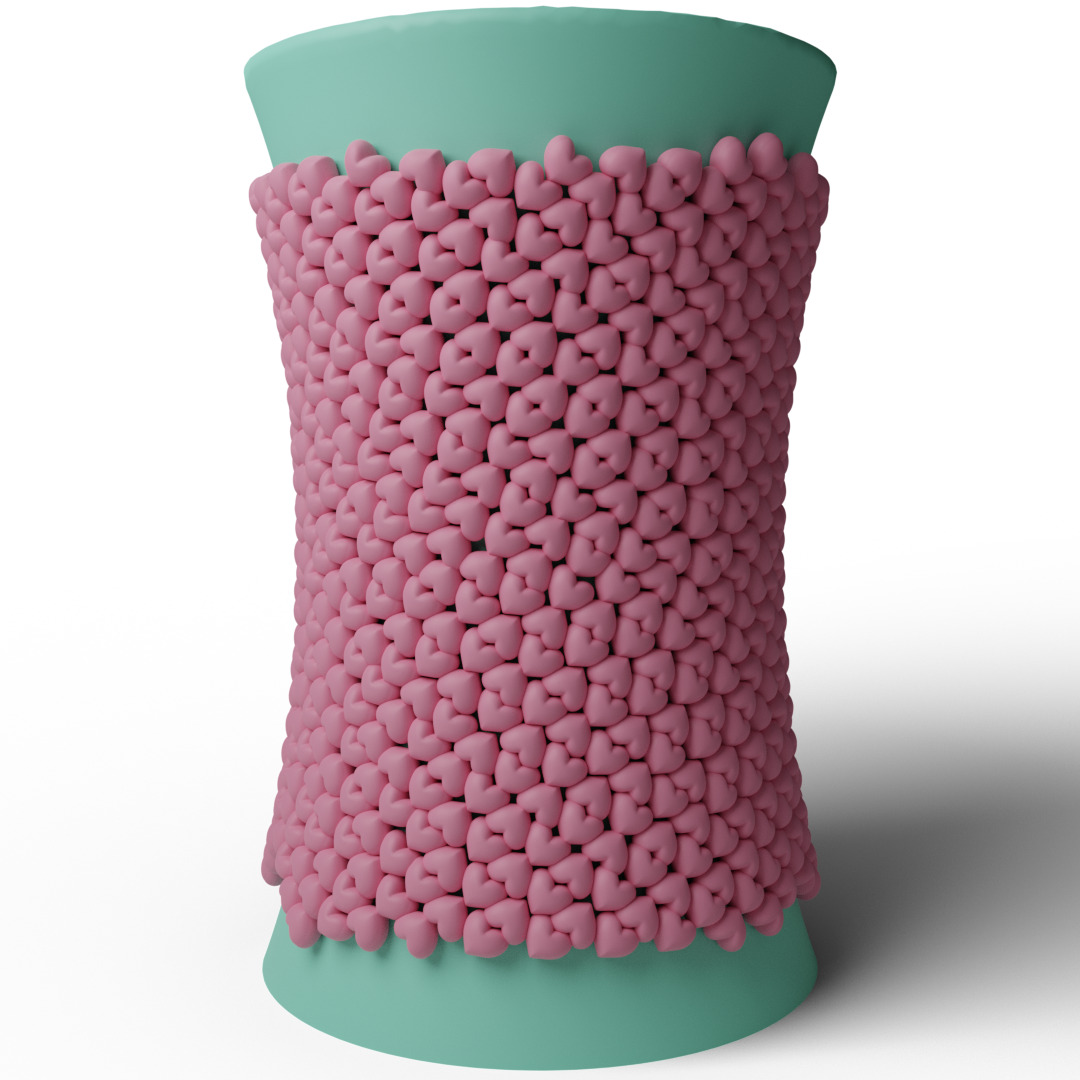} &
\includegraphics[width=0.32\linewidth]{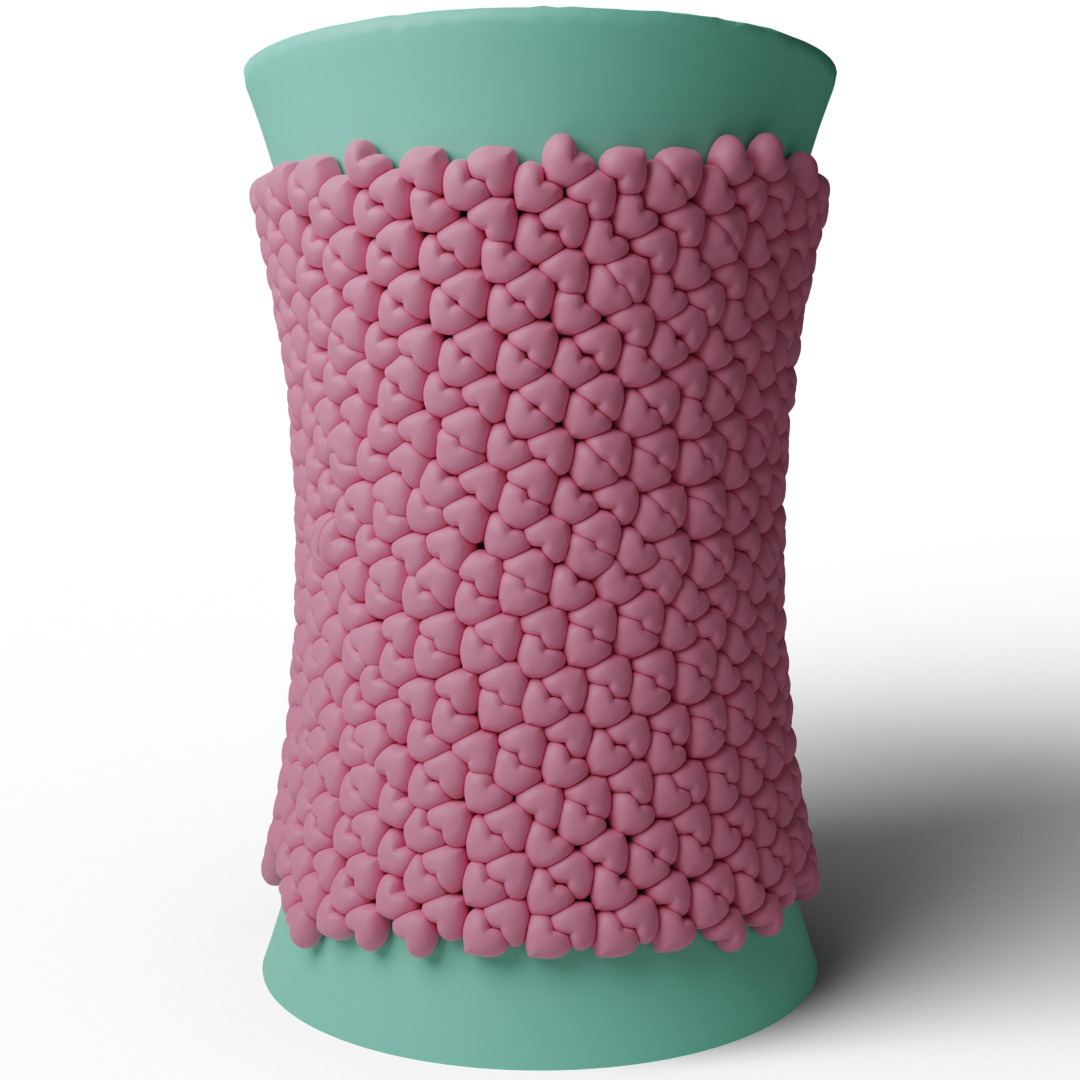} \\
\textsf{Sampled points} & \textsf{Initial placement} & \textsf{Packed decorations} \\
\includegraphics[width=0.30\linewidth]{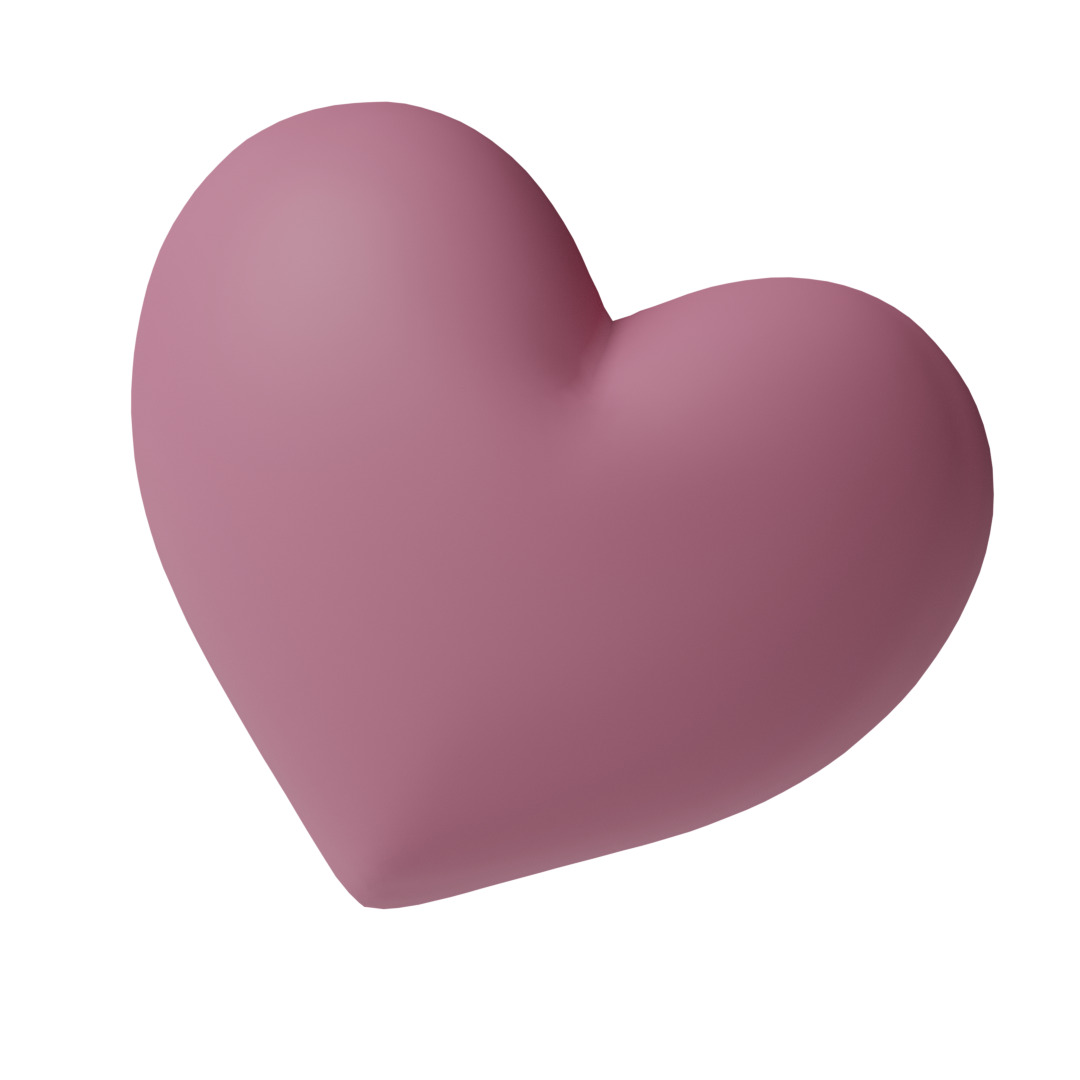} &
\includegraphics[width=0.30\linewidth]{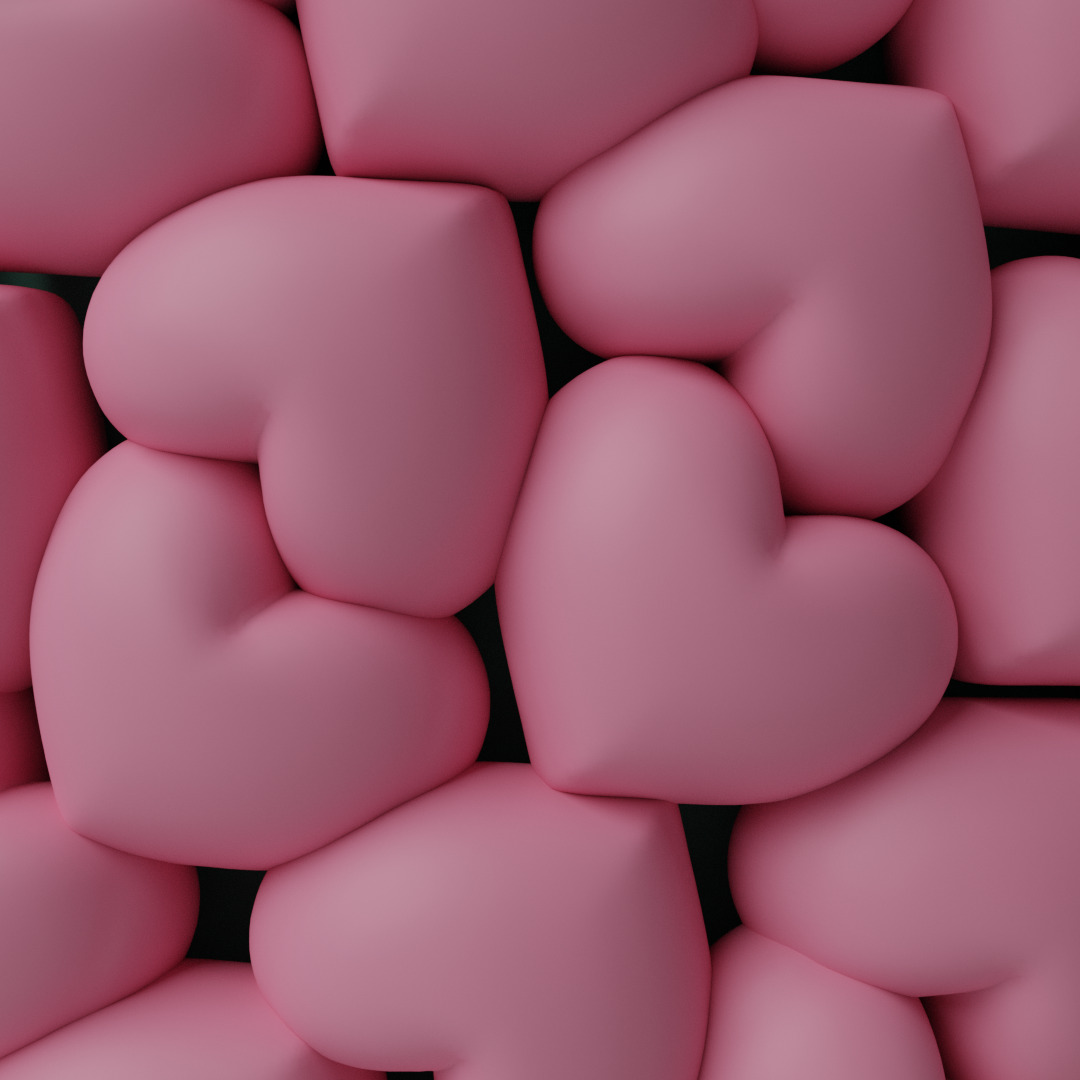} &
\includegraphics[width=0.30\linewidth]{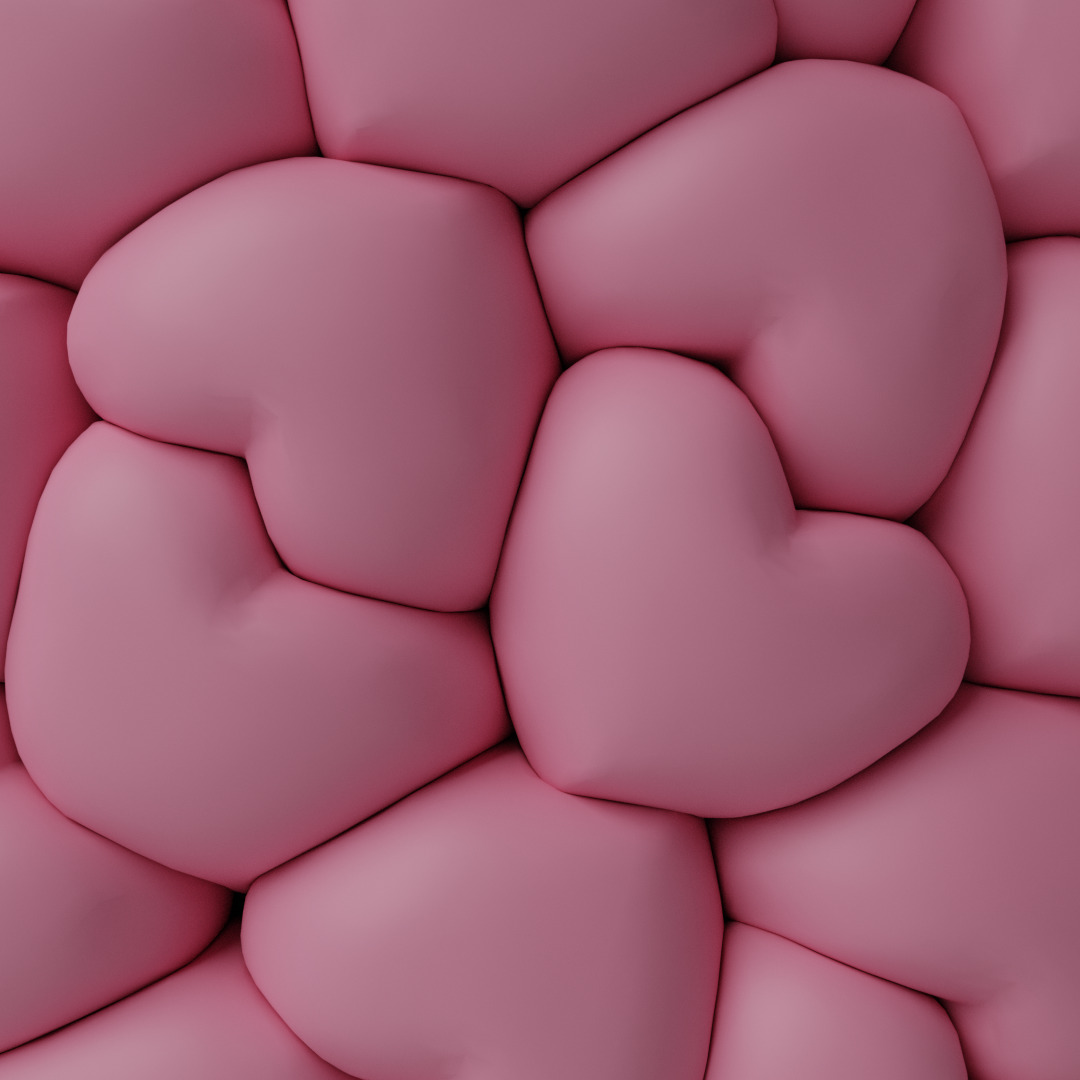} \\
\textsf{Decorative element} & \textsf{Close view (initial)} & \textsf{Close view (packed)} \\
\end{tabular}
\caption{Example of alternate elements' orientation along the main field direction.}
\label{fig:rotdeco}
\end{figure}

In our stripe-based algorithm, we create the sampling points and estimate and store the point connectivity of the sample along with the two stripe directions. In this way, we can develop decorations alternating different elements and elements' orientations, as shown in \cref{fig:rotdeco}.

\paragraph{Elements' perturbation}
Aside from placement, we can easily integrate controls for decorations' rotations and filling different areas with different decorations. The rotation for each element can be set randomly or aligned to a field defined over the object surface, giving the possibility to create different artistic effects that work particularly well when placing elongated objects over a gridded seed distribution. Figure \ref{fig:rotdeco} shows an example of an artistic pattern obtained with alternate rotations of 180 degrees of the elements along the principal stripe field direction.

\begin{figure}[tb]
\centering
\small
\begin{tabular}{@{}c@{\hspace{0.03in}}c@{\hspace{0.03in}}c@{}}
\includegraphics[width=0.32\linewidth]{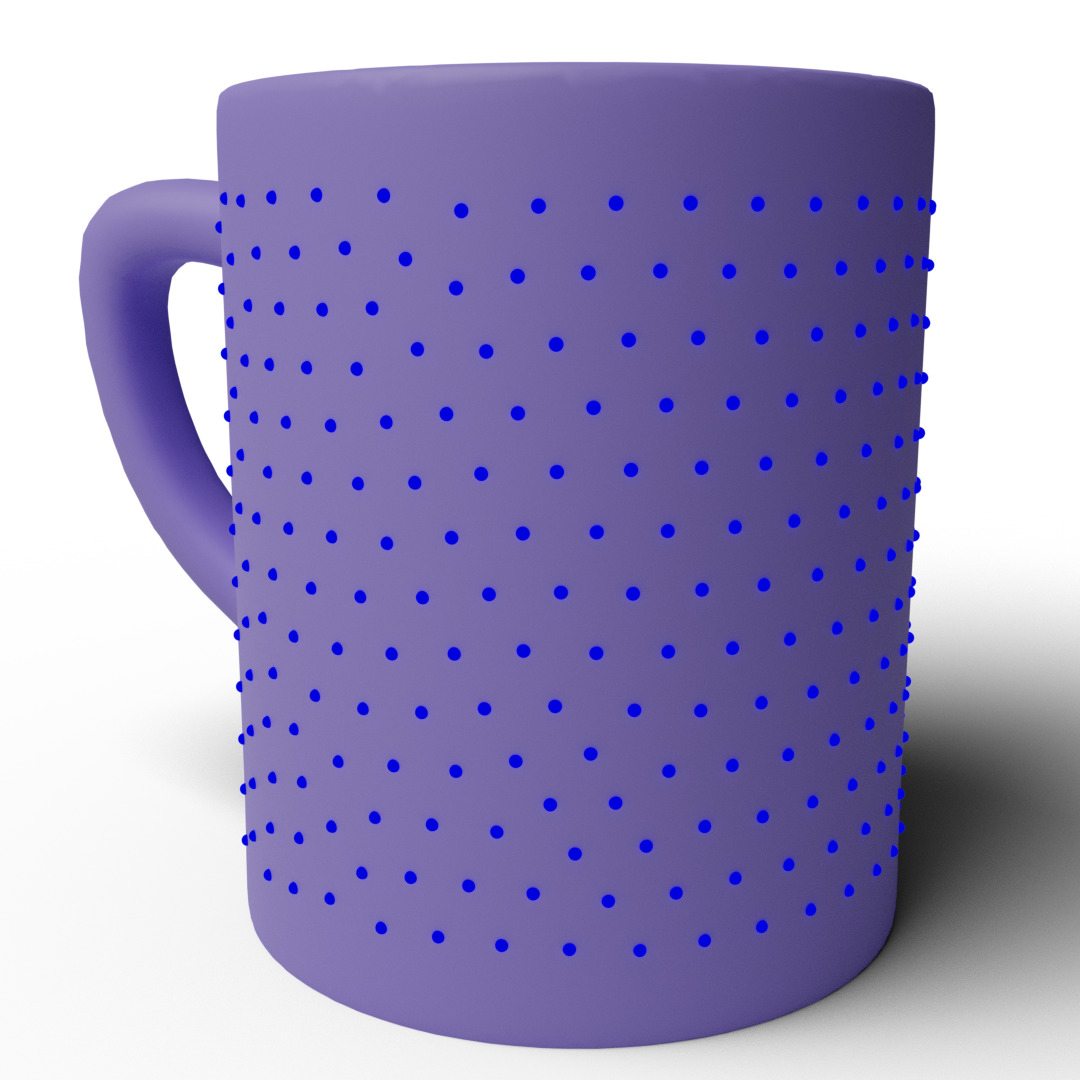} &
\includegraphics[width=0.32\linewidth]{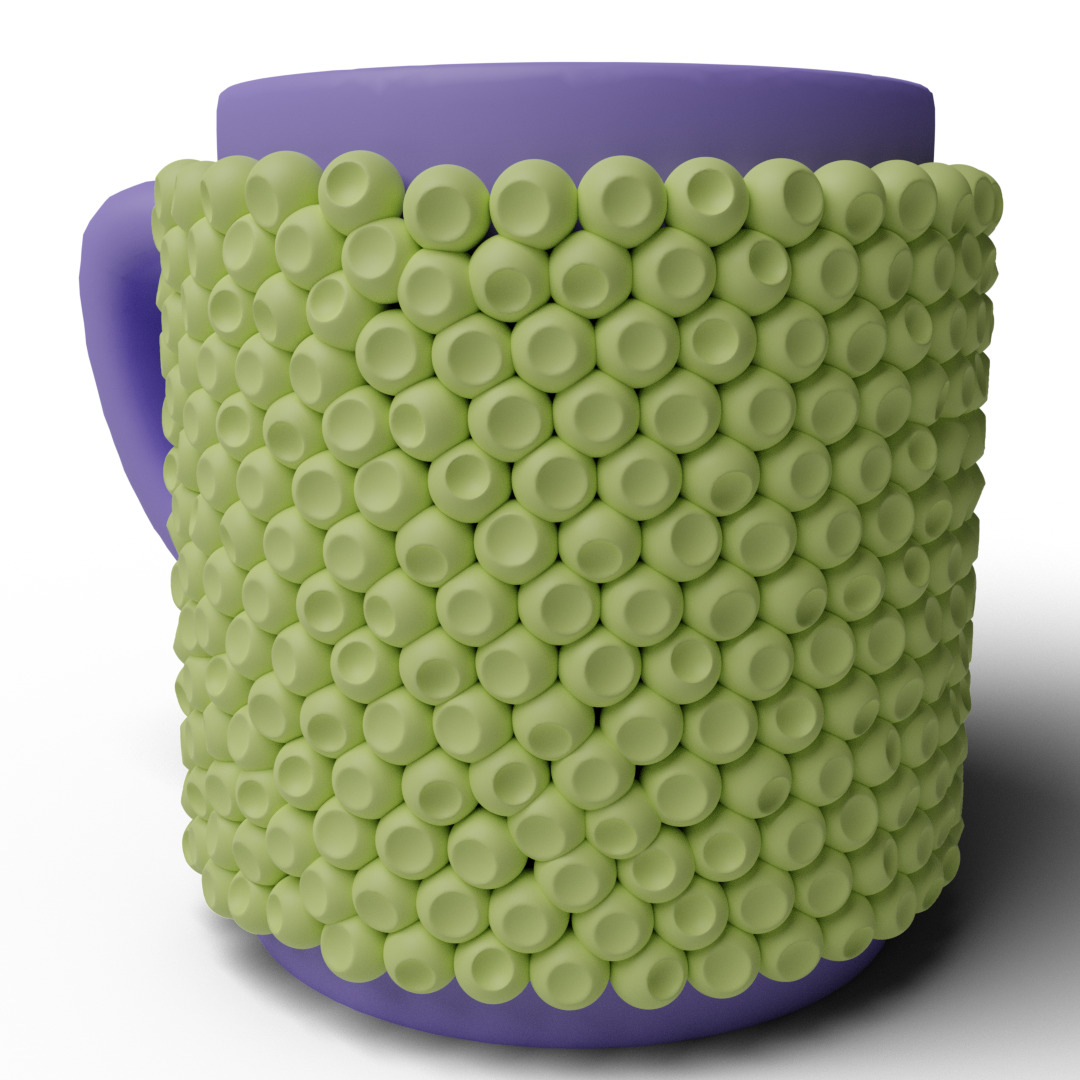} &
\includegraphics[width=0.32\linewidth]{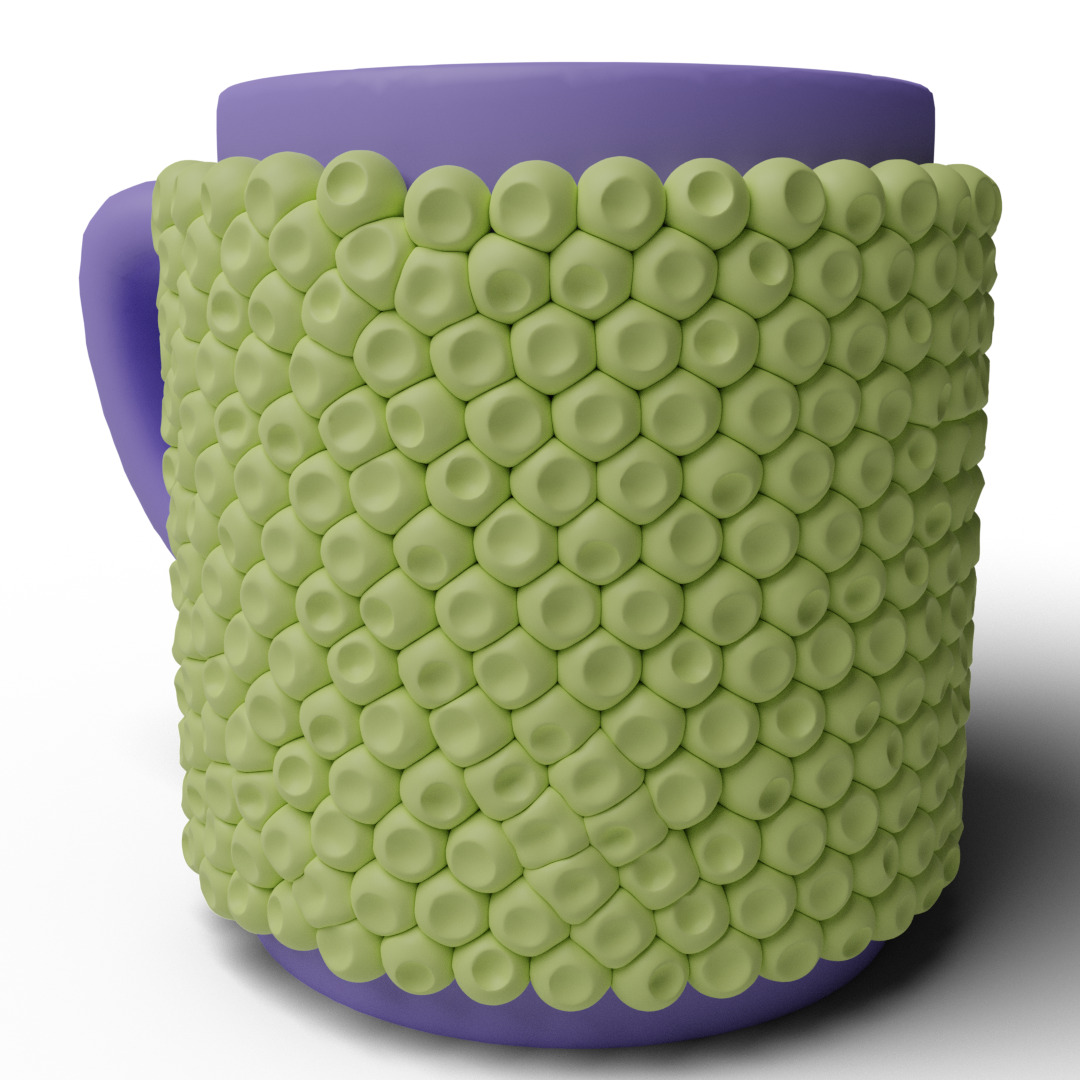} \\
\textsf{Sampled points} & \textsf{Initial placement} & \textsf{Packed decorations} \\
\includegraphics[width=0.30\linewidth]{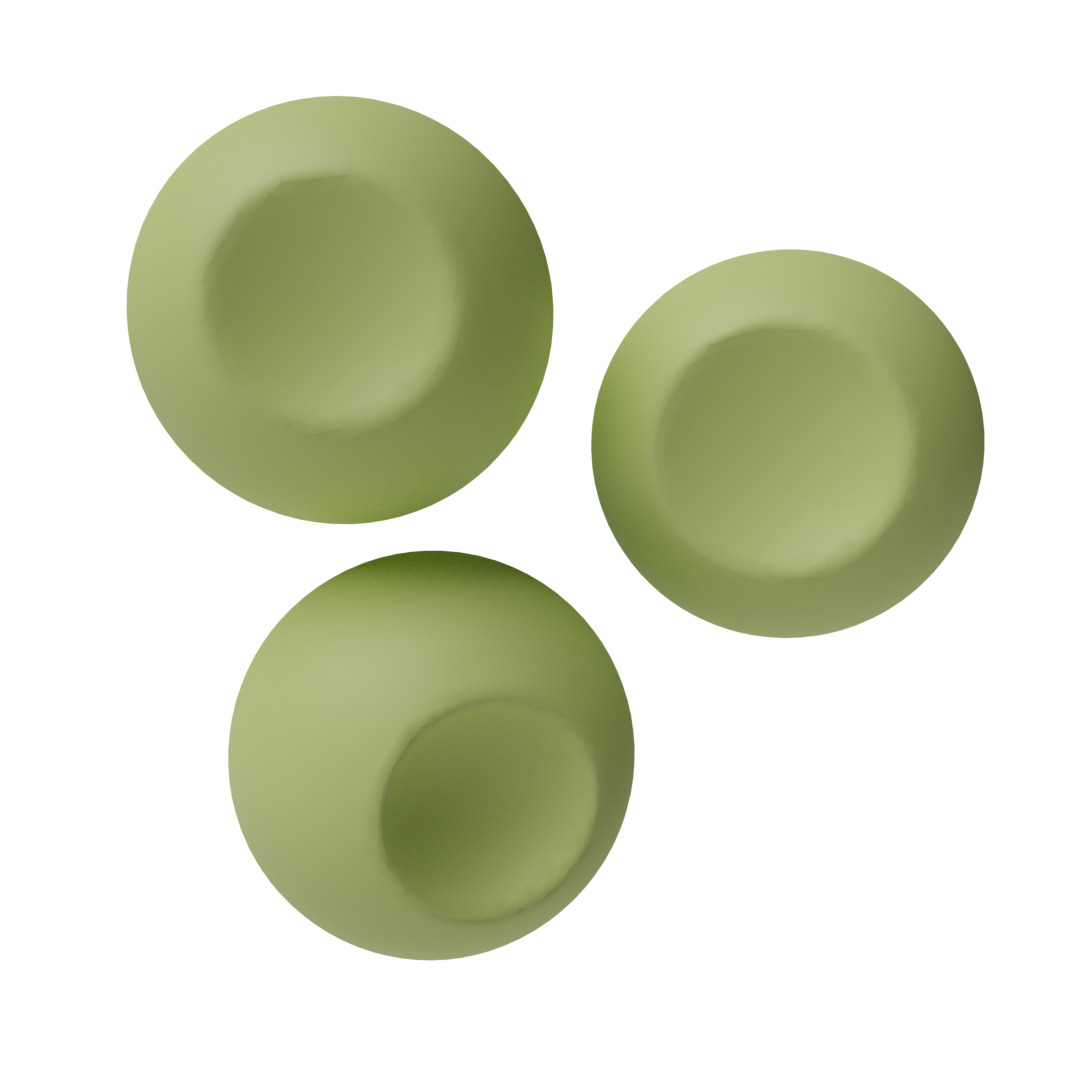} &
\includegraphics[width=0.30\linewidth]{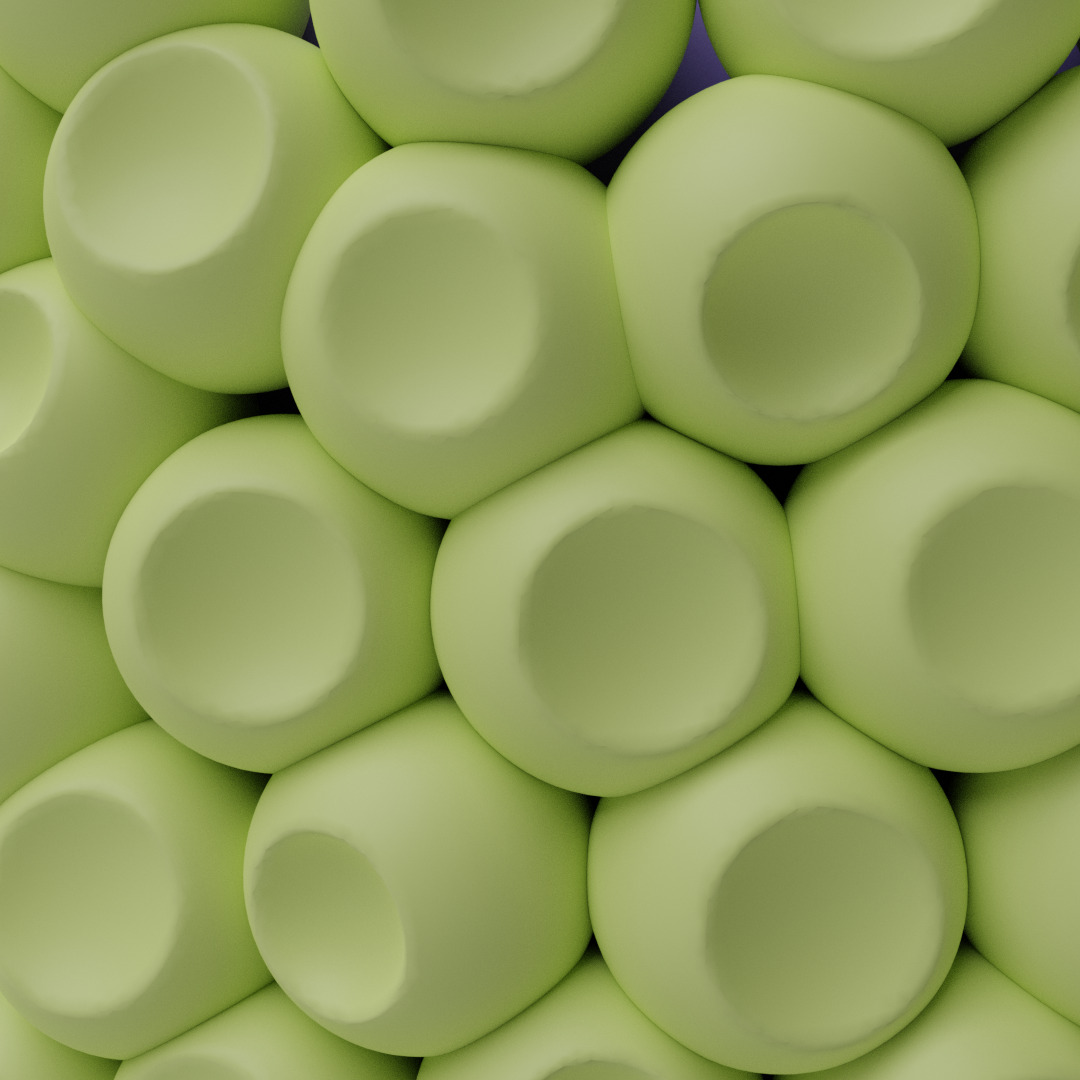} &
\includegraphics[width=0.30\linewidth]{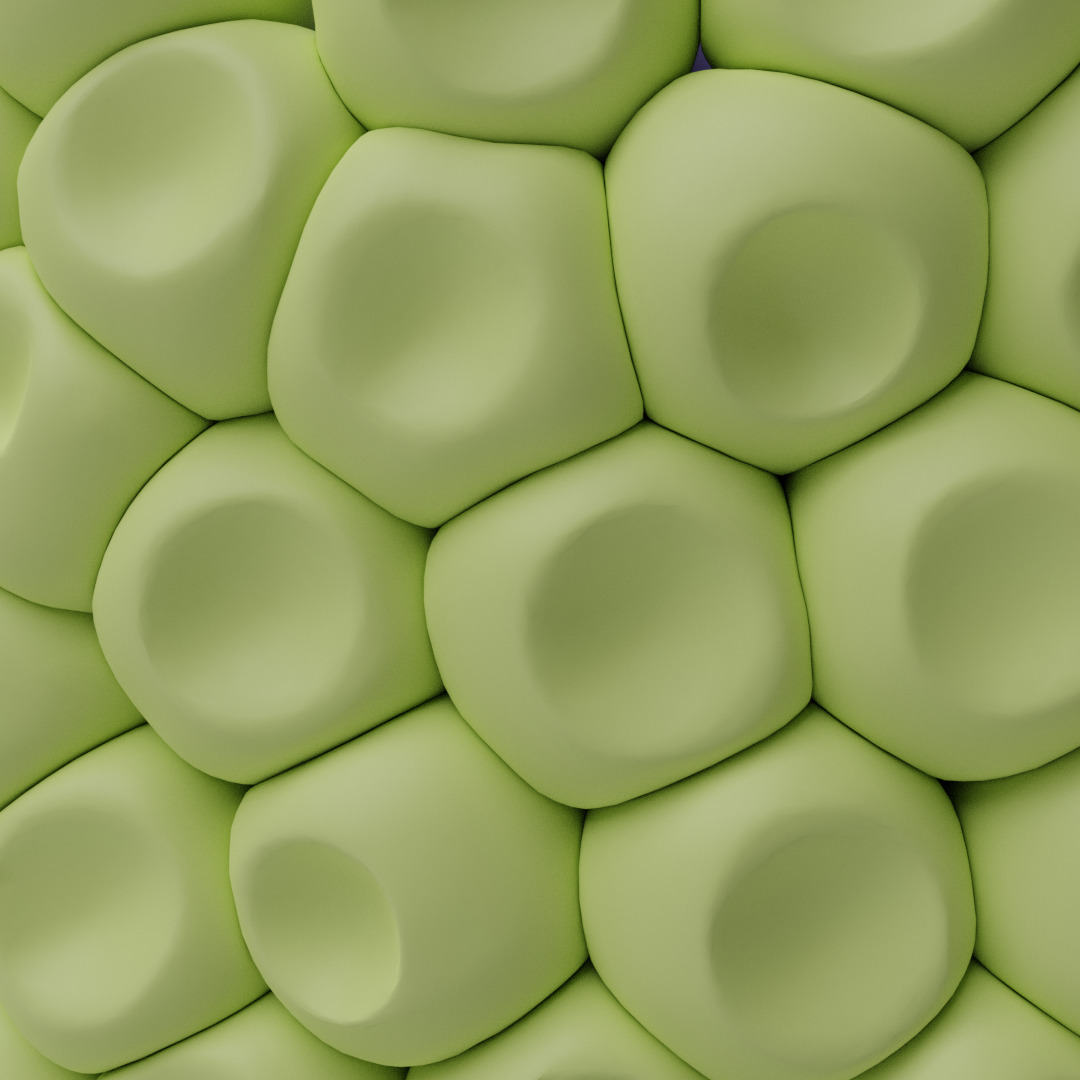} \\
\textsf{Decorative elements} & \textsf{Close view (initial)} & \textsf{Close view (packed)} \\
\end{tabular}
\caption{Using multiple elements' shapes and random orientations it is possible to create a variety of artistic designs.}
\label{fig:vardeco}
\end{figure}

The average elements' size is chosen to guarantee the desired decorative effect, and the number of elements can be tuned to guarantee the desired amount of initial overlap. However, we leave the possibility of applying a further random perturbation to the size of each decoration to increase the number of design options. We allow variations up to $\pm 10\%$, as larger ones would lead to inconsistent overlaps and too many empty spaces.

Lastly, it is possible to load more than one decoration and randomly select one of those for each seed. We can exploit this to increase the final results' variety loading a multitude of slightly different decorations (e.g., stones of various types, flowers, etc.).

\Cref{fig:vardeco} shows an example of decoration mimicking the one in \cref{fig:examples} obtained by using three different elements shapes and random rotations with respect to the main field direction.
\section{Output generation}
\label{sec:output}

The last step we need to perform is to convert the process results into high-quality meshes for visualization or printable shells to decorate real-world objects.

\paragraph{High-quality meshes}
To generate a 3D mesh of the whole object, we first independently create meshes for each deformed decorative element using the Marching Cubes algorithm \cite{MarchingCubes}. Then, we merge the elements and the base object into a single model and remesh the result using the Central Voronoi Tessellation approach \cite{YanLLSW09}, which provides feature-preserving results and is robust and efficient for large meshes.

\begin{figure}[b]
    \centering
    \includegraphics[width=\linewidth]{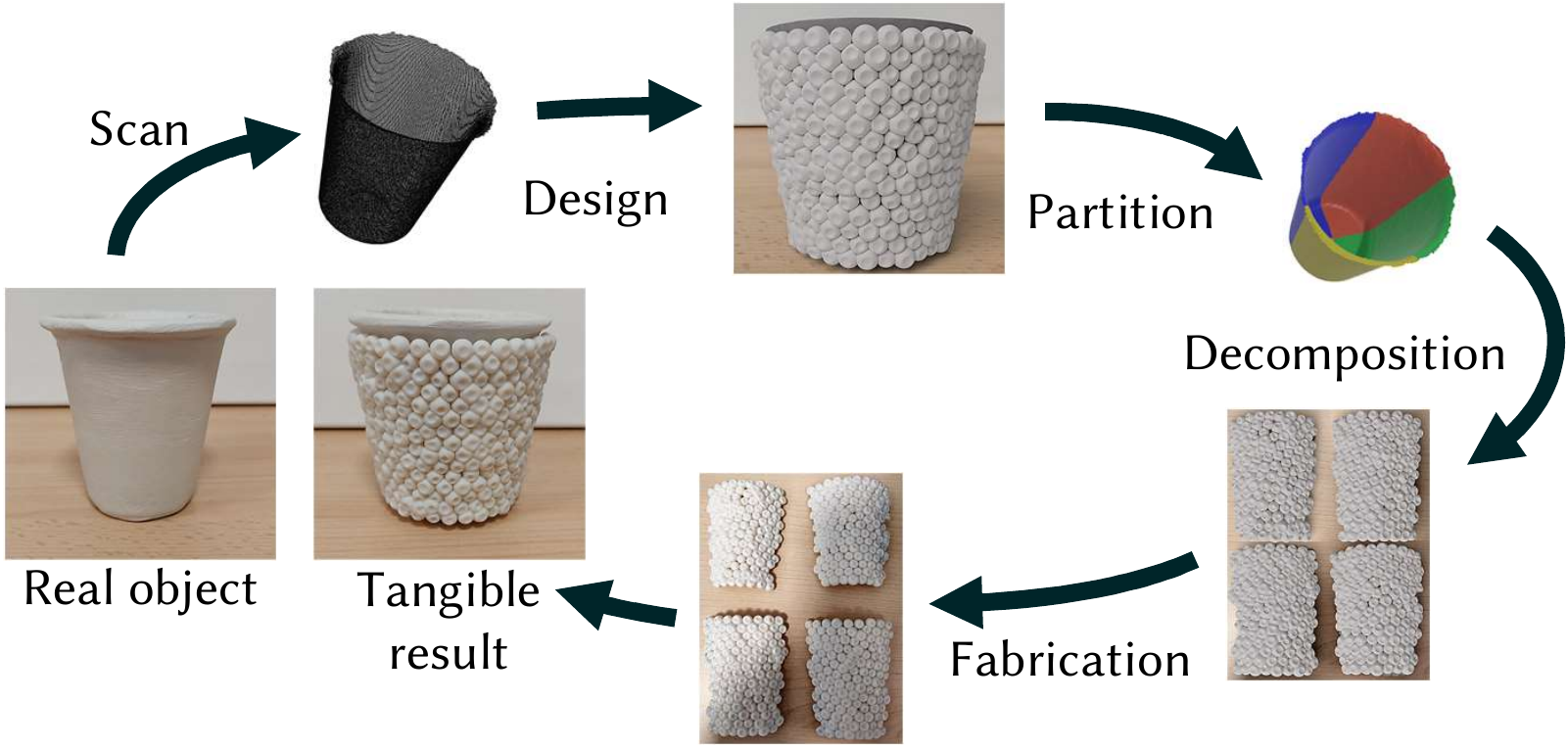}
    \caption{Design and fabrication of tangible decorative shell for a real object. The whole production pipeline is depicted from the real object to decorate to the final tangible result obtained applying the fabricated decomposed shell over the surface.}
    \label{fig:fabricationGlass}
\end{figure}

\paragraph{Printable shells}
It is possible to use our pipeline also to create decorative shells for real-world objects. To decorate an object, we first scan it to obtain an accurate digital model and then create a \emph{PAVEL} decoration for it. This produces a mesh representing all the decorative elements fused together. We then decompose this final mesh into printable parts and wrap it around the original object. \Cref{fig:fabricationGlass} shows the whole process of decorating an object.

For objects with a simple topology, the decomposition works as follows. First, we segment the base object. Any mesh partitioning can be used for this purpose. In the example shown, we cut the object with two mutually perpendicular planes passing through the model's symmetry axis. Given the base segmentation, we cluster the deformed elements placed on each patch and remesh them into single shell elements to be 3D separately printed. One can use a more appropriate decomposition for more complex base objects while the rest of the pipeline remains the same.

\begin{wrapfigure}[9]{l}{.3\columnwidth}
\vspace{-1em}
\includegraphics[width=.35\columnwidth]{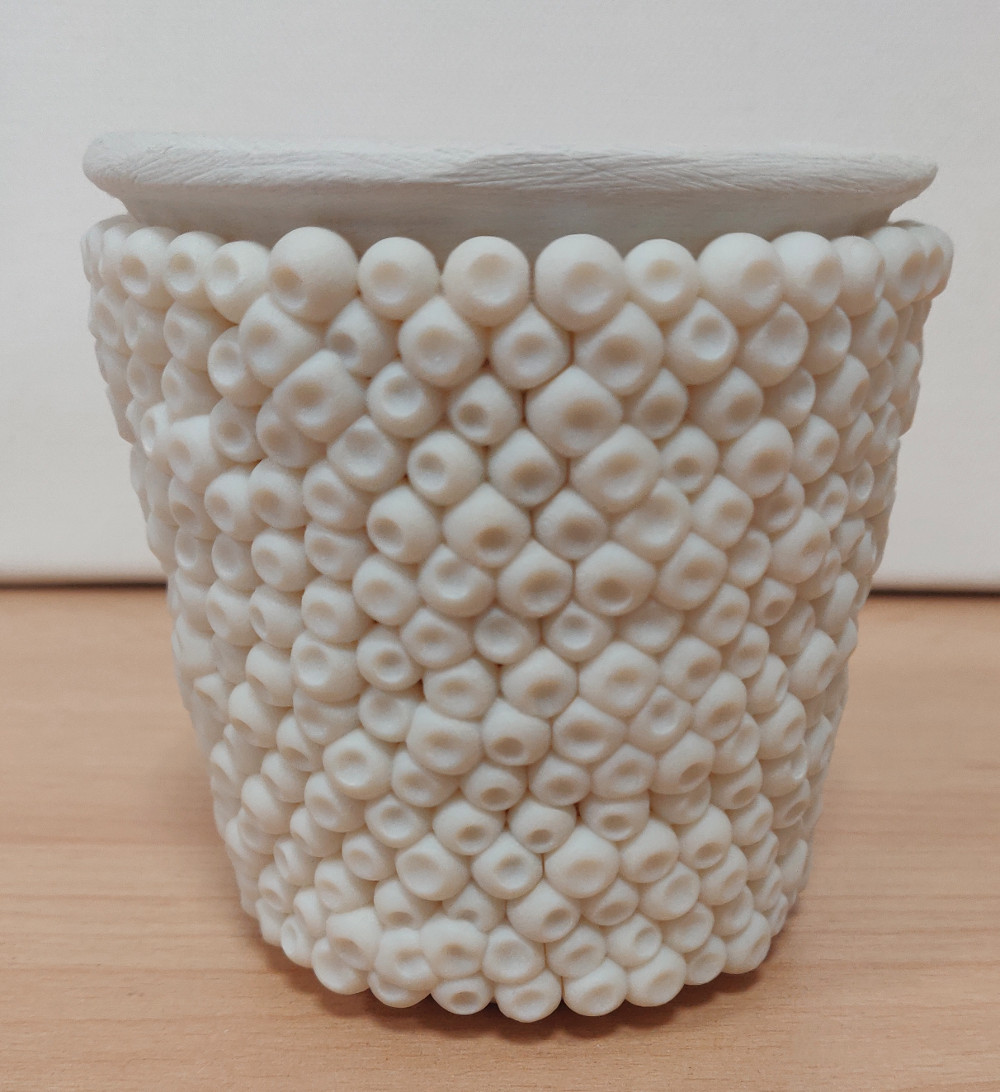}
\end{wrapfigure}

In the inset, we show the final result obtained by gluing the parts of the tangible shell created from the design in \cref{fig:fabricationGlass} over the base object.
In this case, the object was a simple paper cup covered with modeling clay. We scanned it with a low-cost device (Intel Realsense F200) and automatically decorated with an isotropic distribution of elements with different shapes and randomized size and orientation.
We finally printed the shell elements with a Stratasys J826 3D printer.
\section{Results}
\label{sec:results}

\begin{figure}[tb]
\centering
\small
\begin{tabular}{@{}c@{\hspace{0.05in}}c@{}}
\includegraphics[width=.43\linewidth]{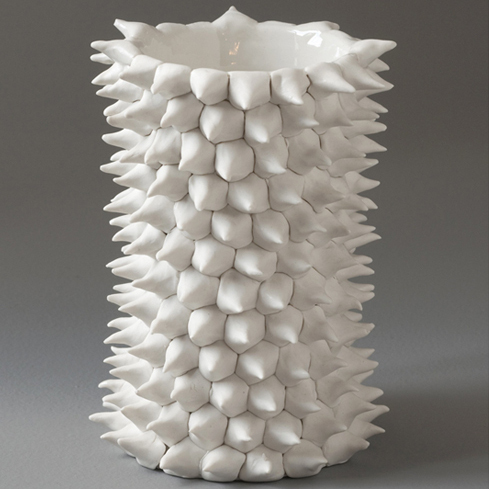} &
\includegraphics[width=.43\linewidth]{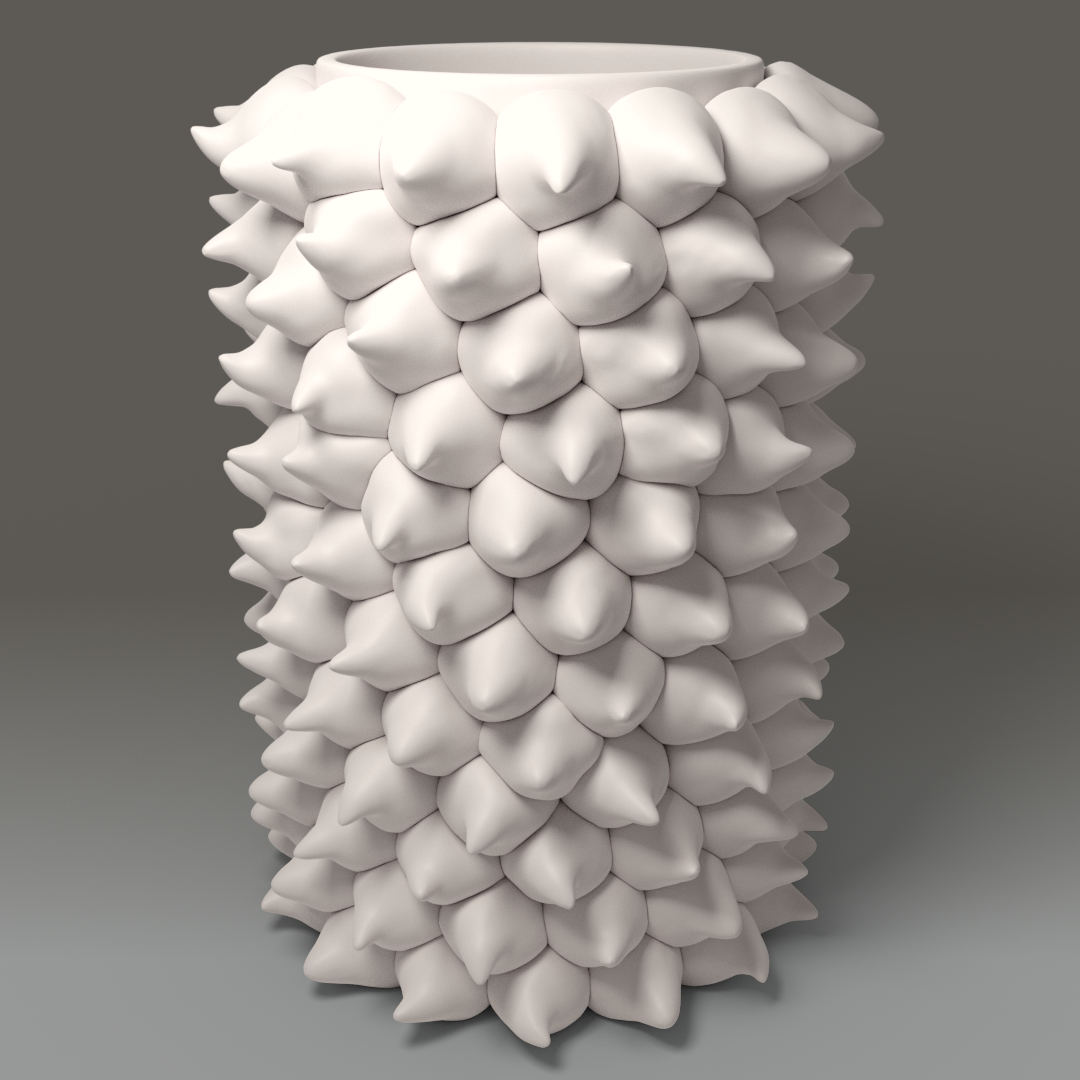} \\
\includegraphics[width=.43\linewidth]{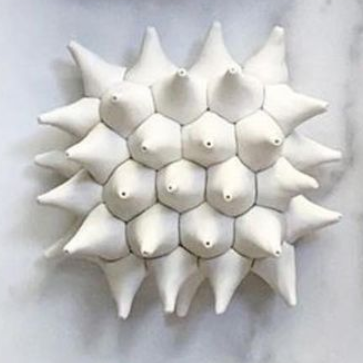} &
\includegraphics[width=.43\linewidth]{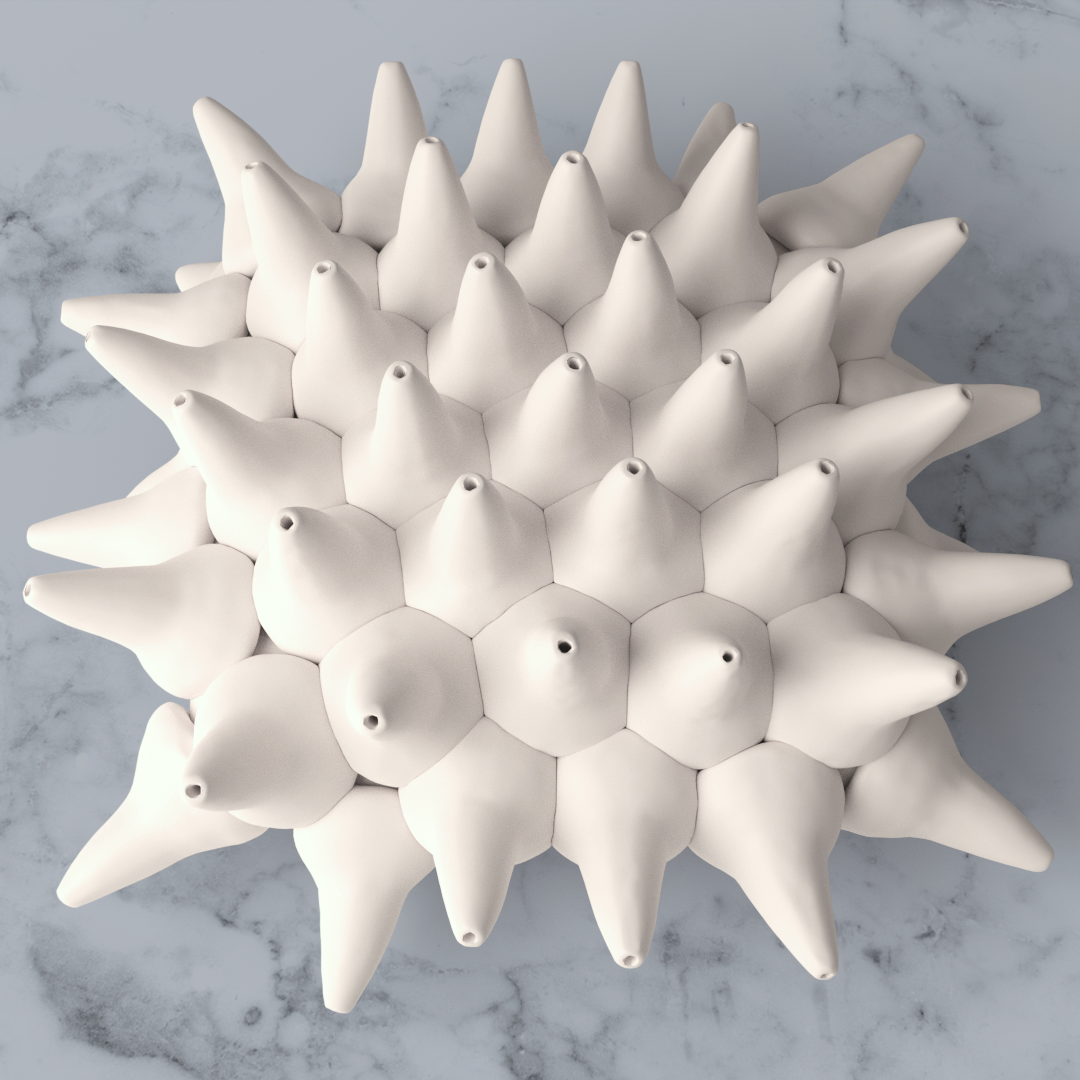} \\
\includegraphics[width=.43\linewidth]{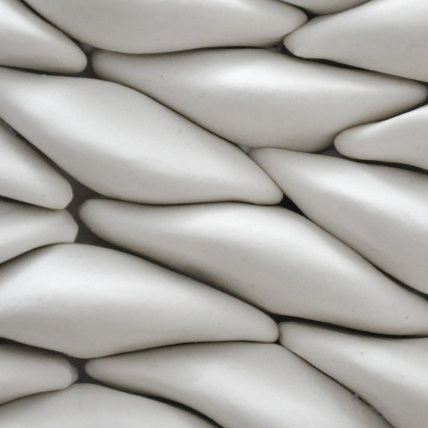} &
\includegraphics[width=.43\linewidth]{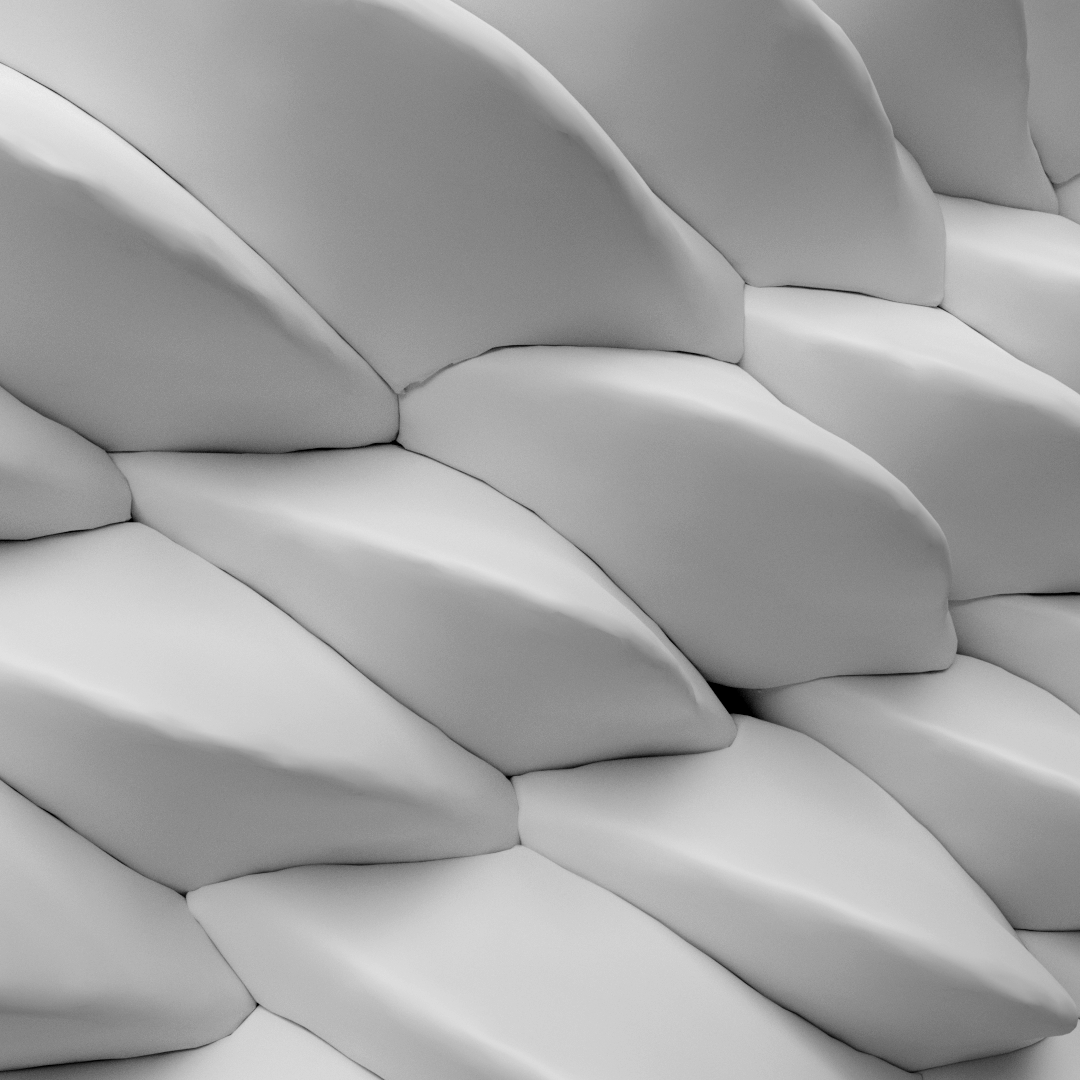} \\
\textsf{Real-world objects} & \textsf{\emph{PAVEL} replica} \\
\end{tabular}
\caption{Comparison of element-based clay handcrafts created by 
an artist (left column) and the corresponding rendered models 
created by \emph{PAVEL} using automatic seeds placement (right column).}
\label{fig:gallery1}
\end{figure}

\paragraph{Reproducing real-world decorations}
Using \emph{PAVEL}, we can create a variety of decorations that mimic real-world artworks. \cref{fig:gallery1} shows side-by-side views of original artworks, courtesy of Heather Knight, and objects generated with our pipeline and rendered with clay-like material properties. 

\begin{figure}
    \centering
    \includegraphics[width=.95\linewidth]{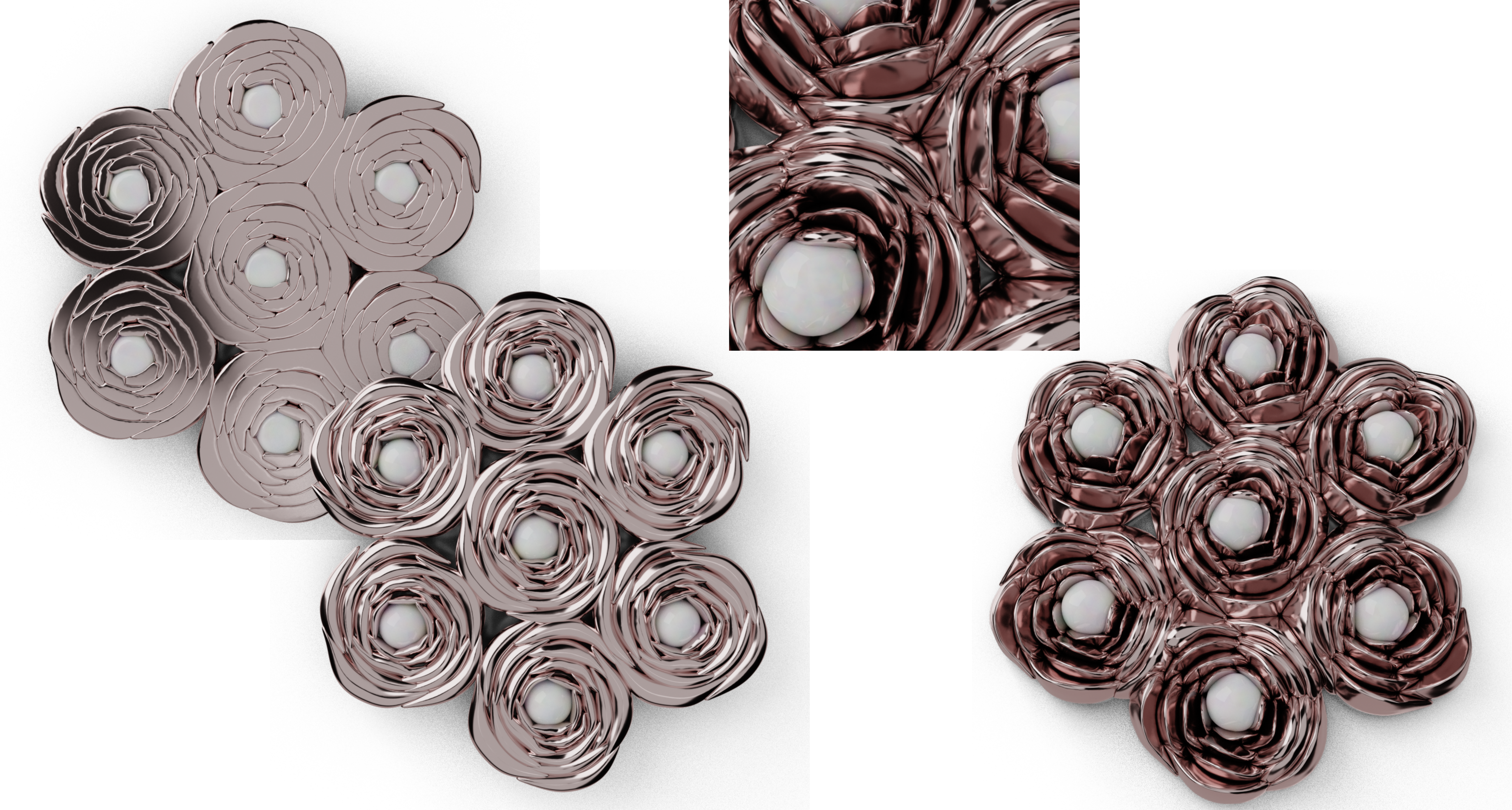}\\
    \includegraphics[width=.95\linewidth]{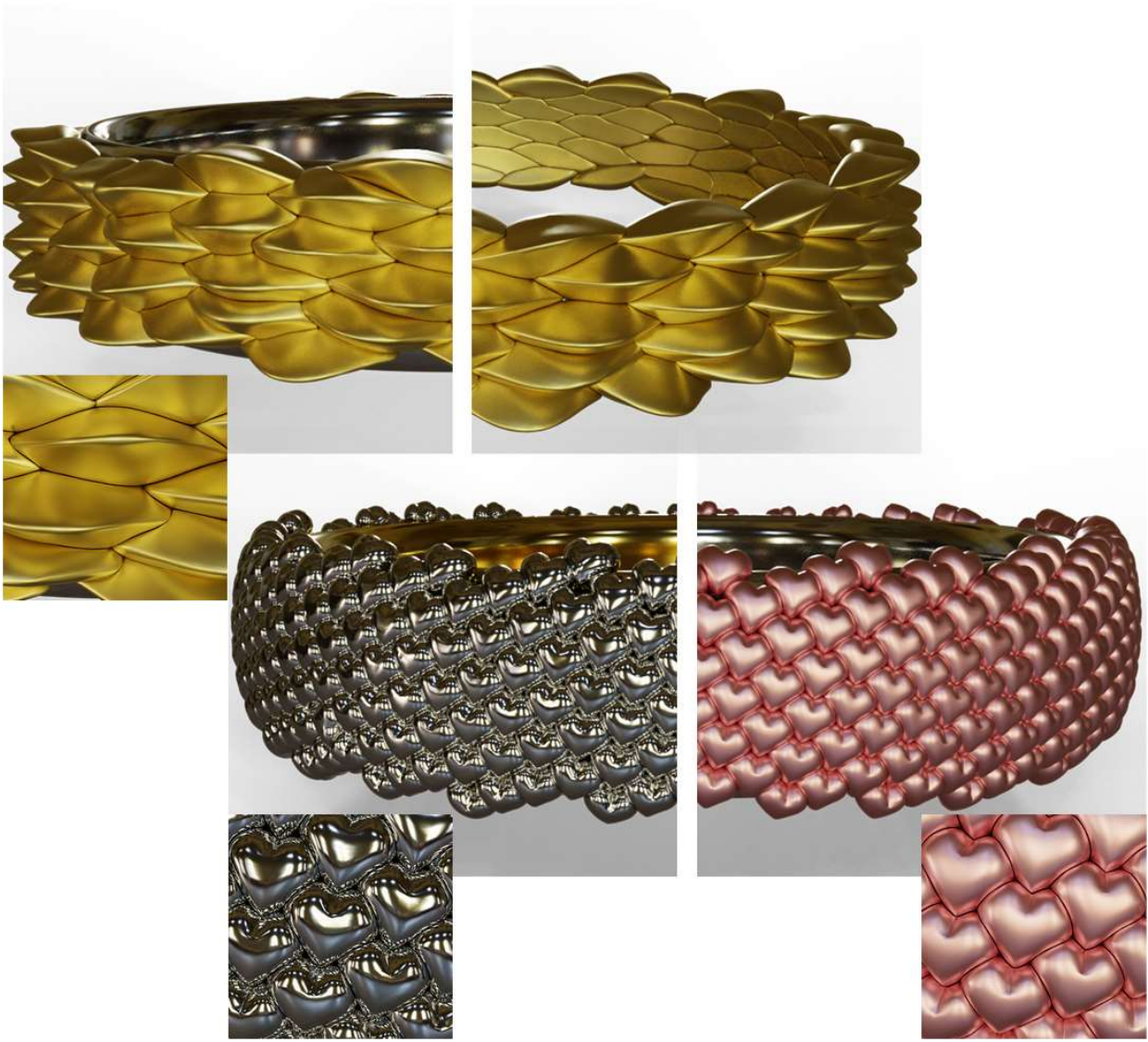}
    \caption{Some jewels designed with \emph{PAVEL}. On top: a decoration made of seven pink gold roses with a pearl in the middle; we use two different metal polishing for the final rendering. On the bottom part, there are three rings. We show details of the decorations and, for one of the rose decorations and the golden ring, also the packing of the decorations on the back, over the surface, visible removing the base shape.}
    \label{fig:jewels}
\end{figure}

\begin{figure}
    \centering
    \includegraphics[width=1\linewidth]{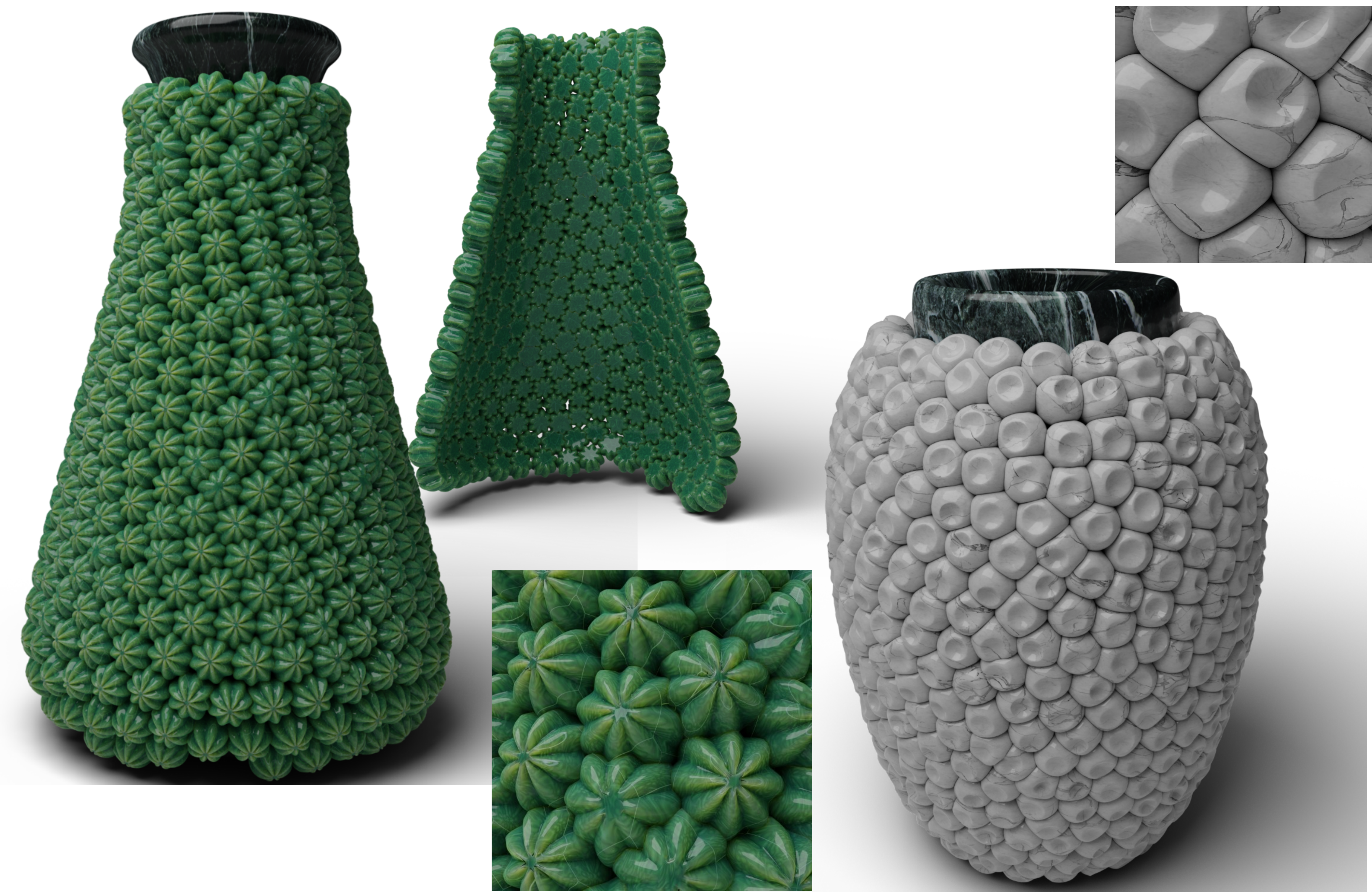}
    \caption{Two vases designed with \emph{PAVEL}. For the jade one we show also the back of one the parts that could be used for fabrication.}
    \label{fig:vases}
\end{figure}

\emph{PAVEL} decorations are not limited to the replicas of clay pieces but can be used to generate various artworks of different kinds. \cref{fig:jewels,fig:vases} shows a few examples, demonstrating potential uses in jewelry and pottery, exploiting manual elements positioning and automatic pattern creation.

\paragraph{Performance}
To understand if PAVEL could be a useful design tool, it is necessary to evaluate the time required by the different steps and the global effort required to create the decorated models. We obtain all our results on a 12-core Ryzen 3900x with 32GB of RAM.
\Cref{tab:sampling} shows the time required by the automatic seeding procedure.
All the times for the non-curvature adaptive methods ranges between 0.6 and 1.5s, with minimal differences between CVT and striped seeding. The most computationally expensive sampling strategy is, as expected, the curvature-dependent one. Our test took 54s to extract the isosurface but only 2.5s to sample it.
This speed allows a reasonably fast interactive design procedure. The digital artist can quickly test different object choices and seeding options visualizing the seeded configurations superimposed on the base model.

\Cref{tab:times} shows the times required by the volume recovery and mesh generation.
While the whole procedure is clearly not interactive, the time required is reasonable for an offline rendering procedure. It is not critically variable varying the surface coverage and the elements' volumes overlap within the parameters ranges discussed in \cref{sec:defo}.
The time required by the volume recovery step is directly dependent on both the number of decorations and their resolution. The times range from 20 to 360 seconds, with the faster times achieved by the models with around 50 decorative elements. Still, the method scales relatively well, handling nearly 2,000 decorations in just under 6 minutes. Another critical factor in performances is the resolution, which is also crucial to maintaining the decoration's details. Our results show how the method can handle millions of voxels to represent the decorations, even in intersections up to 33\% of the volume. 
Lastly, another time-consuming step is CVT smoothing. While it ranges from 10 to 200 seconds to remesh our test models, it's worth noticing that the marching cubes meshes' resolution reaches up to 40 million triangles, and we output smooth meshes with up to 4 million triangles.

\newcommand{\theader}[1]{\multicolumn{1}{c}{#1}}

\begin{table}[tb]
\addtolength{\tabcolsep}{-1pt}
    \centering
    \small
    \begin{tabular}{lrrr}
    \toprule
         \multirow{2}{*}{Model} & \theader{Number of} & \theader{Surface} & \multirow{2}{*}{Time} \\
         & \theader{decorations} & \theader{coverage} & \\
         \midrule
         Amphora, \cref{fig:anisotropicDecos2}& 358 & 1.42 & 0.59s \\
         Sphere, \cref{fig:teaser} left & 400 & 1.43 & 0.69s \\
         Vase, \cref{fig:rotdeco}& 953 & 1.27 & 0.91s \\
         Mug, \cref{fig:vardeco} & 410 & 1.42 & 0.68s \\
         Bunny (adaptive), \cref{fig:samplingcomparison} middle & 1720 & 1.23 & 2.68s \\ % (54.20) \\
         Teapot, \cref{fig:anisotropicDecos} & 642 & 1.26 & 1.82s \\
         Vase, \cref{fig:cvd} & 450 & 1.49 & 0.70s \\
         Glass, \cref{fig:fabricationGlass} & 500 & 1.33 & 0.74s \\
         Vase, \cref{fig:gallery1} top & 182 & 1.48 & 1.43s \\
         Tile, \cref{fig:gallery1} bottom & 65 & 1.32 & 1.30s \\
         Tile, \cref{fig:gallery1} middle & 48 & 1.49 & 0.80s \\
         Ring, \cref{fig:teaser} middle & 529 & 1.30 & 1.00s \\
         Vase, \cref{fig:vases} right & 604 & 1.38 & 0.79s \\
         Vase, \cref{fig:vases} left & 555 & 1.44 & 0.77s \\
         \bottomrule
    \end{tabular}
    \caption{Sampling parameters for the results obtained via automatic decoration placement. The reported time for the bunny is not inclusive of the isosurface generation that is 54.20s.}    \label{tab:sampling}
    \vspace{-1em}
\end{table}

\begin{table*}[tb]
\centering
\addtolength{\tabcolsep}{-1pt}
\small
\begin{tabular}{lrrrrrrrr}
\toprule
     \multirow{2}{*}{Model} & \theader{Deformation} & \theader{Marching cubes} & \theader{Decorations} & \theader{Overlapping}   & \theader{Base} & \theader{Marching Cubes} & \theader{Final} \\
   & \theader{time} & \theader{and Smoothing} & \theader{voxels} & \theader{voxels}  & \theader{voxels} & \theader{triangles} & \theader{triangles} \\
         \midrule
   Amphora, \cref{fig:anisotropicDecos2}             & ~112.5s & ~58.6s  & 31.68M  & ~30.2\%  & 64.71M  & 12340K & 716K   \\ 
   Sphere, \cref{fig:teaser} left                    & ~163.2s & ~124.3s & 49.41M  & ~12.8\%  & 41.39M  & 17862K & 2484K  \\ 
   Vase, \cref{fig:rotdeco}                          & ~208.6s & ~135.4s & 68.29M  & ~16.7\%~ & 418.13M & 25564K & 1985K  \\ 
   Mug, \cref{fig:vardeco}                           & ~121.7s & ~59.8s  & 43.23M  & ~17.0\%  & 200.23M & 14305K & 850K   \\ 
   Bunny (adaptive), \cref{fig:samplingcomparison} middle   & ~357.8s & ~138.6s & 96.24M  & ~13.7\%  & 204.00M    & 38487K & 1745K  \\ 
   Teapot, \cref{fig:anisotropicDecos}               & ~202.4s & ~101.6s & 54.27M  & ~32.3\%  & 414.82M & 22296K & 1386K  \\ 
   Vase, \cref{fig:cvd}                              & ~198.1s & ~158.7s & 67.72M  & ~18.4\%  & 94.80M   & 22616K & 2878K  \\ 
   Glass, \cref{fig:fabricationGlass}                & ~316.5s & ~101.2s & 127.36M & ~14.0\%  & 395.85M & 31279K & 1028K  \\ 
   Vase, \cref{fig:gallery1} top                     & ~68.6s  & ~41.8s  & 25.87M  & ~17.4\%  & 7.08M   &  7960K & 788K   \\ 
   Tile, \cref{fig:gallery1} bottom                  & ~23.7s  & ~13.7s  & 5.27M   & ~11.8\%  & 11.14M  &  2331K & 260K   \\ 
   Tile, \cref{fig:gallery1} middle                  & ~68.2s  & ~19.0s  & 5.71M   & ~13.5\%  & 6.68M   &  2156K & 295K   \\ 
   Mask,  \cref{fig:teaser} right                    & ~322.1s & ~74.7s  & 32.26M  & ~16.6\%  & 13.64M  &  8518K & 472K   \\ 
   Roses, \cref{fig:jewels} top                      & ~68.2s  & ~19.0s  & 5.61M   & ~24.6\%  & 1.64M   &  3405K & 325K   \\ 
   Ring, \cref{fig:jewels} middle                    & ~50.7s  & ~19.4s  & 10.70M   & ~21.1\%  & 10.35M  &  4675K & 287K   \\ 
   Ring, \cref{fig:teaser} middle                    & ~158.0s & ~83.3s  & 48.89M  & ~21.8\%  & 61.40M   & 17877K & 1095K  \\ 
   Vase, \cref{fig:vases} right                      & ~200.6s & ~98.2s  & 72.94M  & ~16.9\%  & 79.91M  & 23445K & 1249K  \\ 
   Vase, \cref{fig:vases} left                       & ~287.5s & ~204.2s & 64.90M   & ~17.1\%  & 61.49M  & 23437K & 3755K  \\ 
\bottomrule
\end{tabular}
\caption{We report the execution times (in seconds) for the two most expensive pipeline steps, deformation and output generation, for the set of models we discuss. For each model, we list the number of voxels of the base shape and the decorations (in millions), the number of overlapping voxels in the decorations, for which our system solves and deform, the number of triangles of the marching cubes' extracted mesh (in thousands), and the size of the final mesh of the decorations alone (in thousand triangles, excluding the base).}
\label{tab:times}
\vspace{-1em}
\end{table*}
\section{Discussion}
\label{sec:disc}

To the best of our knowledge, \emph{PAVEL} is the first attempt to propose a method to create packed volumetric decorations on 3D models. It is impossible to generate these decorations with either sculpting or texture mapping, and we work directly on positioning and deforming volumetric elements. Throughout the paper, we demonstrate that we can visually simulate real-world hand-crafted objects. We do so by placing decorative elements with point-sampled uniform or stripe-align distributions and packing elements with a volume-preserving deformation. The decorated models can be used for virtual applications of 3D printed for tangible reproductions. 

\paragraph{Limitations}
Our method has two significant limitations. First, decorations' automatic placement is limited to approximately isotropic elements or elongated shapes placed on stripes. As stated in Section \ref{sec:intro}, this is a willing design choice to simplify the packing procedure leaving to the deformation step to fill the gaps between elements. However, we plan to improve the initial packing step with novel options, adding a greedy refinement of elements' position and orientation after the initial placement to enhance the quality of the packing of non-radially symmetric elements.

The second limitation is our use of fast marching that limits the complexity of the deformed objects' details since the number of voxels quickly grows w.r.t. surface details. In the future, we plan to explore sparse representations based on spatial hashing and wide-branching trees that are of common use in 3D animation.

\paragraph{Future work}
In addition to the improvements listed above, we plan to improve the pipeline by adding more options to control the design, namely the possibility of performing automatic seeding for specific decorations on different parts of the mesh. 

Another planned work is the optimization of the shell decomposition for printing. We plan to develop methods for the automatic decomposition of the \emph{PAVEL} decoration in patches of controlled size and printer-friendly. We will also investigate the possibility of adding printability constraints in the fast marching evolution.

We also plan to better exploit the elements' deformation's parallelizability to obtain a faster implementation. We believe that it should be possible to parallelize both the computation of the elements deformations and the merging of the final result and obtain significant speed up in the case of large numbers of small decorations.

\bibliographystyle{ACM-Reference-Format}

\bibliography{bibliography}

\end{document}